\documentclass{aa}  

\usepackage{graphicx}
\usepackage{txfonts}
\usepackage{amsmath}
\usepackage[normalem]{ulem}
\usepackage{subcaption}
\usepackage{hyperref}
\usepackage{xcolor}
\usepackage{ulem}
\usepackage{dblfloatfix}
\usepackage[export]{adjustbox}
\usepackage{hyperref}
\hypersetup{
    colorlinks = true,
    citecolor ={cyan}
}
\def\rsun{{~R}_{\odot}}
\def\msun{{~M}_{\odot}}
\def\zsun{{~Z}_{\odot}}

\usepackage{color}
\bibpunct[; ]{(}{)}{;}{a}{}{;}

\definecolor{amber}{rgb}{1.0, 0.49, 0.0}
\definecolor{atomictangerine}{rgb}{1.0, 0.6, 0.4}

\newcommand{\logteff}{\ensuremath{\,{\rm log} (T_{\rm eff} / {\rm K})}}
\newcommand{\logL}{\ensuremath{\,{\rm log} (L / L_{\odot})}}

\usepackage{array}
\newcolumntype{L}[1]{>{\raggedright\let\newline\\\arraybackslash\hspace{0pt}}m{#1}}
\newcolumntype{C}[1]{>{\centering\let\newline\\\arraybackslash\hspace{0pt}}m{#1}}
\newcolumntype{R}[1]{>{\raggedleft\let\newline\\\arraybackslash\hspace{0pt}}m{#1}}

\begin{document} 

   \title{It has to be cool: on supergiant progenitors of binary black hole mergers from common-envelope evolution}
   \titlerunning{Supergiant progenitors of binary black hole mergers from common-envelope evolution}
   \authorrunning{Klencki et al.}

   \author{Jakub Klencki\inst{1}
          \and
          Gijs Nelemans\inst{1,2,3}      
          \and
          Alina G. Istrate\inst{1}
          \and 
          Martyna Chruslinska\inst{1}
   }

   \institute{
   Department of Astrophysics/IMAPP, Radboud University, P O Box 9010, NL-6500 GL Nijmegen, The Netherlands\\
          \email{j.klencki@astro.ru.nl}
         \and
          Institute of Astronomy, KU Leuven, Celestijnenlaan 200D, B-3001 Leuven, Belgium
         \and
         SRON, Netherlands Institute for Space Research, Sorbonnelaan 2, NL-3584 CA Utrecht, The Netherlands
   }
   \date{Received June, 2020; accepted ???}

  \abstract
   {Common-envelope (CE) evolution in massive binary systems is thought to be one of the most promising channels for
   the formation of compact binary mergers. In the case of merging binary black holes (BBHs), the essential CE phase
   takes place at a stage when the first BH is already formed and the companion star expands as a supergiant.
   We study which BH binaries with supergiant companions will evolve through and potentially survive a CE phase. To
   this end, we compute envelope binding energies from detailed massive stellar models at different evolutionary stages
   and metallicities. We make multiple physically extreme choices of assumptions that favor easier CE ejection as well
   as account for recent advancements in mass transfer stability criteria.

    We find that even with the most optimistic assumptions, a successful CE ejection in BH binaries is only possible if
    the donor is a massive convective-envelope giant, i.e. a red supergiant (RSG). The same is true for neutron star
    binaries with massive companions. In other words, pre-CE progenitors of BBH mergers are BH binaries with RSG companions.
    We find that due to its influence on the radial expansion of massive giants, metallicity has an indirect but a very 
    strong effect on the chemical profile, density structure, and the binding energies of RSG envelopes.
    Our results suggest that merger rates from population synthesis models could be severely overestimated, especially at
    low metallicity. Additionally, the lack of observed RSGs with luminosities above $\logL \approx$ 5.6$-$5.8, corresponding
    to stars with $M \gtrsim 40 \msun$, puts into question the viability of the CE channel for the formation of the most
    massive BBH mergers. Either such RSGs elude detection due to very short lifetimes, or they do not exist
    and the CE channel can only produce BBH systems with total mass $\lesssim 50 \msun$. Finally, we 
    discuss an alternative CE scenario, in which a partial envelope ejection is followed by a phase of possibly long
    and stable mass transfer.

   }
  
   \maketitle

\section{Introduction}
\label{sec:intro}

Since the discovery of the first gravitational wave (GW) signal from a binary black hole (BBH) coalescence
by the Advanced LIGO Interferometer in September 2015 \citep[GW150914,][]{Abbott16_firstBBH}, 
the LIGO/Virgo Collaboration has reported the detection of nine further BBH mergers by the end of its second observing run O2 \citep{LIGO_o1o2sum}.
The third observing run O3 has been concluded and 
a large number of publicly issued alerts is an indication that a few tens additional detections of BBH mergers are on the way, 
including the recently published discoveries of GW190412 \citep{LVK_GW190412}, GW190814 \citep{LVK_GW190814} 
(being either a BBH or a black hole-neutron star, BH-NS, merger), and GW190521 \citep{LIGO_GW190521}.
With the growing population of BBHs, the discussion on possible formation scenarios of 
compact binary mergers is as lively as ever. A large number of channels have been put forth, especially in the case of BBHs.
These include but are not limited to the formation from isolated binaries through common envelope (CE) evolution
\citep[][]{Dominik2012,Mennekens2014,Belczynski2016,Eldridge2016,Klencki2018,Mapelli2018,Kruckow2018,Breivik2020} 
or in chemically homogeneous evolution regime \citep{Mandel2016,deMink2016,Marchant2016},
dynamical formation in globular clusters \citep{Rodriguez2016a, Askar2017,Samsing2018}, in nuclear clusters \citep{ArcaSedda2018,Fragione2019},
or in disks of active galactic nuclei \citep{Antonini2016, Stone2017, McKernan2018}, 
as well as formation channels involving triple \citep{Antonini17} or quadruple stellar systems \citep{Fragione2019b}. 
So far, it has not been possible to distinguish between various channels based on the gravitational 
wave information alone. In particular, the promising method of distinguishing between dynamical and isolated binary formation 
based on the BBH spin-orbit misalignment distribution \citep{Farr2017,FarrHolzFarr2018} is hindered by our lack of
knowledge of the natal black hole (BH) spins and, to a lesser extend, the possibility of BH natal kicks changing the spin orientation
\citep{OShaughnessy2017,Gerosa2018,Belczynski2020,Bavera2020}. As a result, the contribution of 
various channels to the entire population of BBH mergers is usually estimated on theoretical grounds \citep{Abadie2010,Barack2019}.
The CE evolution channel is sometimes considered to be especially promising 
thanks to its potential to produce a relatively high merger rate of BBHs compared to other channels, although any 
rate prediction from theoretical population models are highly uncertain.

The essential stage in the CE evolution channel is a dynamically unstable phase of mass transfer
that leads to a rapid spiral-in of the companion object inside the shared envelope originating from the giant donor star \citep{Paczynski1976,Webbink1984,Iben1993, Podsiadlowski2001,Ivanova2013}. 
The drag force is thought to cause a dramatic shrinkage of the binary separation and the dissipated orbital
energy to lead to an ejection of the CE (under the right circumstances) or a merger otherwise.
The huge range in both timescales and length scales involved in this complex process makes hydrodynamic simulations challenging \citep[eg.][]{Ricker2012,Passy2012,Nandez2016,MacLeod2017,Fragos2019}.
As a result, the exact outcome of the CE phase is difficult to predict. There is, however, a substantial amount of
evidence for significant orbital shrinkage in progenitors of various short-period systems such as cataclysmic
variables \citep[e.g.][]{Paczynski1976,Meyer1979}, binary white dwarfs \citep[e.g.][]{Han1998,Nelemans2000,Nelemans2001}, 
or binary neutron stars \citep[BNSs, e.g.][]{vdHeuvel1994,Tauris2006,Chruslinska2018,VignaGomez2020}, which cannot be explained without invoking a mechanism such as the CE phase. 

In the case of more massive stars, the progenitors of stellar BHs ($\gtrsim 20 \msun$), such observational support is more
difficult to find and comes mainly from the population of short-period ($<$ 1 day) X-ray binaries with stellar BH accretors
and low-mass ($\lesssim 1 \msun$) donors \citep[BH-LMXBs;][]{Casares2014}, which are believed to have formed through CE evolution \citep{PortegiesZwart1997,Kalogera1999}.
High-mass X-ray binaries hosting stellar BHs and Wolf-Rayet (WR) companions in short-period orbits of a few hours up to two days such
as Cygnus X-3, IC10 X-1, and NGC300X-1 may be products of the CE evolution as well \citep{Lommen2005,Carpano2007}, although
it is debated whether or not they could originate from stable mass transfer evolution instead \citep{vdHeuvel2017}.
Notably, such systems could be the immediate progenitors of merging BBH or BH-NS systems \citep{Bulik2011,Belczynski2013}.
On top of that, hydrodynamic simulations of the CE phase are particularly challenging in the case of massive stars \citep[see][for details]{Ricker2019} 
and most of the hitherto results have been limited to low-mass giants, WD progenitors. 

Motivated by the LIGO discoveries of the BBH mergers GW150914 and GW151226, \citet{Kruckow2016} analyzed the prospect of 
successful CE evolution in binaries of stellar BHs and massive companions (i.e. potential BBH progenitors) by considering
the energy balance between the envelope binding energy and the available energy sources \citep{vdHeuvel1976,Webbink1984, Ivanova2013}.
They concluded that for the right binary parameters the CE evolution channel may indeed operate for massive stars in a similar
way to low- and intermediate-mass donors,
and that it may produce BBH systems with individual BH 
masses all the way up to lower edge of the pair-instability supernova 
(PISN) BH mass gap (at $\sim 45-55 \msun$, e.g. \citealt{Heger2003,Woosley2017,Leung2019,Renzo2020}, 
although the limit could perhaps be moved to higher masses within the uncertainty of the $^{12}{\rm C} (\alpha , \gamma) ^{16}{\rm O}$ reaction 
rate, \citealt{Farmer2019}).

Here, we extend the analysis of \citet{Kruckow2016} by additionally accounting for conditions necessary for the mass transfer to become 
dynamically unstable (i.e. for the occurrence of CE evolution). Such conditions can usually be formulated 
in the form of a threshold mass ratio, such that for mass ratios $q = M_{\rm donor} / M_{\rm accretor}$ above a critical 
value $q_{\rm crit}$ the mass transfer becomes unstable, whereas for $q < q_{\rm crit}$ the CE evolution is avoided. 
Importantly, the mass ratio $q$ plays a significant role in considerations of the CE energy budget: 
the lower the $q$ (i.e. the more equal-mass system) the easier the CE ejection \citep[e.g. Fig.~6 of][]{Kruckow2016}. 
Recent studies of mass transfer stability from massive giants reveal that the mass transfer 
remains stable for a larger parameter space than previously thought, thus avoiding a CE evolution in the majority of cases \citep{Woods2011,Pavlovskii2017}.
It is therefore essential to consider realistic $q_{\rm crit}$ values when addressing the question of CE ejectability in BBH progenitors. 

In Sec.~\ref{sec:model} we describe the ingredients of our model: (a) detailed stellar models of massive giants for six metallicities 
between $Z = 0.017 = \zsun$ and $Z = 0.00017 = 0.01 \zsun$, (b) conditions for mass transfer instability and the occurrence of CE evolution, as well as (c) assumed contribution of various energy sources and 
sinks to the overall energy budget of the CE evolution. In Sec.~\ref{sec:results} we present our calculated envelope
binding energies, highlighting the substantial 
impact of outer convective envelopes. Furthermore, we compute the parameter space for successful CE ejections and
explore the impact of various assumptions. Finally, we explore the predicted population of CE survivors in the
Hertzprung-Russell (HR) diagram. We discuss various aspects of our findings in Sec.~\ref{sec:discussion} and conclude in Sec.~\ref{sec:conclusions}.


\section{The model}
\label{sec:model}

\subsection{Stellar models of massive giants}

\label{sec.method_models}

We use rotating single stellar models from \citet{Klencki2020} computed with the MESA stellar evolution code \citep{Paxton2011,Paxton2013,Paxton2015,Paxton2018,Paxton2019}.
The use of single models unperturbed by previous binary interactions is a simplification because in most realistic scenarios the stellar companion in a BH binary 
is a mass gainer from a mass transfer phase that was initiated by the BH progenitor \citep[also in the case of BBH merger progenitors, e.g.][]{Belczynski2016}. 
While this has become a common practice, it is an important caveat to mention. We will discuss this topic further in connection to those results that are likely 
the most sensitive to the assumption of single stellar models (see Sec.~\ref{sec:res_details} and Sec.~\ref{sec.disc_uncert}.

The tracks from \citet{Klencki2020} cover a range of masses from $10$ to $80 \msun$ and
six metallicity values $Z$ = 0.017, 0.0068, 0.0034, 0.0017, 0.00068, and 0.00017 (or, in fractions of the solar metallicity: $1.0$, $0.4$, $0.2$, $0.1$, $0.04$, and $0.01 \zsun$, 
where we assume $\zsun = 0.017$ after \citealt{Grevesse1996}), 
with the initial rotation rate set to 40\% of the critical value ($\Omega / \Omega_{\rm crit} = 0.4$).
We note that for higher initial rotation rates massive stars of $M \gtrsim 50 \msun$ at very low metallicities were shown to evolve in a chemically-homogeneous
way \citep[e.g.][]{Marchant2017}. 
Convection was modeled using the mixing-length theory \citep{BohmVitense1958} with a mixing-length of $\alpha_{\rm ML} = 1.5$ 
and following the Ledoux criteria for convection with a high efficiency of semiconvective mixing $\alpha_{\rm SC} = 100$. The models assume
step overshooting above the hydrogen and helium burning cores with an overshooting length of $0.345$ pressure scale heights \citep{Brott2011}. 
The wind mass-loss was modeled following \citep{Brott2011} as a combination of mass-loss recipes from \citet{Nieuwenhuijzen1990},\citet{Hamann1995}, and \citet{Vink2001}.
Notably, the models from \citet{Klencki2020} were computed under a set of assumptions that maximizes the potential of massive stars to reach the
sizes of red-supergiants and to develop deep outer convective envelopes: with no LBV-like mass-loss above the Humphreys-Davidson limit
as well as without preventing the formation of density inversions by the use of MLT++ in MESA (see Appendix~A in \citealt{Klencki2020}). Such an approach 
allows us to explore the maximum potential parameter space for binary interactions in wide binaries and,
consequently, the maximum parameter space for CE evolution. In the case of convective-envelope stars with initial masses
$\gtrsim 40 \msun$ the true parameter space might be significantly smaller, as discussed in Sec.~\ref{sec.disc_Lrsgmax}.
See \citet{Klencki2020} for a description of other physical ingredients of the models and \url{https://zenodo.org/record/3740578#.X1s2cnVfguw} for the MESA input files.
In App.~\ref{sec.App_CE_acc} we demonstrate the models are numerically robust for the purposes of this work.

\subsection{Conditions for mass transfer instability}

\label{sec.method_qcrit}

Mass transfer in a binary can become dynamically unstable when either the donor or the accretor (or both) expands too much
in size with respect to its Roche-lobe. It was shown that if the timescale of mass transfer 
is short compared to the thermal timescale of the mass gainer then the accreting star could be driven out of thermal equilibrium
and expand significantly \citep[e.g.][]{Benson1970,Neo1977}, filling its Roche-lobe and leading to the formation of a contact binary \citep{Pols1994,Wellstein2001}
and potentially a CE evolution \citep{deMink2007,Marchant2016}. Details of this process remain largely uncertain, 
partly because of the complicated gas dynamics in semi-detached binaries \citep{Lubow1975}, poorly constrained specific entropy of the accreted 
material \citep{Shu1981}, and unknown efficiency of accretion by stars rotating near their breakup limit \citep{Popham1991}, see also discussion by \citet{deMink2013}.

The focus of this study, however, is mass transfer in the case of compact-object accretors: stellar BHs or NSs.\footnote{We note that
double-core CE evolution, when both the donor and the accretor are evolved giants, could also lead to the formation of compact binary mergers, 
although this channel requires both stars to be of very similar mass \citep[e.g.][]{Bethe1998,Dewi2006}.}
Based on X-ray binaries such as Cyg X-2 \citep{King1999} and SS433 \citep{Fabrika2004}, 
it is believed that unlike a stellar accretor, a BH or a NS can deal with very high mass transfer rates via jets and disk outflows
which prevent the non-accreted matter from piling up and eventually overflowing their Roche-lobes. 
For that reason, for the rest of this section we will discuss mass transfer instability due to an increasing 
Roche-lobe overflow by the donor star. As explained above, in the case of stellar accretors
the mass transfer could be unstable in many more cases.

Donor stars with outer radiative envelopes (the case that dominates the parameter space) respond to mass loss by 
contraction on the adiabatic time-scale \citep{Hjellming1987,Soberman1997}. In other words, they are characterized by a 
positive value of $\zeta_{\rm ad} = (\partial \, {\rm log \,} R / \partial \, {\rm log \,} M)_{\rm ad}$. In such a 
case, 
the mass transfer can become unstable on the adiabatic timescale only when the size of the Roche-lobe is shrinking more 
quickly than the size of the radiative donor, which is possible if the mass ratio $q = M_{\rm donor} / M_{\rm accretor}$ is 
higher than some critical value $q_{\rm crit;rad}$. Traditionally, radiative-envelope 
post-main sequence (MS) giants were assumed to be described by $\zeta_{\rm ad} \approx 6.5$, which corresponds to $q_{\rm crit} \approx 
3.5-4$ \citep{Tout1997,Hurley2002}: a value that is often assumed \citep[eg.][]{Schneider2015,vdHeuvel2017,VignaGomez2018}. Detailed models of mass transfer from 
intermediate-mass radiative-envelope donors ($\sim 1$ to $6 \msun$) also found stability for mass ratios of up to about 
$4$ \citep[eg.][]{Tauris2000,Chen2002,Ivanova2004}. Observationally, the well-studied SS433 system, that 
consists of a Roche-lobe filling A-type supergiant ($\sim 12.3 \msun$) and a stellar BH ($\sim 4.3 \msun$) in a $13.1$ 
day period orbit \citep{Hillwig2008}, is a likely case of stable mass transfer evolution that has been continuing for 
$\sim 10^5$ yr already \citep[with the likely initial value of $q \approx 3-4$][]{King2000,Begelman2006,vdHeuvel2017}.
Notably, mass transfer from MS donors was found to be less stable with typical $q_{\rm crit;MS} \approx 1.5$ \citep{deMink2007}.

A single value of $\zeta_{\rm ad}$ across the entire mass and radius spectrum of post-MS stars is a serious
simplification. \citet{Ge2015} computed adiabatic responses for a large grid of models (donor masses between $0.1$ 
and $100 \msun$), also taking into account the possibility of a delayed dynamical instability \citep{Ge2010}. The 
authors found that the value of $\zeta_{\rm ad} = 6.5$ and the critical mass 
ratio $q_{\rm crit;rad}$ between $3$ and $5$ could be a good approximation in the case of stars below $10 \msun$. 
However, for more massive post-MS stars they find a higher $q_{\rm crit;rad}$  of at least $5$, and even $q_{\rm 
crit;rad} > 10$ for $R > 300 \rsun$ (going up to extreme $q_{\rm crit;rad} > 20$ for $R > 1000 \rsun$). 

Donor stars with outer convective envelopes, on the other hand, are expected to respond to mass loss by expanding on 
the adiabatic time-scale. This makes the mass transfer from convective donors much more
prone to dynamical instability than from radiative donors. Many authors have relied on condensed polytrope models of 
\citet{Hjellming1987} to obtain the value of $\zeta_{\rm ad}$ for convective-envelope donors, which typically leads to 
critical mass ratios $q_{\rm crit;conv}$ of about $0.8$.

It is important to realize the caveats behind the numbers cited above.
The value of $\zeta_{\rm ad}$ derived in the adiabatic 
approximation \citep[eg.][]{Hjellming1987,Tout1997,Ge2015} is valid for predicting the donor behavior only if the 
outer envelope thermal timescale is much longer than adiabatic timescale (i.e. the entropy profile can 
be considered constant). This is not necessarily the case for outer envelopes of massive stars with large radii. 
\citet{Woods2011} showed that in the case of convective donors the thermal timescale in the envelope's outermost 
superadiabatic layer (which forms on top of the convective zone) can become even smaller than the dynamical timescale. 
They found that thermal relaxation of such outer layers can effectively increase the stability of mass transfer (by up 
to a factor of $\sim 1.7$ in $q_{\rm crit;conv}$ in the limited number of cases the authors have 
investigated).\footnote{Another issue with applying condensed polytropes to compute $\zeta_{\rm ad}$ of 
convective donors as in \citet{Hjellming1987} is that in massive stars the superadiabatic layer is large enough to 
have to be taken into account and consequently the n = 3/2 polytrope is no longer a valid approximation of the entropy 
profile \citep{Woods2011,Pavlovskii2015}.} In the case of massive radiative donors, a qualitatively opposite 
effect of decreased stability is expected \citep[see Sec.~4.1 of][]{Ge2015}, which is why the values of $q_{\rm crit;rad} > 10$ 
are most likely overestimated. 

The fact that thermal relaxations needs to be taken into account means that mass transfer stability from 
massive giants cannot be reliably predicted without detailed evolutionary calculations. Some of such results are 
already available. Using 1-D stellar codes with hydrodynamic terms to probe the rapid donor response, both 
\citet{Passy2012} and \citet{Pavlovskii2015} found that the initial reaction of convective-envelope stars to mass loss
is a slight contraction, contrary to what the adiabatic models predict. This increases the stability, in line with the 
conclusions of \citet{Woods2011}. Taking overflow through the outer Langrangian point as instability criteria, 
\citet{Pavlovskii2015} obtained revised critical mass ratios for convective donors $q_{\rm crit;conv} = 1.5-2.2$, 
computed for models of up to $50 \msun$ in which the outer convective envelope is well developed ($M_{\rm conv} \gtrsim 0.3 M_{\rm donor}$). 
For donors with less developed convective envelopes \citet{Pavlovskii2015} obtain stability 
up to the highest mass ratio in their grid, $q = 3.5$, though they do not specify the exact $M_{\rm conv}$ value. 
In a recent work focused on intermediate-mass donors ($M_{\rm donor}$ between 1 and 8 $\msun$) \citet{Misra2020} 
found stable mass transfer evolution from convective-envelope donors for mass ratios up to $q \sim 2$.

In the case of massive radiative-envelope donors with $M_{\rm donor}$ between $20$ and $80 \msun$ (or donors with 
shallow outer convective envelopes), \citet{Pavlovskii2017} found that the previously accepted stability criteria 
should also be revised. By computing detailed models of mass transfer in binary systems with BH accretors, they 
consistently obtained stable mass transfer for mass ratios as high as $6-8$ from $M_{\rm donor} \geq 40 \msun$ 
donors.\footnote{In the case of $M_{\rm donor} < 40 \msun$, \citet{Pavlovskii2017} only tested mass ratios up to $4$.} 
This further supports the trend that $q_{\rm crit;rad}$ is typically larger 
for more massive stars ($M > 10 \msun$) than the $3.5-4$ critical mass ratios for intermediate mass giants. 
In an upcoming paper (Klencki et al. in prep.) with detailed binary models, we obtain stable mass transfer from massive radiative donors for mass ratios up to $q \approx 5$. 

It is clear that the problem of mass 
transfer stability from massive giants is far from being fully understood. In particular, detailed 1-D mass transfer 
simulations have to rely on approximate methods for computing the mass transfer rate through the $L_1$ nozzle 
\citep[eg.][]{Kolb1990}. One obstacle going forward is that long-term 3-D mass transfer simulations (i.e. not just the 
initial donor response) are unlikely to be possible in the near future. At the same time, however, there is a 
growing number of results indicating that mass transfer from massive radiative-envelope donors might be
stable for mass ratios at least as high as $\sim 5$.

\subsection{Energy budget criterion for a successful CE ejection}

\label{sec.method_energy_budget}

We estimate the fate of the CE phase based on the energy formalism (\citealt{vdHeuvel1976,Tutukov1979,Iben1984,Webbink1984,Livio1988},
see also \citealt{deMarco2011,Ivanova2013} for more recent reviews of this formalism), 
in which the difference in orbital energies between the initial and the final state, $\Delta E_{\rm orb}$, multiplied 
by an efficiency parameter $\alpha_{\rm CE}$ is equated to the energy needed to unbind the envelope $E_{\rm bind}$:
$E_{\rm bind} \approx \alpha_{\rm CE} \Delta E_{\rm orb}$. We extend this energy budget by including an additional term 
of energy feedback from accretion of matter by the spiraling in BH: $\Delta E_{\rm acc}$. The full equation takes the 
following form.

\begin{equation}
\label{eq.energy_budget}
\begin{aligned}
 E_{\rm bind} & = \Delta E_{\rm acc} + \alpha_{\rm CE} \Delta E_{\rm orb} \\
	      & = \Delta E_{\rm acc} + \alpha_{\rm CE} \Bigg( -\frac{G M_{\rm donor} M_{\rm BH}} {2 a_{\rm i}} + \frac{G M_{\rm core} M_{\rm BH}} {2 a_{\rm f}}  \Bigg)
 \end{aligned}
\end{equation}
Here, $M_{\rm donor}$ is mass of the giant donor that initiates the CE phase, 
$M_{\rm core}$ is the mass of its core that becomes the remnant after a successful envelope ejection, $M_{\rm BH}$ is the mass of the BH accretor,
$a_{\rm i}$ and $a_{\rm f}$ is the separation at the onset and at the end of the CE phase, respectively.
The $\alpha_{\rm CE}$ parameter accounts for the fact that not all the orbital energy can be deposited into the envelope
without any losses and as such takes a value between $0$ and $1$.
Notably, $\alpha_{\rm CE} = 1.0$ is quite extreme as it assumes no energy loss in any form from the system.
For comparison, in their 3-D hydrodynamic simulations of the dynamical phase of CE evolution from low-mass giants
(typically spanning several years) \citet{Nandez2015,Nandez2016} found that usually about 30\%$-$40\% of the orbital energy leaves the system as residual 
(mainly kinetic) energy of the unbound ejecta. In 1-D synthetic models of the potential longer-lasting (10$-$1000 years)
self-regulated phase, during which most of the input heat is radiated away, \citet{Clayton2017} found $\alpha_{\rm CE}$ values in the range 0.046$-$0.25.
In the same time $\alpha_{\rm CE} \approx 2$ was inferred from 1-D simulation by \citet{Fragos2019} in the case of NS accretors 
and $\alpha_{\rm CE}$ as large as $5$ (or even larger) is often used in population synthesis, especially that of compact binary mergers 
\citep[e.g.][]{Mapelli2018}. In Sec.~\ref{sec.disc_uncert} we further discuss the uncertainties in the value of $\alpha_{\rm CE}$ together with possibilites of 
$\alpha_{\rm CE} > 1$ considered in the literature.

Following \citet{Voss2003} and \citet{Kruckow2016}, we estimate the energy feedback from accretion as the Eddington 
luminosity of the BH ($\dot{L}_{\rm Edd}$) multiplied by the CE duration $\tau_{\rm CE}$:
\begin{equation}
\label{eqn.Eacc}
\Delta E_{\rm acc} = \dot{L}_{\rm Edd} \tau_{\rm CE} \approx 1.6 \times 10^{48} {\rm erg} \, \Bigg(\frac{M_{\rm BH}}{M_{\odot}}\Bigg) \, \Bigg(\frac{\tau_{\rm CE}}{1000 {\rm yr}}\Bigg)
\end{equation}
We discuss this approach in view of hydrodynamic models of accretion during CE in Sec.~\ref{sec.disc_uncert}.

To compute the energy needed to unbind the envelope $E_{\rm bind}$ we take three terms into account:
the energy needed to overcome the gravitational potential of the envelope $-E_{\rm grav}$, lowered by the
thermal energy stored in the envelope $U_{\rm th}$ \citep[kinetic energy of particles and the energy stored in radiation,][]{Han1994}
as well as the recombination energy $E_{\rm rec}$ available 
if all the ions recombine into atoms and atoms associate into molecules \citep[the latter is mostly relevant for $H_2$, ][]{Ivanova2015}.
Therefore, $E_{\rm bind}$ is calculated as:

\begin{equation}
\label{eq.Ebind}
\begin{aligned}
 E_{\rm bind} & =  -E_{\rm grav} - U_{\rm th} - E_{\rm rec} \\
	      & = - \int_{core}^{surface} \Bigg( - \frac{G M(r)}{r} + u \Bigg) \,{\rm d}m
 \end{aligned}
\end{equation}
where $u$ is the internal energy (both thermal and recombination energy) per unit mass.
In MESA the value of $u$ in each layer is taken from the tabulated equation of state \citep[Sec. 4.2 of][]{Paxton2011}.
We note that it might be more physically accurate to treat the energy from recombination as a separate 
energy source rather then to include it into the envelope binding energy because this energy is not available immediately, 
and its release must be triggered at a later stage of the CE phase \citep{Ivanova2015}. 
It is also not clear how much of this energy can in practice be used to help eject the envelope \citep{Nandez2015}.
However, similarly to \citet{Wang2016}, we opt to include $E_{\rm rec}$ in $E_{\rm bind}$ so that 
the binding energy computed this way combines all the terms derived from the structure of the giant donor.
We note that the contribution of the recombination energy to the total
internal energy of massive giant envelopes is relatively small \citep[$< 10\%$, see ][]{Kruckow2016}.

An important choice that has to be made when computing $E_{\rm bind}$ is a choice of the
boundary between the core (which will become the remnant of the giant donor)
and the ejected envelope, the so called bifurcation point. \citet{Tauris2001} showed that because most of the binding energy is located
in deep envelope layers close to the helium core, the exact choice of the bifurcation point
can have a substantial impact on the calculated binding energy. A number of criteria for the core-envelope boundary 
during the CE ejection have been proposed in the literature \citep[see Sect.~4 of ][]{Ivanova2013}. In particular, 
\citet{Ivanova2011} suggested that the bifurcation point location can be associated with the point of maximum 
compression $M_{\rm cp}$ within the H-burning shell, i.e. where the value of $P/\rho$ has a local maximum.
They argued that for masses $M > M_{\rm cp}$ the remnant would re-expand on a thermal timescale after the CE ejection 
or possibly still during the CE phase itself, either causing another phase of mass transfer or prolonging the CE evolution, 
ultimately leaving behind a remnant with $M \approx M_{\rm cp}$.\footnote{It is worth pointing out that \citet{Ivanova2011}
emulated the CE evolution by studying adiabatic mass-loss sequences, thus preventing any thermal relaxation of the donor star. If, however, 
a CE phase is characterized by the thermal-timescale then the maximum compression point could move during the CE event with respect to its pre-CE location.}
\citet{Kruckow2016} found that in most giants
the location of the maximum-compression point can be approximated by the mass coordinate where $X_{\rm H} = 0.1$ and used that
as their criterion for the bifurcation point. Here we follow their approach. See Sect.~\ref{sec.disc_composite_scenario} for a further discussion.

Having computed $E_{\rm bind}$ from stellar models and assumed the value for $\alpha_{\rm CE}$, we can use Eqn.~\ref{eq.energy_budget} 
to calculate the post-CE separation $a_{\rm f}$ of a binary with a given BH mass. A successful CE ejection takes place if 
the remnant core of the giant donor is smaller then its Roche lobe in the post-CE binary. We estimate the size of the remnant $R_{\rm remnant}$
to be equal to the outer radial coordinate of the core in the pre-CE model of the giant donor multiplied by a factor of 2
as guided by \citet[][]{Ivanova2011} to account for the fact that the compressed core expands in radius during the CE phase itself.

Finally, for practical purposes, the envelope binding energy is often written as \citep{deKool1990}:
\begin{equation}\label{eq: binding energy}
  E_{\rm bind} = \frac{G M_{\rm donor} M_{\rm env}} {\lambda_{\rm CE} R_{\rm donor}}
\end{equation}
where $R_{\rm donor}$ is the radius of the giant donor, $M_{\rm env}$ is the mass of its envelope 
($M_{\rm env} = M_{\rm donor} - M_{\rm core}$), and $\lambda_{\rm CE}$ is a parameter 
that depends on the detailed envelope structure. In Appendix~\ref{app:fits} 
we provide fits to $\lambda_{\rm CE}$ values computed in this work, with application to population synthesis.

\subsection{Model summary: the 'optimistic' choice of assumptions}

\label{sec.model_summary}

Our goal is to find to what extend can BH binaries with massive stellar companions evolve through and survive a CE phase. Therefore, whenever possible, we make
choices of assumptions that would facilitate an easier CE ejection. For clarity, these choices are summarized below
(see Sec.~\ref{sec.method_qcrit} and Sec.~\ref{sec.method_energy_budget} for details). 
We refer to Sec.~\ref{sec.disc_uncert} for a discussion of different assumptions and other caveats of the method.

\begin{itemize}
 \item For a binary with a giant donor of a given mass and
a BH accretor that undergoes a CE evolution, the final separation will be larger (and therefore the CE survival more likely)
for larger BH masses, as can be deduced from Eqn.~\ref{eq.energy_budget}. Therefore, we assume optimistically low values for the
critical (minimum) mass ratios $q_{\rm crit} = M_{\rm donor} / M_{\rm BH}$ that are required for the CE evolution to occur:
$q_{\rm crit; conv} = 1.5$ for convective-envelope donors and $q_{\rm crit; rad} = 3.5$ for post-MS radiative-envelope donors. 
Additionally, we classify a giant star as convective already when the outermost 10\% of the envelope (in mass) is convective 
(disregarding the sometimes present insignificantly small surface radiative layers).
\item We compute envelope binding energies $E_{\rm bind}$ under assumptions that all the internal energy as well as
the entire energy released during recombination can be used to help unbind the envelope (Eq.~\ref{eq.Ebind}).
\item We assume that orbital energy from the binary orbit shrinking within the CE can be transferred to the envelope without any losses,
i.e. $\alpha_{\rm CE} = 1.0$ in Eqn.~\ref{eq.energy_budget}. 
\item We assume that the envelope acceleration is perfectly fine-tunned to reach exactly the local escape velocity, i.e. there is no term for residual 
kinetic energy at infinity in Eqn.~\ref{eq.energy_budget}.
\item We include energy feedback from accretion 
$\Delta E_{\rm acc} = L_{\rm Edd} \tau_{\rm CE}$ and assume a relatively long duration of the CE phase $\tau_{\rm CE} = 1000$ yr \citep{Ivanova2013}.
For instance, for a $1.4 \msun$ NS this yields $\Delta E_{\rm acc} \approx 2 \times 10^{48} {\rm erg}$, whereas for a $20 \msun$ BH about 
$\Delta E_{\rm acc} \approx 3 \times 10^{49} {\rm erg}$. The relevance of this values compared to the envelope binding energy can be seen in the next section \ref{sec.res_Ebind}.
\end{itemize}

\section{Results}
\label{sec:results}

\subsection{The envelope binding energy -- impact of the outer convective layer}

\label{sec.res_Ebind}

\begin{figure*}
\centering
    \includegraphics[width=0.97\textwidth]{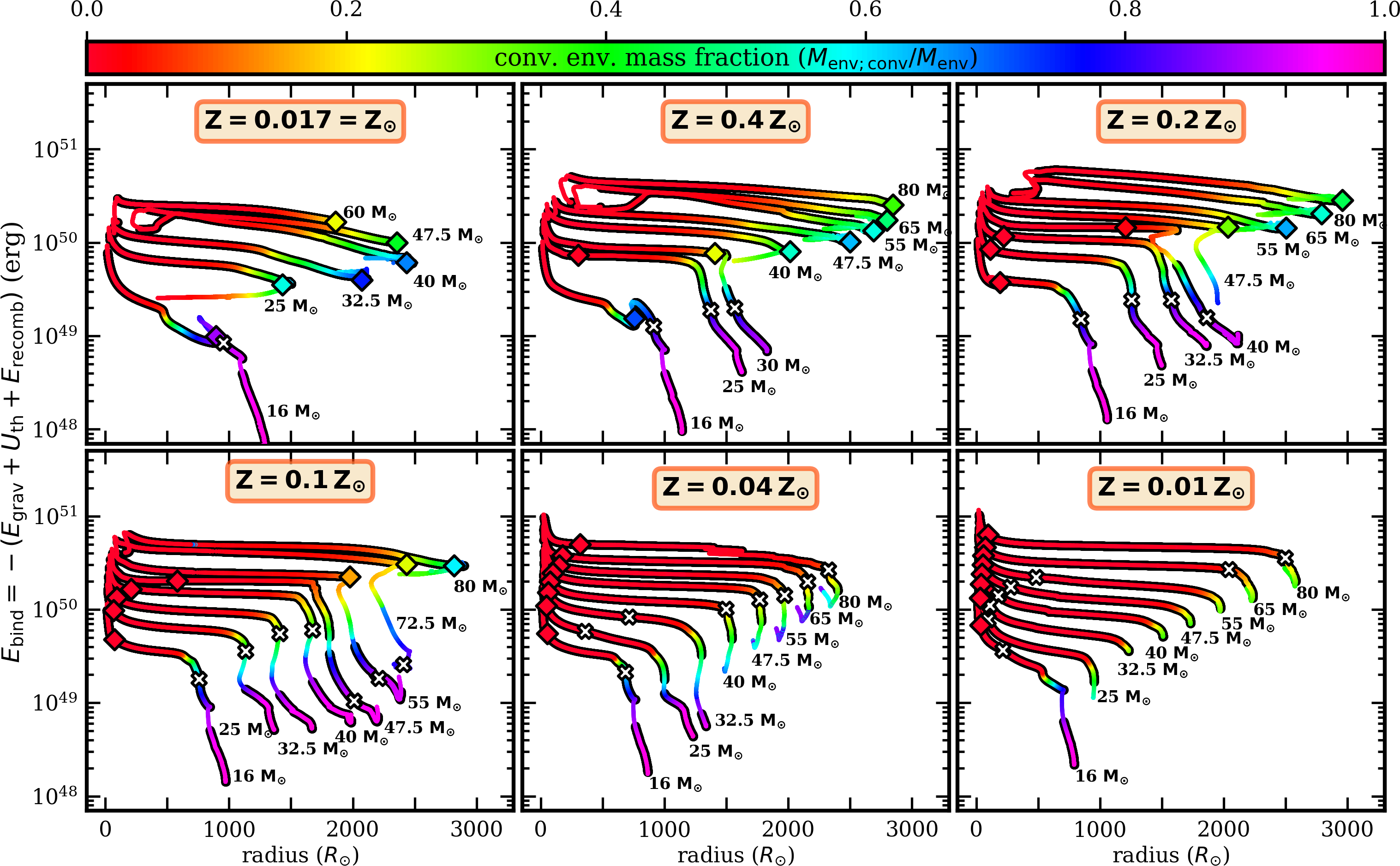}
    \caption{Envelope binding energies $E_{\rm bind}$ of massive giants (i.e. the post-MS part of the evolution)  
    as a function of their radius for six different metallicities
    (selected models from \citealt{Klencki2020}). See Fig.~\ref{fig.Ebind_6_all} for a figure including all the
    masses in the grid and Fig.~\ref{fig.App_lambda_6} for a figure showing all the corresponding $\lambda_{\rm CE}$ values.
    $E_{\rm bind}$ combines the gravitational potential energy as well as the internal energy (including 
    the recombination terms), see Eqn.~\ref{eq.Ebind}. Only post-MS evolution is shown. Colors indicate what fraction of the envelope mass 
    is in the outer convective zone (i.e. $M_{\rm env;conv} / M_{\rm env}$). Diamonds (same coloring) and white crosses mark the onset
    and end of core-helium burning, respectively. Part of the evolution when the radius is increasing 
    beyond the previously largest radius that has been reached, i.e. the part relevant for RLOF and mass transfer, is shaded in black.
    Development of a deep outer convective envelope layer can reach to a very significant decrease of the binding energy, unless the giant 
    is a HG star (i.e. before the core-helium ignition), see Sec.~\ref{sec:res_details} for details.
}
    \label{fig.Ebind_6}
\end{figure*}

In Figure~\ref{fig.Ebind_6} we plot the envelope binding energy $E_{\rm bind}$ of massive giants (i.e. the post-MS part of the evolution) 
of several chosen masses at each of the six studied metallicities. 
Colors in Fig.~\ref{fig.Ebind_6} indicate what fraction $f_{\rm conv}$ of the envelope mass is in the outer convective zone
i.e. $f_{\rm conv} = M_{\rm env;conv} / M_{\rm env}$, where $M_{\rm env;conv}$ is the mass of the outer convective zone and $M_{\rm env}$ is the envelope mass.
In particular, $f_{\rm conv} = 0.0$ indicates a giant with a radiative outer envelope.
The binding energies were computed from stellar models described in Sect.~\ref{sec.method_models} under 
an assumption that all of the thermal energy as well as the recombination energy can be used to help eject the envelope during a CE phase (see Sect.~\ref{sec.method_energy_budget} for details).

Diamonds in Fig.~\ref{fig.Ebind_6} mark the onset of core-helium burning (defined as the point when central helium abundance $Y_{\rm C}$ drops below $0.975$),
meaning that all the previous evolution is the Hertzprung-Gap (HG) phase. 
White crosses correspond to the central helium depletion ($Y_{\rm C}<10^{-3}$). 
At solar metallicity (the top left panel), most of the radial expansion of a giant happens already during the HG phase for all the masses considered here. 
As the metallicity decreases, more and more massive stars stay relatively compact during their HG evolution and most of the expansion in their case takes place during 
core-helium burning. Massive models at the lowest metallicity in our grid ($Z = 0.01\zsun$, the lower right panel) stay compact for even longer and 
expand significant only after the central helium depletion. This is a well known relation between metallicity and the degree of post-MS expansion of massive giants,
details of which are sensitive to a number of poorly constrained ingredients of stellar models such as convective boundary mixing, semiconvection, or rotational mixing
\citep[e.g.][]{Georgy2013,Schootemeijer2019,Higgins2019,Klencki2020,Kaiser2020}. We note that some of the most massive models in Fig.~\ref{fig.Ebind_6} for metallicities $Z \geq 0.1 \zsun$
were not evolved all the way until the central helium depletion but only to the point when strong stellar winds started to significantly shed the giant's envelope, 
causing it to decrease in radius and begin a blue-ward evolution in the HR diagram towards the WR stage. 

The overall behavior of $E_{\rm bind}$ as a function of radius in Fig.~\ref{fig.Ebind_6} is similar to what was found in previous 
studies \citep{Dewi2000,Podsiadlowski2003,Loveridge2011,Wang2016}. Notably, some of the models reveal 
a significant decrease in the envelope binding energy when approaching their largest radii, in some cases by more than an order of magnitude. 
The sudden drop in $E_{\rm bind}$ coincides with the point when the outer envelope of the giant becomes convective. 
This is because the entire outer convective zone of an envelope is located at relatively large radial coordinates and, 
as a result, its density is small and it is very loosely bound to the star \citep[see Fig.~2 of][]{Podsiadlowski2001}.

However, development of a deep outer convective layer does not always lead to a significant decrease of the binding energy. 
The most massive giants at metallicities $Z = 1.0 \zsun$, $0.4\zsun$, $0.2 \zsun$, and $0.1 \zsun$ in Fig.~\ref{fig.Ebind_6}
expand enough to become convective (up to $f_{\rm conv} \approx 0.6$), yet their $E_{\rm bind}$ hardly decreases
as a result of that. Those are the models in which the outer layers become convective already during the HG phase, 
i.e. before the onset of core-helium burning, as indicated by the diamonds in Fig.~\ref{fig.Ebind_6}. As a result, 
because of its impact on the radial expansion of massive giants, metallicity has an indirect but very strong effect on the binding energy of a convective-envelope giant.
The reason why in HG giants the binding energy 
does not significantly decrease when the envelope becomes convective requires a detailed explanation, see Sec.~\ref{sec:res_details}. 

\subsection{CE ejectability in progenitors of BBH mergers}

\label{sec.res_CEeject}

\begin{figure*}
\centering
    \includegraphics[width=0.97\textwidth]{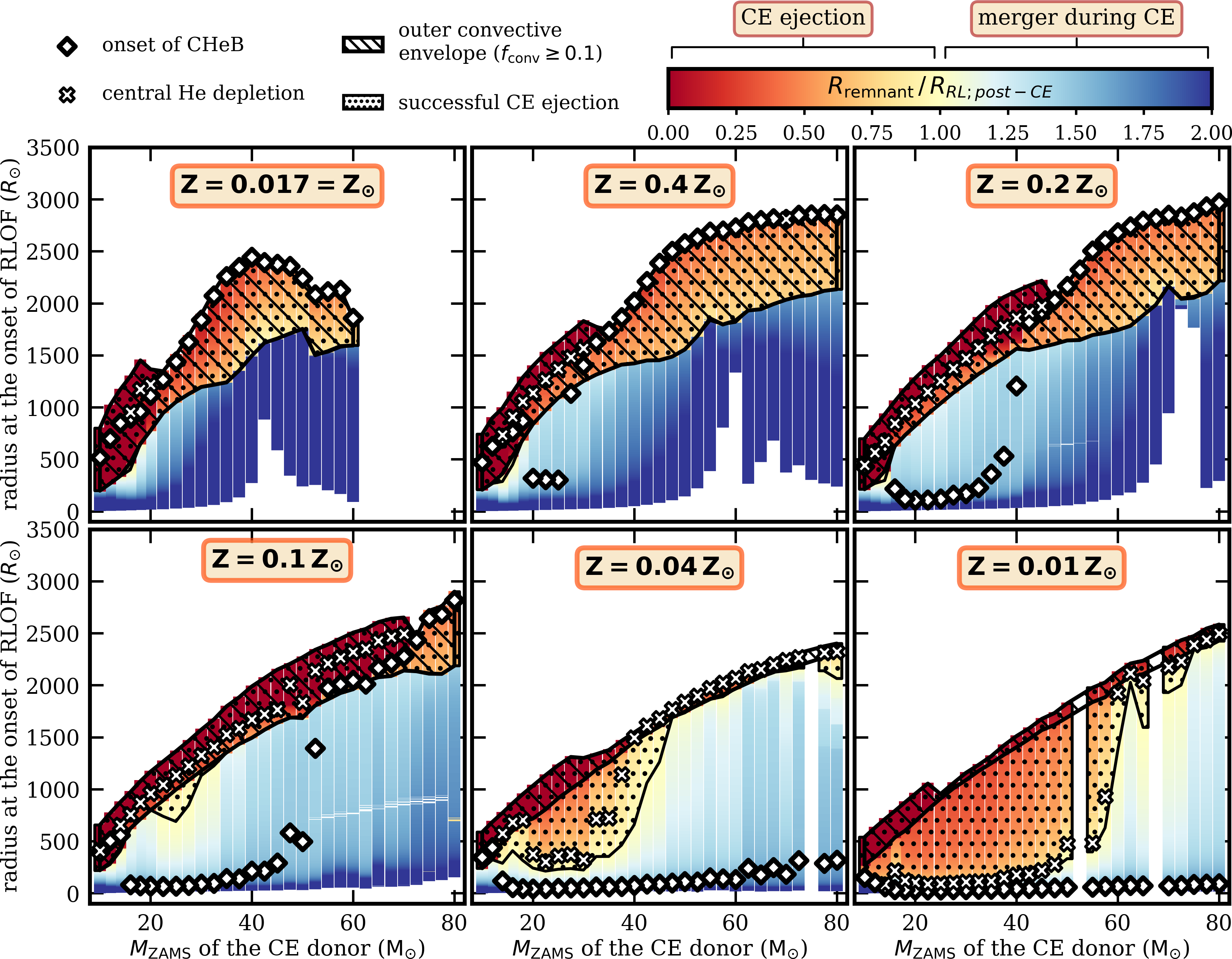}
    \caption{CE ejectability in binaries with massive giant donor stars and BH companions. For each evolutionary track 
    of a massive giant with $M_{\rm ZAMS}$ between $10$ and $80 \msun$,
    we compute what would be the outcome of a CE phase initiated by that star at any given point during its post-MS evolution 
    in a circular binary with a BH companion with mass $M_{\rm BH} = M_{\rm donor} / q_{\rm crit}$ (see text for detailed assumptions). The resulting ratio of 
    the size of the remnant core of the CE donor $R_{\rm remnant}$ and the size of its Roche lobe $R_{\rm RL;post-CE}$ is color-coded 
    in the figure as a function of $M_{\rm ZAMS}$ and the radius of the giant at the onset of the CE phase. Radii above the white diamonds (crosses) indicate 
    giants that are past the onset (end) of core-helium burning in their evolution. Hatched region marks the parameter space where the giant donor 
    has an outer convective envelope with $f_{\rm conv} = M_{\rm conv} / M_{\rm env} \geq 0.1$. Dotted region marks the parameter 
    space were the CE ejection is possible (i.e. $R_{\rm remnant} / R_{\rm RL;post-CE} < 1.0$).
}
    \label{fig.sep_6}
\end{figure*}

Here we study the possibility of ejecting envelopes with $E_{\rm bind}$ computed in the previous section during the CE evolution
(i.e. the CE ejectability) in binaries with stellar BH companions, potential progenitors of BBH mergers. 
Our goal is to find the maximum parameter space in which the CE ejection is feasible according to energy budget 
considerations introduced in detail in Sec.~\ref{sec.method_energy_budget}. To this end, we make a number of optimistic assumptions 
summarized in Sec.~\ref{sec.model_summary}, each of them making CE ejection easier.

For each evolutionary track of a massive giant analyzed in this work, we compute what would be the outcome separation of a CE phase initiated by that star 
at any given point during its post-MS evolution assuming a circular binary 
in which the companion is a BH with mass $M_{\rm BH} = M_{\rm donor} / q_{\rm crit}$ (note that $M_{\rm donor} < M_{\rm ZAMS}$
due to wind mass loss, where ZAMS stands for zero-age main sequence).
Survival of the CE evolution is then determined based on the size of the remnant core of the CE donor $R_{\rm remnant}$ 
(estimated as twice the radial coordinate of the outer core boundary in the pre-CE structure of the donor, see Sec.~\ref{sec.method_energy_budget})
compared to the size of its Roche lobe in the post-CE binary $R_{\rm RL;post-CE}$. 
Values $R_{\rm remnant}/R_{\rm RL;post-CE} > 1$ indicate a merger during the CE phase, whereas $R_{\rm remnant}/R_{\rm RL;post-CE} \leq 1$ indicates a successful CE ejection.

The result is showed in Fig.~\ref{fig.sep_6} where, as a function of $M_{\rm ZAMS}$ and the radius of the giant 
at the onset of the CE phase, we color-code what would be the resulting ratio $R_{\rm remnant}/R_{\rm RL;post-CE}$ in a post-CE binary. 
Note that at solar metallicity (top left panel) stars above $60 \msun$ do not expand in their post-MS evolution
due to strong mass-loss in winds, hence they do not show in Fig.~\ref{fig.sep_6} (this result depends on the assumed mass-loss recipe
and could to some extend be affected by the numerical accuracy of MS models with inflated envelopes, see App.~\ref{sec.App_CE_acc}).
Missing masses (one at $Z = 0.04 \zsun$ and two at $Z = 0.01 \zsun$) are non-converging MESA models.

The most striking finding is that the parameter space for successful CE ejections (marked as the dotted area) is almost exclusively limited 
to cases in which the CE evolution is initiated by a giant donor with an outer convective envelope (marked as the hatched area). The only 
exception are some of the radiative-envelope 
donors with $M_{\rm ZAMS} \lesssim 40 \msun$ at $Z = 0.04\zsun$ and $M_{\rm ZAMS} \lesssim 60 \msun$ at $Z = 0.01\zsun$.
This is the result of convective-envelope giants having smaller envelope binding energies compared to radiative-envelope stars 
(see Fig.~\ref{fig.Ebind_6} and the associated text), combined with the fact that $q_{\rm crit; conv} < q_{\rm crit; rad}$, i.e.
the BH accretors can be more massive in CE events initiated by convective-envelope donors. 

One can also notice in Fig.~\ref{fig.sep_6},
that among the potential CE survivors the $R_{\rm remnant}/R_{\rm RL;post-CE}$ ratio is either only slightly smaller 
than $1.0$ (yellow to orange colors) or significantly smaller than $1.0$ (dark red colors). 
Generally, 
the first group corresponds to convective giants that are still at the HG phase
(below diamonds in Fig.~\ref{fig.sep_6}) and their $E_{\rm bind}$ did not decrease 
as significantly with the increasing convective-envelope mass fraction $f_{\rm conv}$ (even for $f_{\rm conv} \gtrsim 0.6$), as also seen in Fig.~\ref{fig.Ebind_6}.
Given a number of optimistic assumptions in our model, it is uncertain whether models from the first group, with the relatively
high binding energies, can survive the CE phase (see also Fig.~\ref{fig.sep_6_real} below).
The second group are cases in which the CE evolution is initiated by a much more evolved star, with a helium-depleted core
(above the crosses in Fig.~\ref{fig.sep_6}) and a deep outer-convective envelope $f_{\rm conv} > 0.5$. The envelopes of such
giants are very loosely bound (see Sec.~\ref{sec.res_Ebind}) and not much orbital shrinkage is needed to provide energy for 
their ejection. As a result, the energy budget method predicts $R_{\rm remnant}/R_{\rm RL;post-CE}$ ratios that are below $0.1$. 
In reality, it seems unlikely for the donor's remnant to fill such a small fraction of its Roche lobe in the post-CE system because
in that case the companion would never be at the bottom layers of the donor's envelope, thus being unable to help in
their ejection. As such, very small $R_{\rm remnant}/R_{\rm RL;post-CE}$ ratios in Fig.~\ref{fig.sep_6} should rather be 
interpreted as cases in which the CE ejection can be easily achieved in terms of the energy budget, not as a reliable prediction of the post-CE binary separation.

The above distinction between the two types of convective-envelope donors also shows that the evolutionary stage of the donor can have
a significant impact on the outcome of the CE phase. We illustrate the differences between envelope structures of giants that become 
convective during the rapid HG expansion and those that reach the convective-envelope stage later during their evolution in
Sec.~\ref{sec:res_details}. Notably, whether a star becomes convective during the HG expansion or only later during its evolution
is an uncertain outcome of detailed stellar models but its relation to metallicity appears robust \citep[e.g.][]{Schootemeijer2019,Klencki2020}.

Even though BH binaries are the main focus of this study, we also consider CE ejectability in systems in which the accretor is a 
NS with $M_{\rm NS} = 2 \msun$ (with all the other assumptions being the same). The result is plotted in Fig.~\ref{fig.sep_6_NS} 
in the Appendix. Because of the lower accretor mass, the parameter space for CE ejection in NS binaries is smaller than in the 
case of BH binaries. Interestingly, it is limited to cases in which the donor is already a core-helium depleted star. In such 
systems there might be no additional case BB mass transfer phase after a successful CE ejection, in contrast 
to what has been advocated in the classical picture of the double NS binary formation \citep[for a recent review see][]{Tauris2017}. 
We leave a further discussion of this finding for future studies.

\begin{figure}
\centering
    \includegraphics[width=1.0\columnwidth]{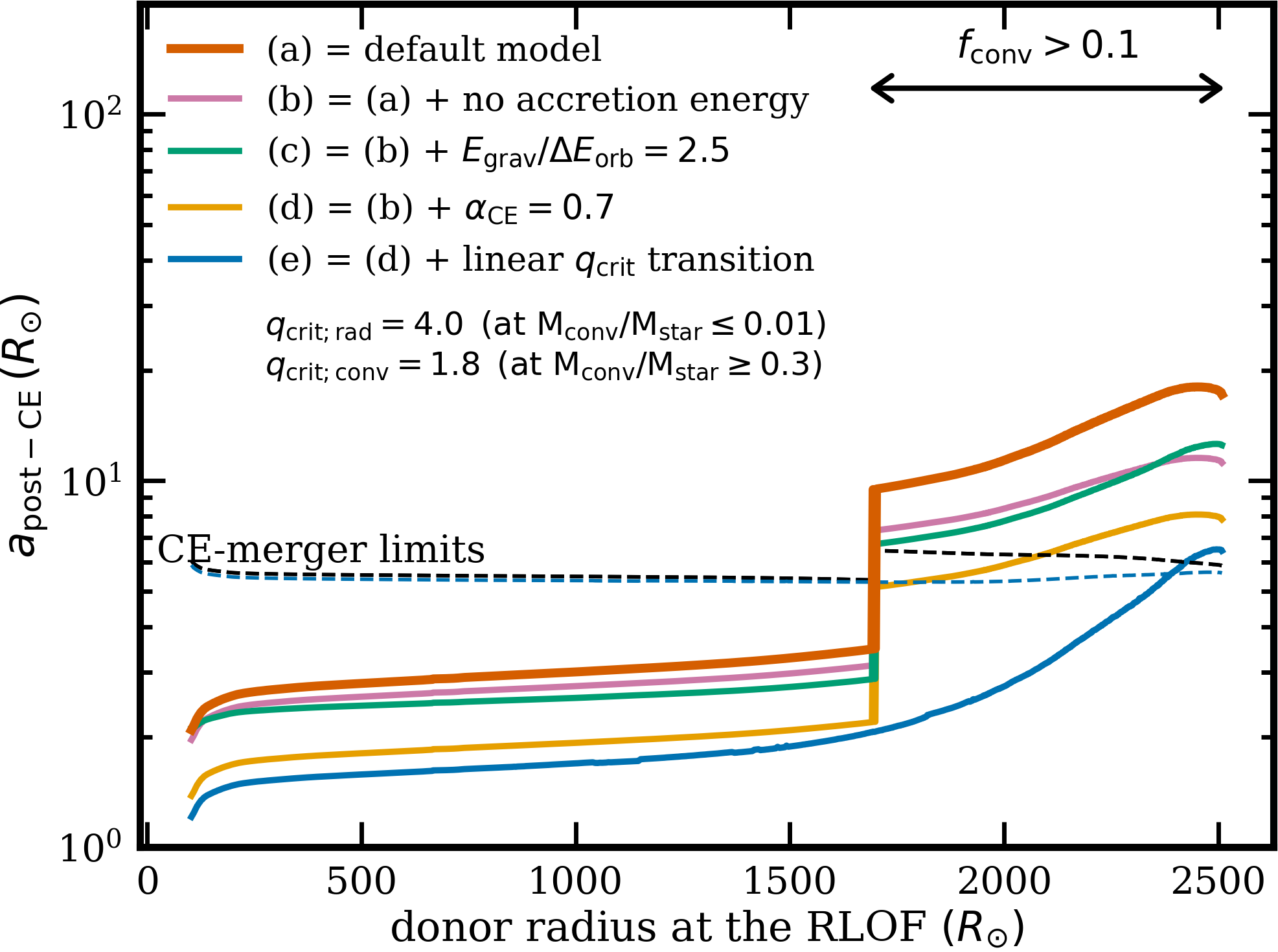}
    \caption{Exploring the CE ejectability for a BH binary with a $M = 55 \msun$ giant donor at $Z = 0.2 \zsun$ metallicity as a 
    function of the giant's radius at the onset of a CE phase. See text for the explanation of different models (a)-(e). Cases when
    the post-CE separation $a_{\rm post-CE}$ is greater then the minimum separation required to avoid a merger (marked as a black
    dashed line for models (a)-(d) and a dashed purple line for model (e)) survive the CE phase.
}
    \label{fig.var_plot}
\end{figure}

In Fig.\ref{fig.var_plot}, similar to Fig.~6 of \citet{Kruckow2016}, we explore how several variations in the assumptions of the default
model would affect the separation of the post-CE binary $a_{\rm post-CE}$ and the CE ejectability. We examine the case of a giant CE donor with $M_{\rm ZAMS} = 55 \msun$  
at $Z = 0.2\zsun$ metallicity, showing $a_{\rm post-CE}$ as a function of its radius at the moment of Roche-lobe overflow (RLOF).
The default 'optimistic' model is plotted in blue as model (a), see Sec.~\ref{sec.model_summary} for the summary of its assumptions.
Cases when $a_{\rm post-CE}$ is greater then the minimum separation required to avoid a merger (marked as a black dashed line) is
the parameter space for CE ejectability. Note that it is limited to convective-envelope donor cases ($f_{\rm conv} > 0.1$).
Slow changes in the dashed line are related to the evolution of the core radius (contraction) as well as to a transition from 
the radiative to the convective envelope stage at around $R_{\rm donor} \approx 1700 \rsun$, at which point the assumed binary mass ratio decreases
from $q_{\rm crit;rad} = 3.5$ to $q_{\rm crit;conv} = 1.5$.

\begin{figure}
\centering
    \includegraphics[width=1.0\columnwidth]{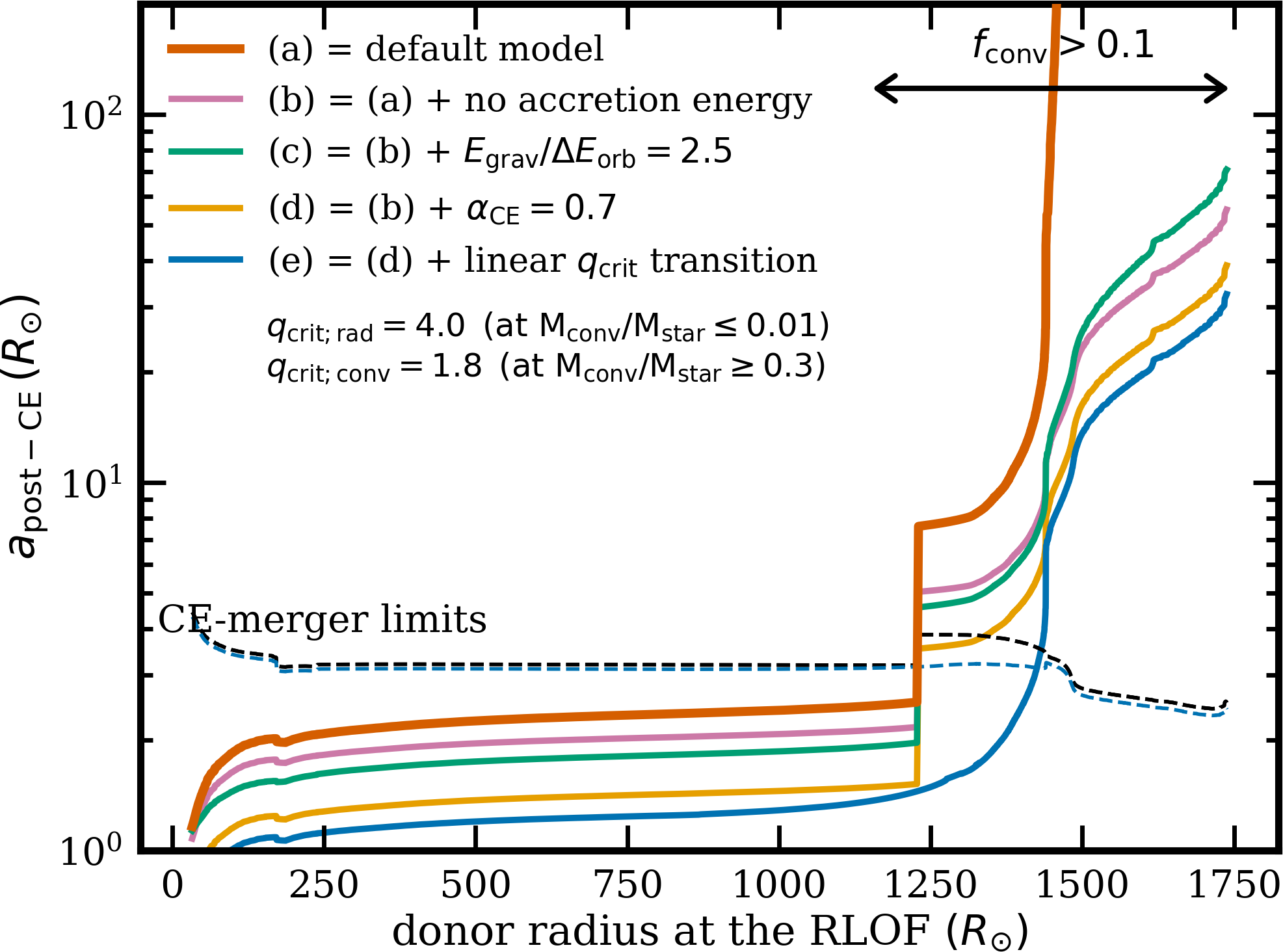}
    \caption{Same as Fig.~\ref{fig.var_plot} but for a $30 \msun$ model at $0.2 \zsun$ metallicity. For 
    the donor radii $R \gtrsim 1400 \rsun$ the outer envelope becomes deeply convective during a core-helium burning stage,
    which causes a significant decrease in the binding energy of the model (Fig.~\ref{fig.Ebind_6}) and an increase in the 
    post-CE separation $a_{\rm post-CE}$.
}
    \label{fig.var_plot_30}
\end{figure}

In green we plot a model (b), which is different from the default model by not including the $\Delta E_{\rm acc}$ term in 
Eqn.~\ref{eq.energy_budget}. It corresponds to a situation when the energetic feedback from accretion is significantly smaller 
then the Eddington luminosity or if the duration of the CE phase is much shorter then the assumed 1000 yr. Excluding $E_{\rm acc}$ 
from the energy budget affects the post-CE separation by less than a factor of two for the $55 \msun$ donor in Fig.~\ref{fig.var_plot}. 
The effect is larger for lower mass stars, see for example a $30 \msun$ donor case in Fig.~\ref{fig.var_plot_30}, because of 
their lower binding energies. Note that $\Delta E_{\rm acc} \propto M_{\rm accretor}$ (Eqn.~\ref{eqn.Eacc}), 
so for a given mass ratio $M_{\rm donor} / M_{\rm accretor} = q_{\rm crit}$ the energy from accretion
scales linearly with the donor mass. We discuss super-Eddington accretion during a CE phase in Sec.~\ref{sec.disc_uncert}.

\begin{figure*}
\centering
    \includegraphics[width=0.85\textwidth]{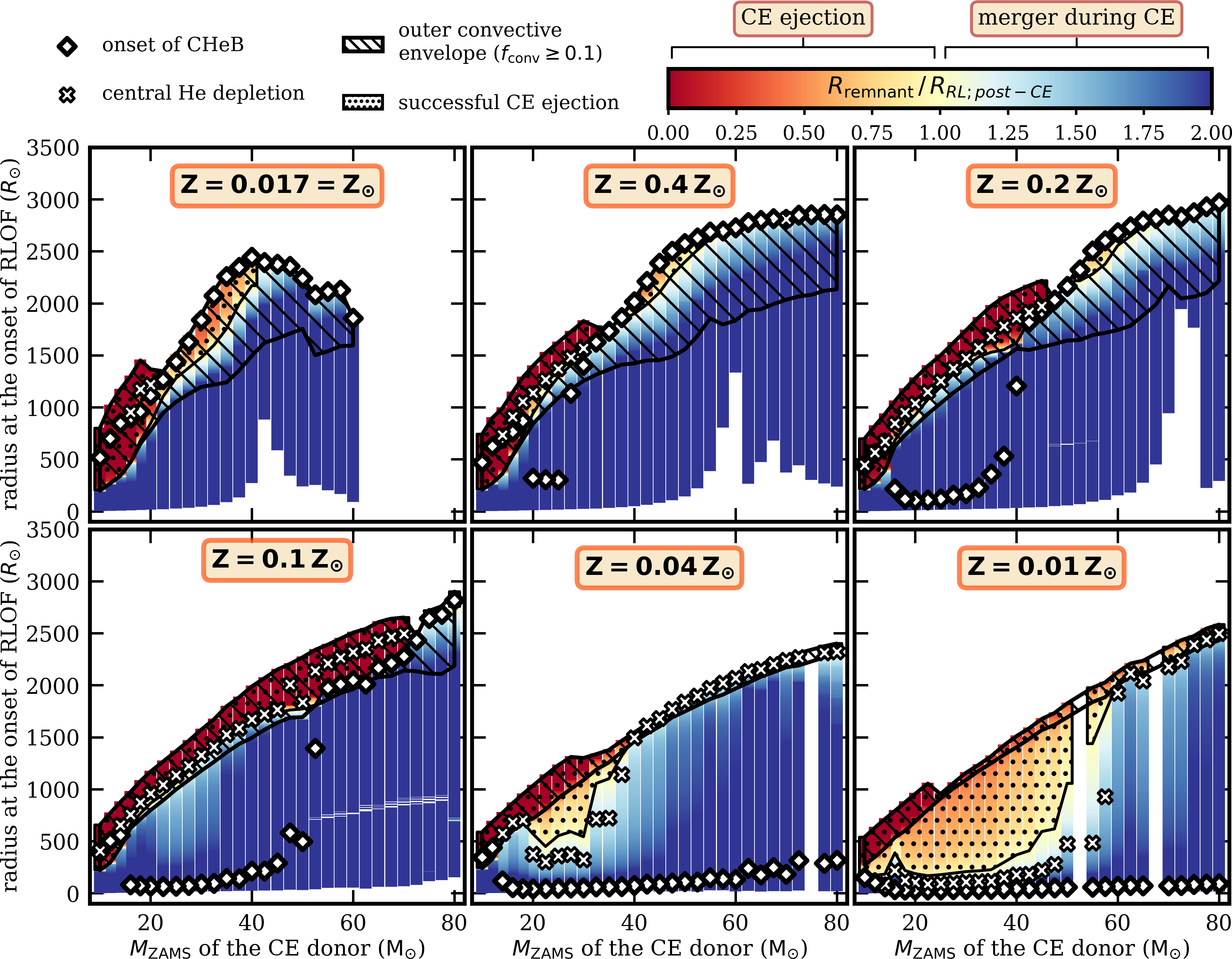}
    \caption{Same as Fig.~\ref{fig.sep_6} but assuming 
    a more realistic energy budget and the mass transfer stability criteria: $\alpha_{\rm CE} = 0.7$ instead of $\alpha_{\rm CE} = 1.0$ 
    and a smooth transition of $q_{\rm crit}$ from $q_{\rm crit} = 4.0$ for radiative donors with $M_{\rm conv}/M_{\rm star} \leq 0.01$ 
    up to $q_{\rm crit} = 1.8$ for convective-envelope giants with $M_{\rm conv}/M_{\rm star} \geq 0.3$
    as a linear function of $M_{\rm conv}/M_{\rm star}$. Note that $f_{\rm conv} = M_{\rm conv} / M_{\rm env}$, and $f_{\rm conv} > 0.1$ 
    region is hatched in the figure.
}
    \label{fig.sep_6_real}
\end{figure*}

Instead of relating the change in orbital energy $\Delta E_{\rm orb}$ to the envelope binding energy $E_{\rm bind}$ via the $\alpha_{\rm CE}$
parameter, one can relate $\Delta E_{\rm orb}$ to just the gravitational potential component of the 
binding energy: $\Delta E_{\rm grav} / \Delta E_{\rm orb} = \alpha_{\rm grav}$. In one-dimensional hydrodynamic simulations
of a CE in-spiral inside the envelope of a $12 \msun$ red supergiant,  \citet{Fragos2019} find $\alpha_{\rm grav} \approx 2.7$. 
Three-dimensional models of the CE phase from low-mass giants by \citet{Passy2012} and \citet{Ohlmann2016} find similar values of 
$\alpha_{\rm grav} \approx 2.0$ and $\approx 2.5$, respectively. Model (c) in  Fig.~\ref{fig.var_plot} (in red) assumes 
$\alpha_{\rm grav} = 2.5$ and is in fact very similar to the same model but with $\alpha_{\rm CE} = 1.0$ (in green).
This can be expected based on the typical ratio of thermal internal energy $U_{\rm th}$ and the gravitational potential energy
$E_{\rm grav}$ in a giant's envelope, see section 3.2 in \citet{Fragos2019}.

Importantly, the assumption of $\alpha_{\rm CE} = 1.0$ does not only imply a perfect energy transfer (eg. no radiative losses),
but also a perfect fine-tuning when the ejected envelope becomes accelerated to gain precisely the local escape velocity. 
In reality, this is unlikely to be the case. \citet{Nandez2015,Nandez2016} carried out three dimensional hydrodynamic simulations
of CE events with low-mass giants and found that typically about 30\%$-$40\% of the orbital energy leaves the system as residual 
(mainly kinetic) energy of the unbound ejecta. This corresponds to the model (d) in Fig.~\ref{fig.var_plot}, with
$\alpha_{\rm CE} = 0.7$. Note that in this model only half of the convective-envelope donor cases survive the CE inspiral
(the orange line rises above the black dashed threshold in the middle of the $f_{\rm conv} > 0.1$ region).

Finally, in model (e) plotted in purple we also apply a smooth transition of the critical mass ratio from $q_{\rm crit} = 4.0$ 
for radiative donors with $M_{\rm conv}/M_{\rm star} \leq 0.01$ up to $q_{\rm crit} = 1.8$ for convective-envelope giants with $M_{\rm conv}/M_{\rm star} \geq 0.3$
as a linear function of $M_{\rm conv}/M_{\rm star}$. This makes it consistent with \citet{Pavlovskii2015}, who find $q_{\rm crit;conv}$ 
between $1.5$ and $2.2$ for giants with deep convective envelopes of $M_{\rm conv}/M_{\rm star} \geq 0.3$ and $q_{\rm crit} > 3.5$ for
less convective giants. Our default classification criteria for convective-envelope donors ($f_{\rm conv} = M_{\rm conv} / M_{\rm env} \geq 0.1$)
also includes the in-between cases of mostly radiative-envelope donors with relatively shallow convective envelopes, for which the mass transfer
is more stable than for giants with deep convective envelopes. The CE ejection in model (e) is only possible in a small number of cases when
the $55 \msun$ giant is close to its maximum size and its outer convective envelope is the most extended.

In Fig.~\ref{fig.var_plot_30} we plot the same model variations as in Fig.\ref{fig.var_plot} but for a lower mass $30 \msun$ giant 
(also at $Z = 0.2\zsun$ metallicity). The main difference is that for donor radii above $\sim 1450 \rsun$ all 
explored variations result in a successful CE ejection. This is because at that point the $30 \msun$ model 
develops a deeply convective envelope
at late stages of core-helium burning, which leads to a significant decrease of its binding energy for $R \gtrsim 1400 \rsun$ 
(see Fig.~\ref{fig.Ebind_6}). In fact, at those stages the envelope is so loosely bound that the accretion term $\Delta E_{\rm acc}$ 
alone is larger than $E_{\rm bind}$. Fig.~\ref{fig.var_plot_30} illustrates that in models with loosely bound deep convective envelopes
the estimated post-CE separation is often significantly larger then the CE-merger limit. In those cases (appearing in dark red in 
Fig.~\ref{fig.sep_6}) the CE ejectability does not significantly depend on the particular choice of assumptions for the energy budget.

Models (b)-(e) in Fig.~\ref{fig.var_plot} show that our reference model (a) yields a reasonable but very optimistic result in terms of
CE ejectability showed in Fig.~\ref{fig.sep_6}. For comparison, in Fig.~\ref{fig.sep_6_real} we make less optimistic but likely more realistic assumptions of 
$\alpha_{\rm CE} = 0.7$ and a smooth linear transition of $q_{\rm crit}$ between radiative and convective envelope donors as in model
(e) above. The result is a significantly reduced parameter space for CE survival, mostly limited to the cases of loosely bound convective
envelopes in core-helium burning or more evolved giants.

\subsection{It has to be cool: tight link between the CE donors and red supergiants}

\label{sec.HRD}

\begin{figure*}
\centering
    \includegraphics[width=0.8\textwidth]{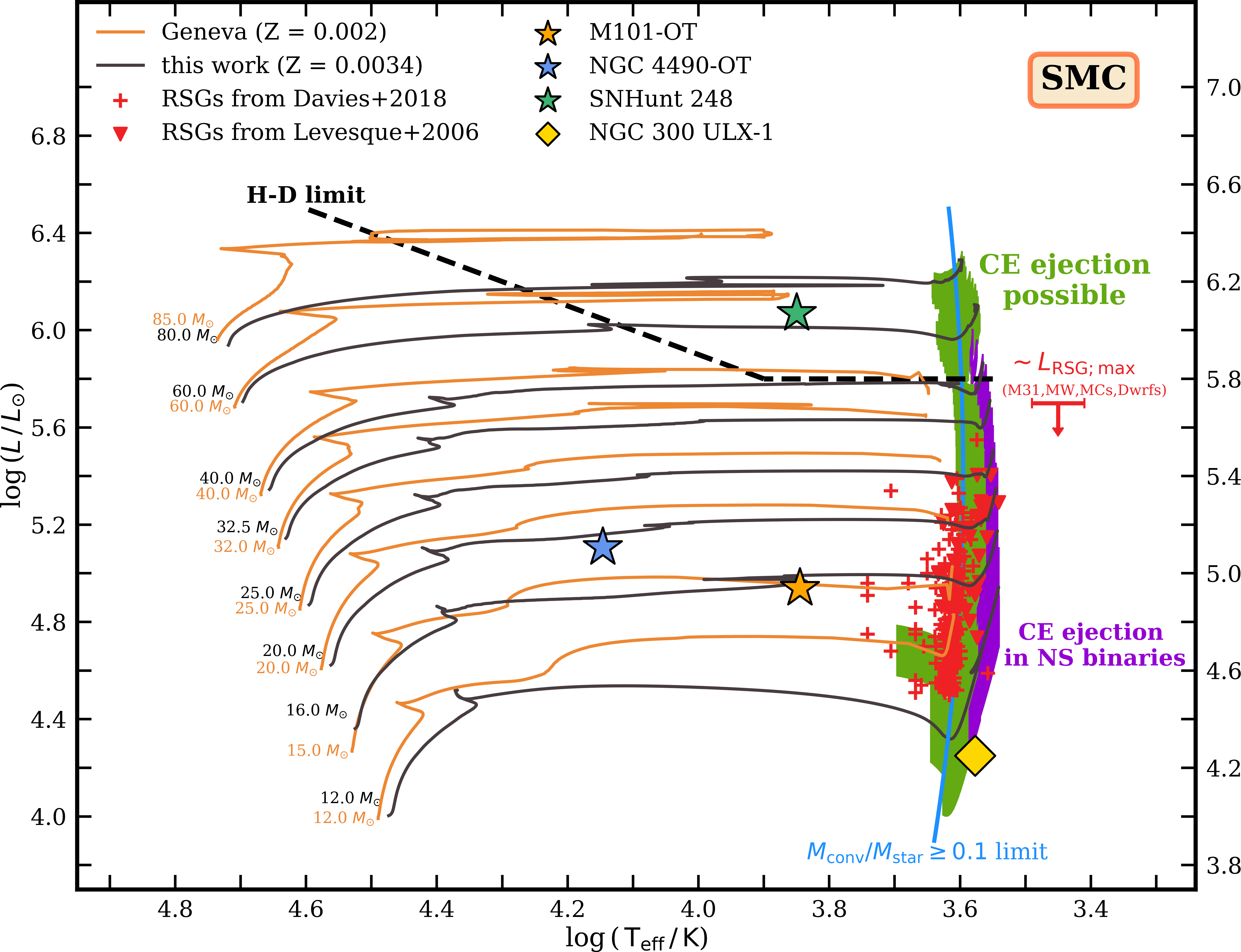}
    \caption{HR diagram for the SMC-like metallicity \citep[$Z \approx 0.2 \zsun$][]{Venn1999} in which the green area marks the position of donor stars at the onset of RLOF 
    for which a CE evolution with a successful CE ejection in BH binaries is possible. A smaller violet area marks the subset of donors 
    for which CE ejection is also possible in the case of NS accretors ($M_{\rm NS} = 2 \msun$). A definitive association of donors to 
    successful CE evolution events with cool (red) supergiants ($\logteff \lesssim 3.7$) is clear. Two sets of selected stellar tracks 
    are plotted: models from \citet{Klencki2020}, which are the 
    main focus of this paper, as well as models from \citep{Georgy2013} computed with the GENEVA code. 
    Physical parameters of RSGs observed in the SMC are taken from \citet{Levesque2006} and \citet{Davies2018}.
    The black dashed line marks the empirical Humphreys-Davidson (H-D) limit, beyond which almost no stars in the Milky Way or in the LMC are observed 
    \citep{Humphreys1979, Humphreys1994,Ulmer1998}. The light-blue line shown the threshold effective temperature 
    below which at least 10\% of the mass of models from \citet{Klencki2020} is in the outer-convective envelope. Colored stars mark
    the inferred physical parameters of three different massive star progenitors of LRNs \citep{Smith2016,Blagorodnova2017,Mauerhan2018}. 
    The yellow diamond shows the RSG donor in a pulsating ULX source NGC 300 ULX-1 \citep{Heida2019_ngc300}.
}
    \label{fig.HRD}
\end{figure*}

In the previous section we showed that unstable mass transfer evolution followed by a CE phase and a successful CE ejection in BH/NS
binaries is only energetically feasible (even under optimistic assumptions) if the donor star is a massive supergiant with an outer 
convective envelope, see Fig.~\ref{fig.sep_6} and Fig.~\ref{fig.sep_6_NS}. Observationally, such stars appear as red supergiants (RGSs)
with effective temperatures of about $\logteff \lesssim 3.7$  (depending on the exact position of the Hayashi line at a given metallicity). 

We illustrate this in an HR diagram for the SMC-like metallicity \citep[$Z \approx 0.2 \zsun$][]{Venn1999} in Fig.~\ref{fig.HRD}, 
in which the estimated region of CE ejectability is marked in green (this corresponds to the dotted area in the top-right panel of 
Fig.~\ref{fig.sep_6}). The part of that parameter space marked in violet are cases in which the CE ejection is also possible with NS accretors.

For comparison, we mark the positions of known RSGs in the SMC from 
the samples of \citet[][with effective temperatures based on spectral-energy density fits]{Davies2018} and \citet[][with effective temperatures from atmospheric model fits]{Levesque2006}. 

For reference, the light-blue line around $\logteff \approx 3.65$ marks the threshold temperature $T_{\rm eff;th}$ below which at 
least 10\% of the mass of the star is in the outer-convective envelope \citep[from Sec.~3.4 of][]{Klencki2020}.
The fact that many of the RSGs in the sample collected by \citet{Davies2018} are hotter than $T_{\rm eff;th}$
and fall outside of the predicted temperature range of RSGs in the green area in Fig.~\ref{fig.HRD} is a likely indication that 
the value of the mixing-length parameter $\alpha_{\rm ML} = 1.5$ in the models from \citet{Klencki2020} is somewhat 
too small.\footnote{Mixing length of around $2$ would probably result in a better agreement with the observed RSG samples \citep{Chun2018}.} 

The brightest RSG in the SMC has a luminosity of $\logL = 5.55 \pm 0.1$ \citep{Davies2018}. A similar empirical 
upper luminosity limit for RSGs of $\logL_{\rm RSG;max} \approx 5.6-5.8$ has been found also for other local galaxies of different types and compositions: 
the Milky Way \citep[$Z = 0.02$][]{Levesque2006}, M31 \citep[$Z = 0.04$][]{Neugent2020}, 
the Large Magellanic Cloud \citep[$Z = 0.007$][]{Davies2018}, 
as well as several dwarf irregular galaxies ($\rm [Fe/H]$ between $-1.0$ and $-0.4$, \citealt{Britavskiy2019}). It is marked in Fig.~\ref{fig.HRD}
at an approximate luminosity $\logL_{\rm RSG;max} = 5.7$, which corresponds to stars with initial masses $M_{\rm ZAMS} \approx 40 \msun$. 
The existence of the empirical upper luminosity limit $\logL_{\rm RSG;max}$ could be an indication that more massive stars never expand to 
reach the RSG stage during their evolution, for instance due to extensive mass loss at the uncertain luminous blue variable stage. In view
of our findings, this would likely prevent donors with $M_{\rm ZAMS} \gtrsim 40 \msun$ from forming BBH mergers in the CE evolution channel.

Lack of RSGs above a certain luminosity limit has been purposefully recovered in various stellar tracks of massive stars. As an example, in
Fig.~\ref{fig.HRD} we plot several rotating stellar tracks computed with the GENEVA code for $Z = 0.002$ \citep{Georgy2013}.
For reference, we also show a few corresponding tracks from \citep{Klencki2020} that were used to compute envelope binding energies in this 
work.\footnote{The fact that GENEVA models are hotter and more luminous during the MS is most likely a result of a higher efficiency of 
rotational mixing in GENEVA models compared to models computed by \citet{Klencki2020}.}
The lack of most luminous RSGs in the GENEVA tracks is most likely a result
of increasing mass-loss rates in stars with super-Eddington layers \citep{Ekstrom2012} as well as of suppressing density inversions in 
radiation-dominated stars by using a mixing length taken on the density rather than on the pressure scale height \citep[App.~A of][]{Klencki2020}.
We discuss the empirical $\logL_{\rm RSG;max}$ limit in the context of CE evolution and the formation of BBH mergers in Sec.~\ref{sec.disc_Lrsgmax}.

It is worth noting that CE evolution in a binary with a BH/NS accretor is preceded by a phase of essentially stable mass transfer that 
takes place between the moment of RLOF and the onset of the CE phase \citep{MacLeod2018,MacLeod2018b,MacLeod2020b}. During this stage the mass transfer rate gradually increases at a 
rate which is primarily dictated by the thermal timescale of the donor's envelope 
\citep[also in the case of convective-envelope donors, see][]{Woods2011,Pavlovskii2015}. In the case of massive donors, this indicates a 
duration of the order of $\sim 10^4$ yr, and a loss of even $\sim 30\%$ mass from the donor's envelope \citep[][also Ivanova, private 
communication]{MacLeod2020}.\footnote{An exception are systems that become unstable on a shorter timescale due to Darwin instability \citep{Darwin1879,Eggleton2001},
which could predominantly occur in extreme mass ratio cases \citep[$q \gtrsim 10$][]{Rasio1995}.} Observationally, a system at this stage 
would likely appear as a bright X-ray binary: an ultra-luminous X-ray source (ULX).
In the case of potential CE survivors among BH and NS binaries this corresponds to ULXs with RSG donor stars.  Several such systems with 
donor stars spectroscopically confirmed to be RSGs are known to date \citep{Heida2015,Heida2016,Lau2019,Heida2019}. In the case of one of 
the sources, a pulsating ULX dubbed NGC 300 ULX-1, \citet{Heida2019} were able to model the spectrum of the optical counterpart as a sum 
of three components: a blue excess (likely due to X-ray irradiation), a red excess (likely due to dust), and a stellar atmosphere model 
of an RSG with $T_{\rm eff} = 3650$\,-\,$3900$ K and $\logL = 4.25 \pm 0.1$ (marked in Fig.~\ref{fig.HRD}). \citet{Lopez2020} completed 
a systematic search for near-infrared candidate counterparts to nearby ULXs and concluded that ULXs with RSG donors constitute about 
$\sim 4\%$ of the observed ULX population, which is about four times more than predicted by population synthesis models \citep{Wiktorowicz2017}. 
One reason for this discrepancy could be that rapid binary evolution codes assume that the moment of RLOF is also the onset of CE evolution, 
without modelling the intermediate mass transfer phase.

Observational signature of the CE phase itself are luminous red-novae (LRN), red transients characterized by a rapid rise in luminosity 
followed by a lengthy plateau \citep{Soker2003,Kulkarni2007,Tylenda2011,Ivanova2013Sci,Kochanek2014,Howitt2020}. In several cases, archival multiband observations 
from before the LRN have revealed the progenitors to be consistent with massive giants: notably a $\sim 18 \msun$ progenitor to 
M101-OT \citep{Blagorodnova2017} and a $\sim 60 \msun$ progenitor to SNHunt 248 \citep{Mauerhan2018}, as well as a putative single-band 
detection of an LBV-like progenitor to NGC 4490-OT \citep{Smith2016}. In none of these cases was the progenitor consistent with a RSG, 
see Fig.~\ref{fig.HRD}. According to our findings, this might be an indication that these CE events finished in mergers, although our 
results may not be directly applicable to CE cases with stellar accretors. 
As a final remark, it was recently shown by \citet{Ginat2020} that inspiral of compact accretors during the final stages of CE evolution 
could produce a GW signature which might be detectable by the next-generation GW detectors such as LISA.

\subsection{A detailed look at the structure and the binding energy of convective-envelope giants}

\label{sec:res_details}

In Sec.~\ref{sec.res_Ebind} we found that the envelope of a massive giant can become significantly less bound when it transitions 
from a radiative to a convective state, with $E_{\rm bind}$ decreasing by an order of magnitude or more. An exception to this rule 
are giants that become convective already during the HG phase, for which the decrease in the binding energy is much less significant.
In this section we explain the origin of this behavior.

\begin{figure}
\centering
    \begin{subfigure}{1.0\columnwidth}
    \includegraphics[width=1.0\columnwidth]{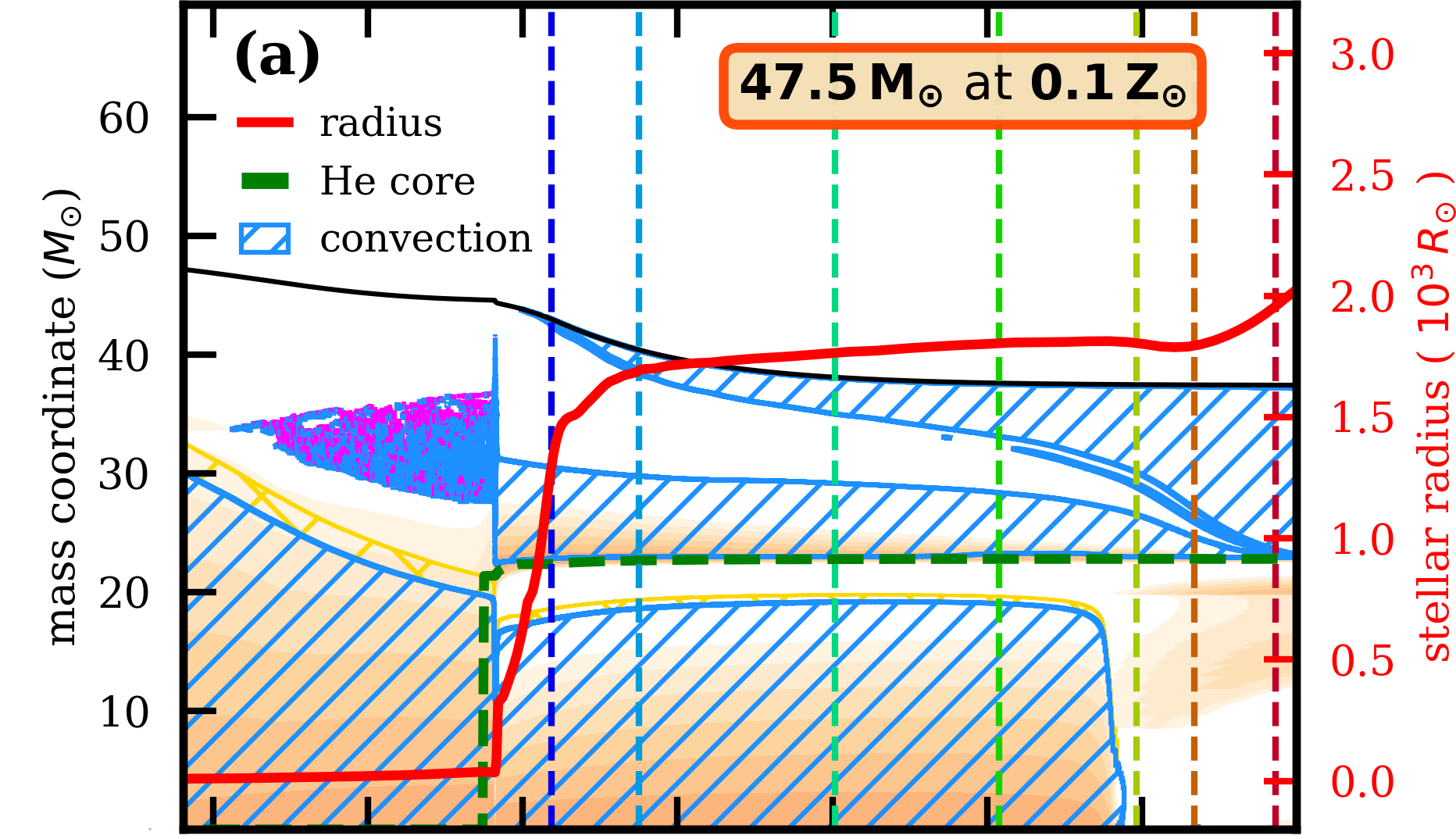}
    \phantomsubcaption
    \vspace{-1.2\baselineskip}
    \label{fig.Ebind_detail_16_01_a}
    \end{subfigure}
    \begin{subfigure}{1.0\columnwidth}
    \includegraphics[width=1.0\columnwidth]{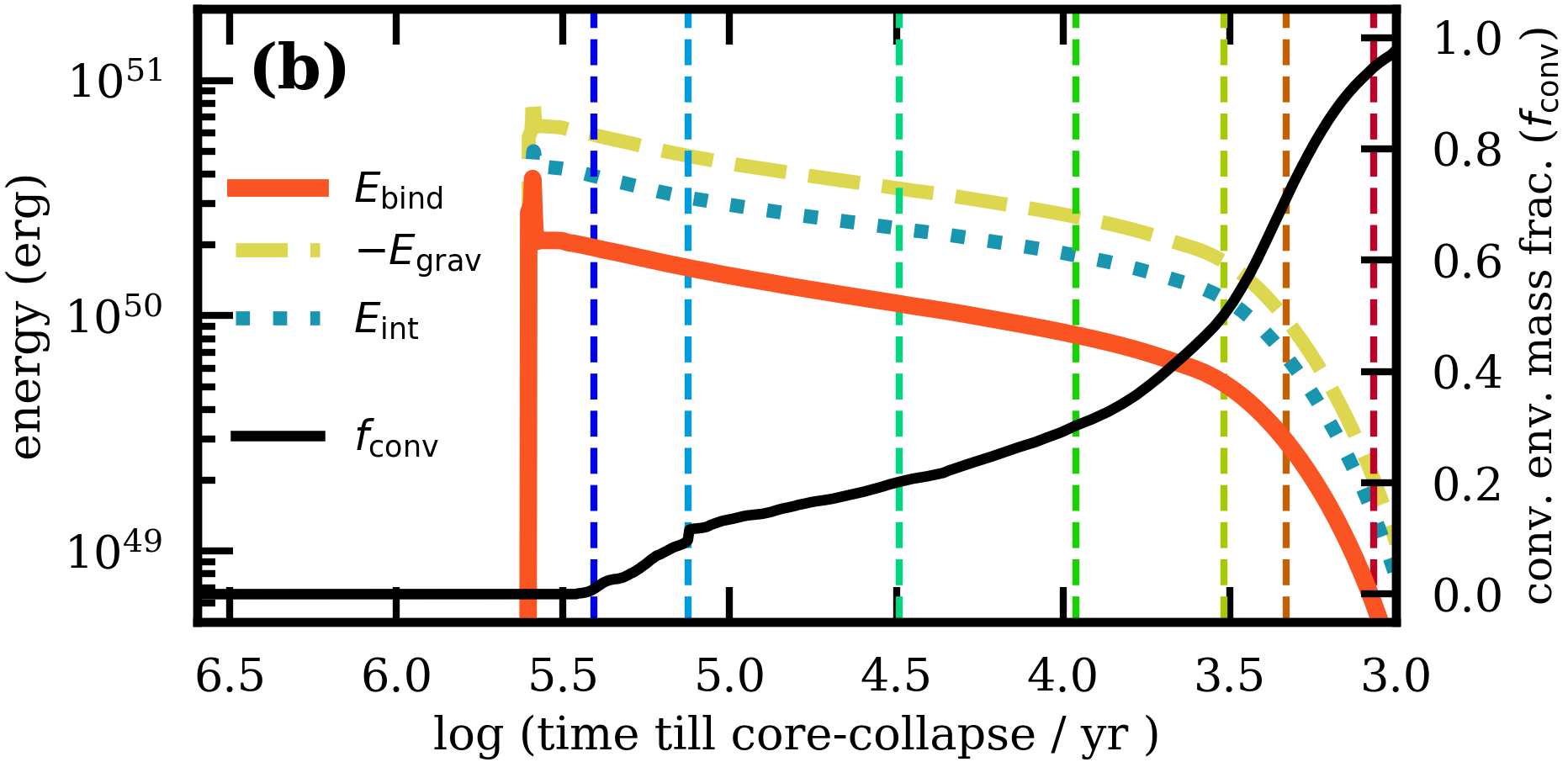}
    \phantomsubcaption
    \vspace{-0.7\baselineskip}
    \label{fig.Ebind_detail_16_01_b}r
    \end{subfigure}
    \begin{subfigure}{1.0\columnwidth}
    \includegraphics[width=1.0\columnwidth]{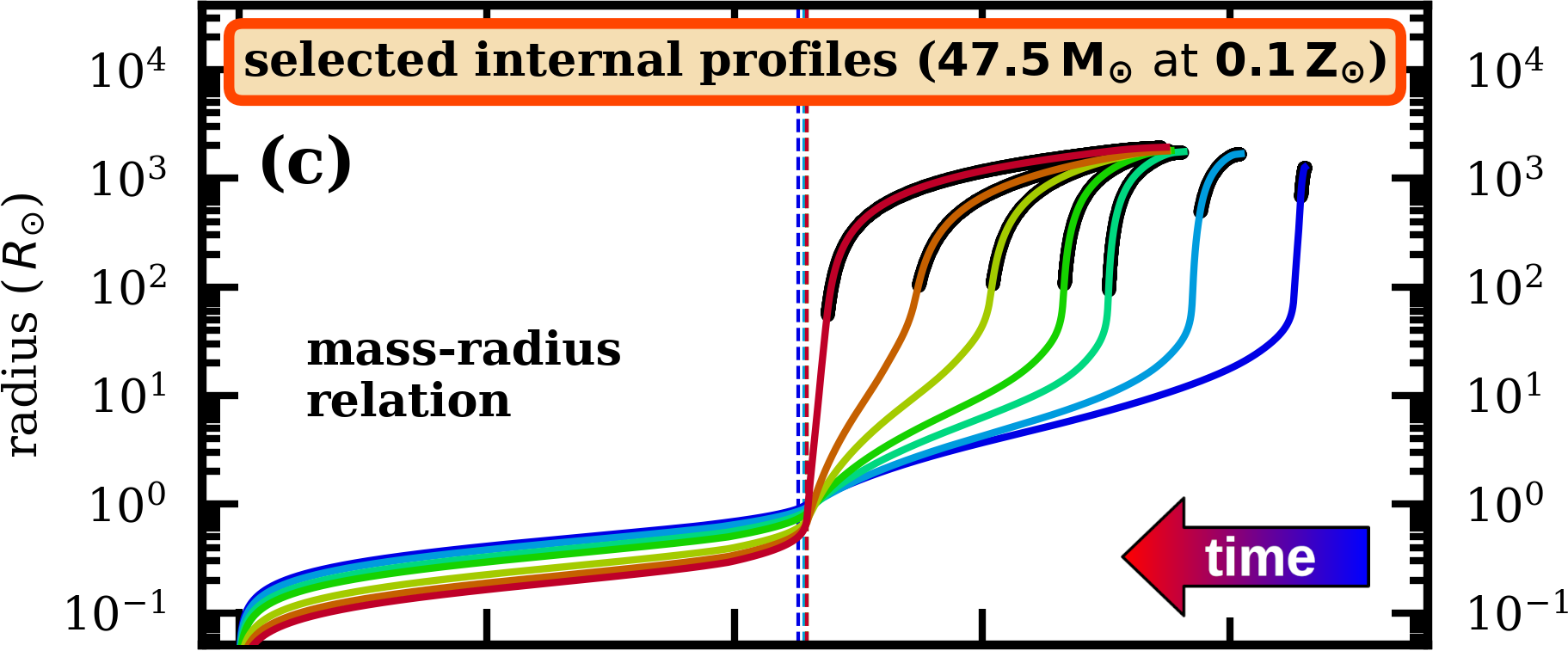}
    \phantomsubcaption
    \vspace{-1.2\baselineskip}
    \label{fig.Ebind_detail_16_01_c}
    \end{subfigure}
    \begin{subfigure}{1.0\columnwidth}
    \includegraphics[width=1.0\columnwidth]{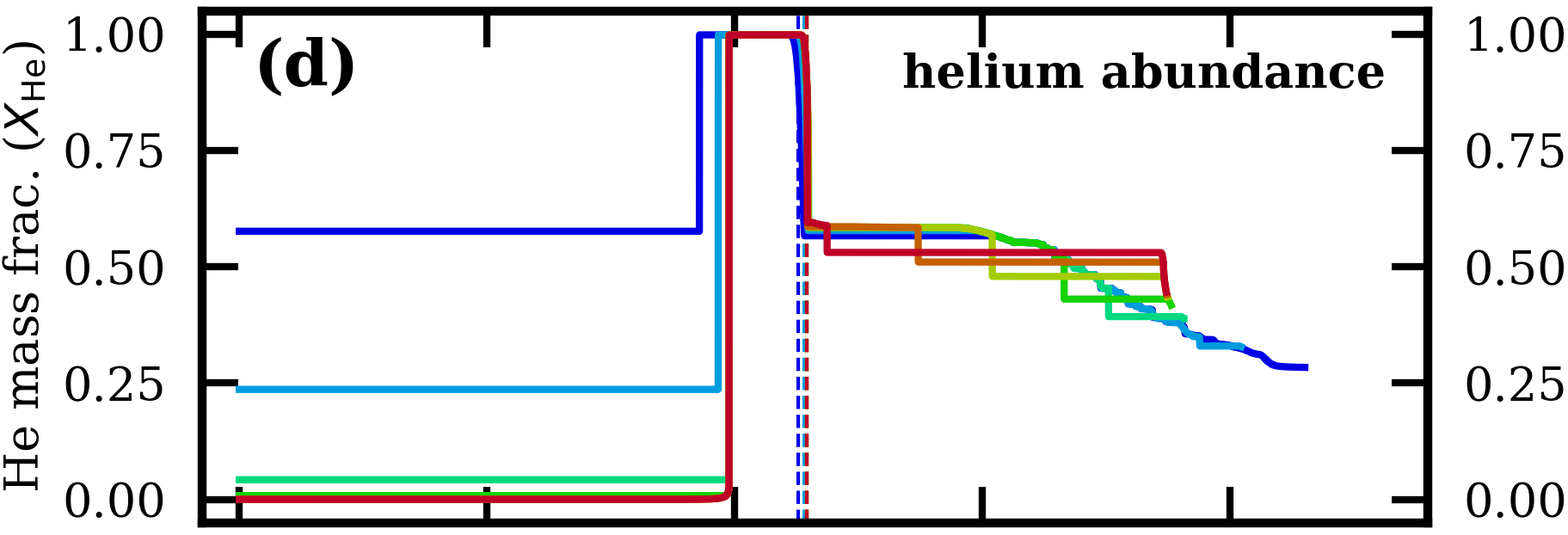}
    \phantomsubcaption
    \vspace{-1.2\baselineskip}
    \label{fig.Ebind_detail_16_01_d}
    \end{subfigure}
    \begin{subfigure}{1.0\columnwidth}
    \includegraphics[width=1.0\columnwidth]{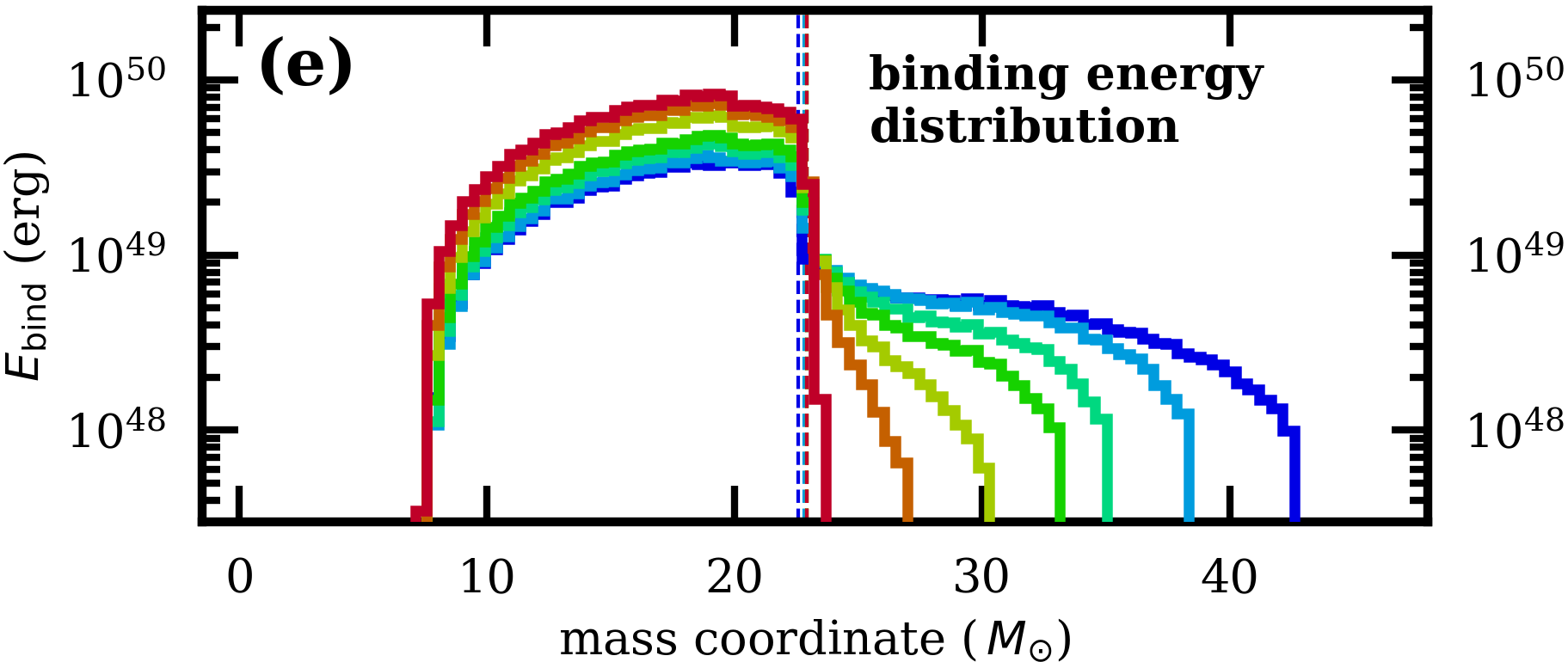}
    \phantomsubcaption
    \label{fig.Ebind_detail_16_01_e}
    \end{subfigure}
    \caption{Detailed look at the evolution of a $M_{\rm ZAMS} = 47.5 \msun$ model at $Z = 0.1\zsun$ metallicity. 
    The upper panel \ref{fig.Ebind_detail_16_01_a} shows the Kippenhahn diagram with stellar radius over-plotted in red.
    Note that the purple color shows regions of semiconvective mixing. 
    Panel \ref{fig.Ebind_detail_16_01_b} shows time evolution of the binding energy $E_{\rm bind}$ and its components: 
    $E_{\rm grav}$ and $E_{\rm int} = U_{\rm th} + E_{\rm rec}$ (see Eqn.~\ref{eq.Ebind}) together with the convective 
    envelope mass fraction $f_{\rm conv}$. In panels \ref{fig.Ebind_detail_16_01_c}, \ref{fig.Ebind_detail_16_01_d}, 
    and \ref{fig.Ebind_detail_16_01_e} we take a look at several internal profiles of the model, the position of which are marked in 
    panels \ref{fig.Ebind_detail_16_01_a} and \ref{fig.Ebind_detail_16_01_b} with dashed vertical lines in corresponding colors.
    The outer convective zone is marked in bold in panel~\ref{fig.Ebind_detail_16_01_c}.
    Solid vertical lines in panels \ref{fig.Ebind_detail_16_01_c}, \ref{fig.Ebind_detail_16_01_d}, and \ref{fig.Ebind_detail_16_01_e}
    mark the core-envelope boundary for the CE evolution (i.e. the bifurcation point at $X_{\rm H} = 0.1$).}
    \label{fig.Ebind_detail_16_01}
\end{figure}

In Fig.~\ref{fig.Ebind_detail_16_01} we illustrate this transition for a model with $M_{\rm ZAMS} = 47.5 \msun$ and $Z = 0.1\zsun$, 
which develops a deep outer convective envelope during late stages of core-helium burning and is an example of a model for which
the associated decrease in $E_{\rm bind}$ is very significant. Panel \ref{fig.Ebind_detail_16_01_a} shows the Kippenhahn diagram 
of the model (excluding 
the initial part of the MS when $X_{\rm H;C} > 0.55$) with stellar radius plotted in red.
In panel \ref{fig.Ebind_detail_16_01_b}, one can see that the binding energy $E_{\rm bind}$
slowly decreases during the core-helium burning phase and then decreases very strongly as the envelope
becomes deeply convective (and $f_{\rm conv}$ increases).

\setlength{\footskip}{50pt}

\begin{figure}
\centering
    \begin{subfigure}{1.0\columnwidth}
    \includegraphics[width=1.0\columnwidth]{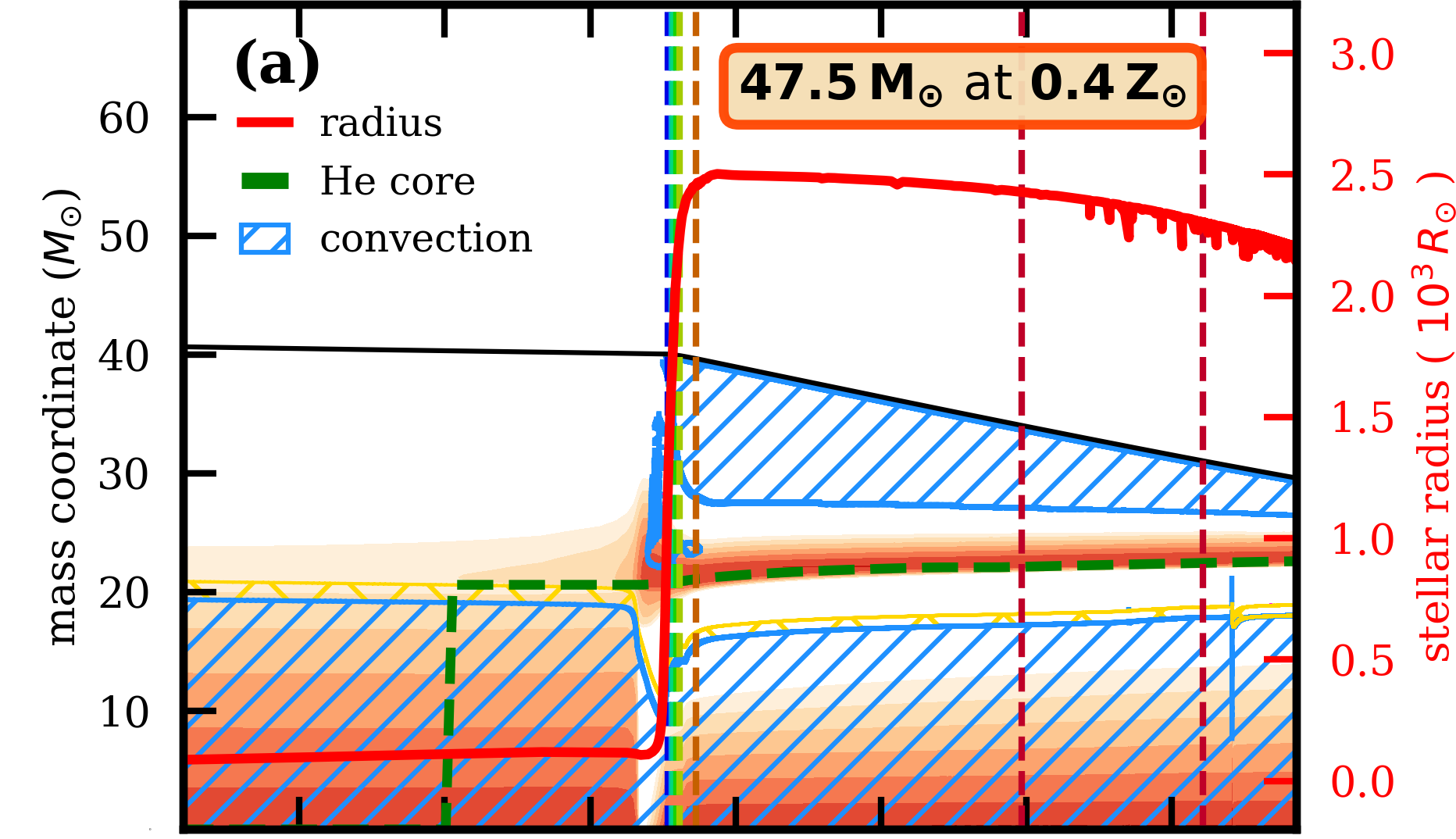}
    \phantomsubcaption
    \vspace{-1.2\baselineskip}
    \label{fig.Ebind_detail_16_04_a}
    \end{subfigure}
    \begin{subfigure}{1.0\columnwidth}
    \includegraphics[width=1.0\columnwidth]{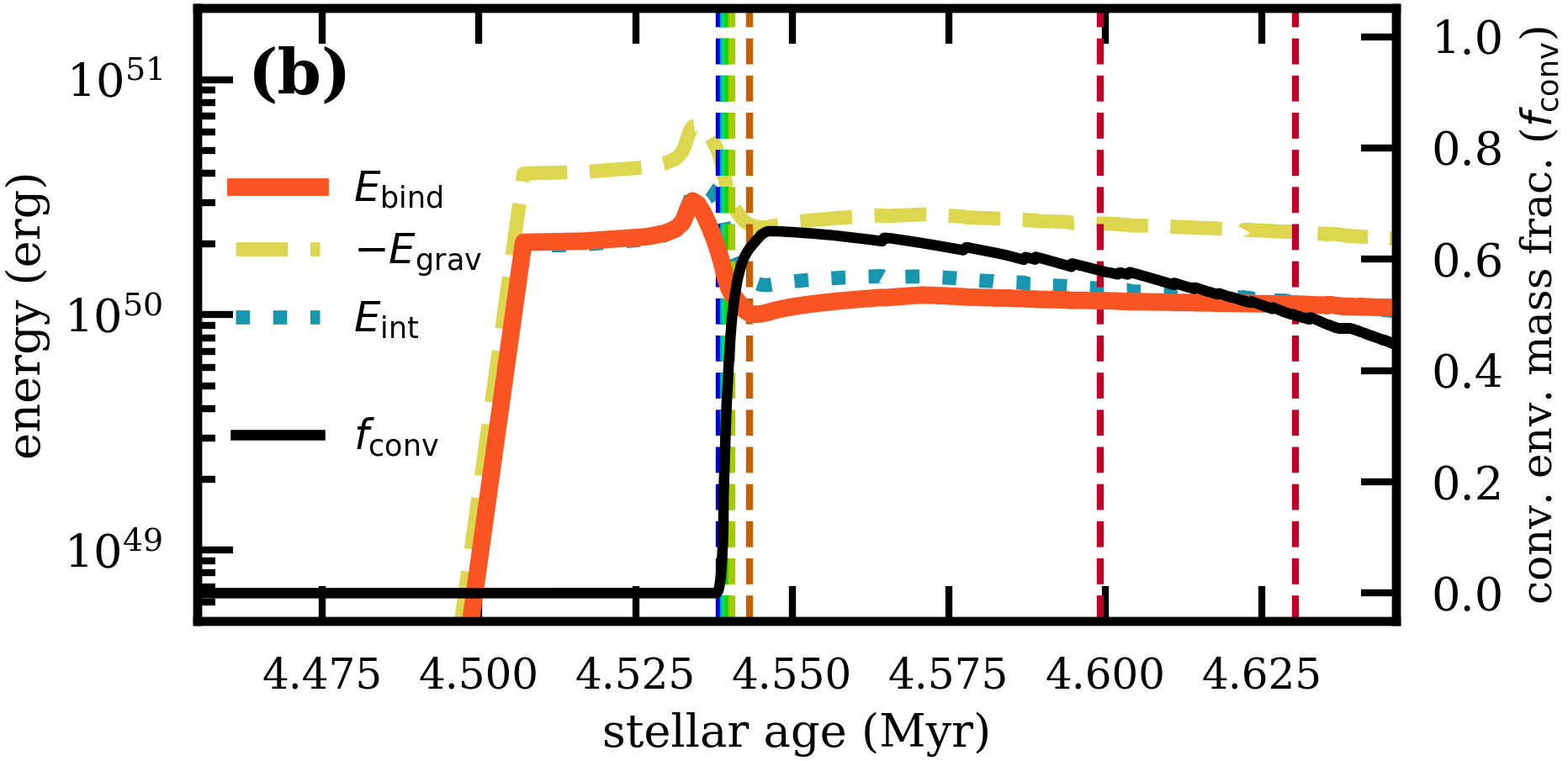}
    \phantomsubcaption
    \vspace{-0.68\baselineskip}
    \label{fig.Ebind_detail_16_04_b}
    \end{subfigure}
    \begin{subfigure}{1.0\columnwidth}
    \includegraphics[width=1.0\columnwidth]{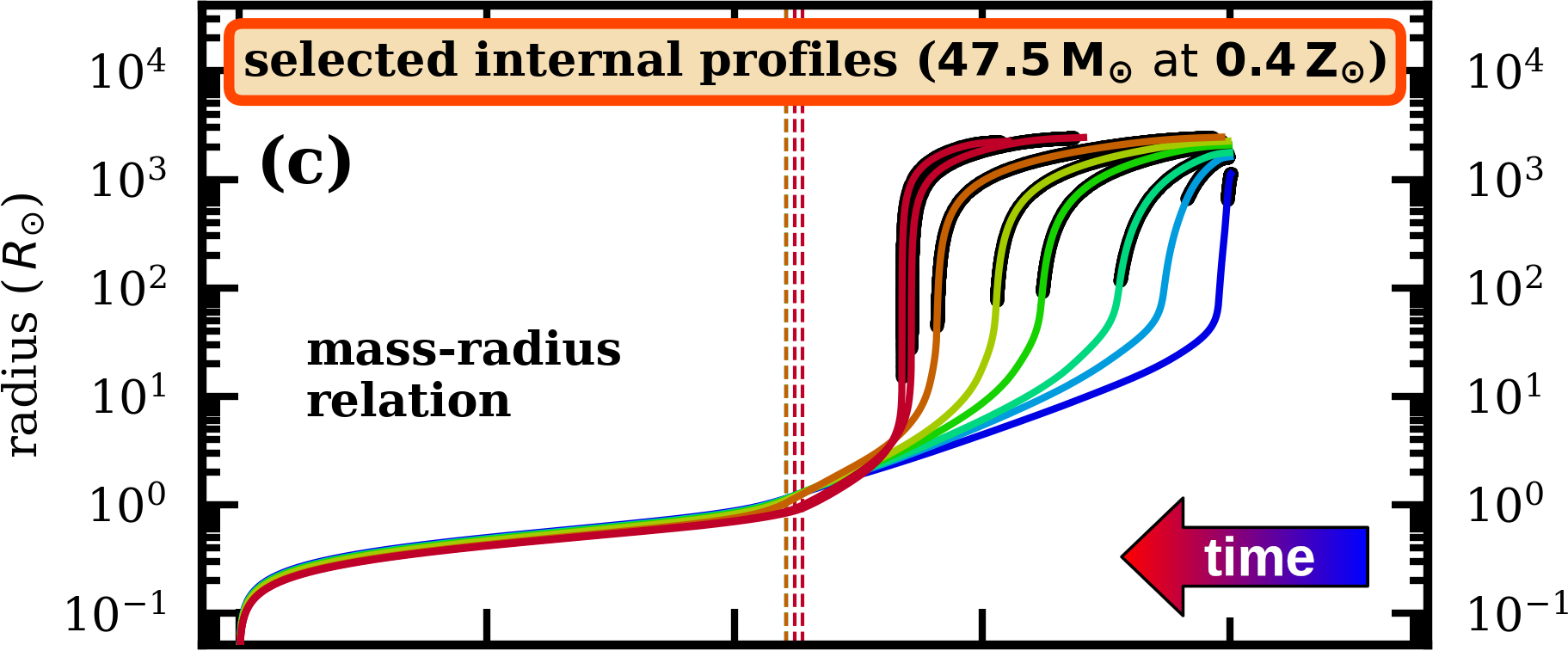}
    \phantomsubcaption
    \vspace{-1.2\baselineskip}
    \label{fig.Ebind_detail_16_04_c}
    \end{subfigure}
    \begin{subfigure}{1.0\columnwidth}
    \includegraphics[width=1.0\columnwidth]{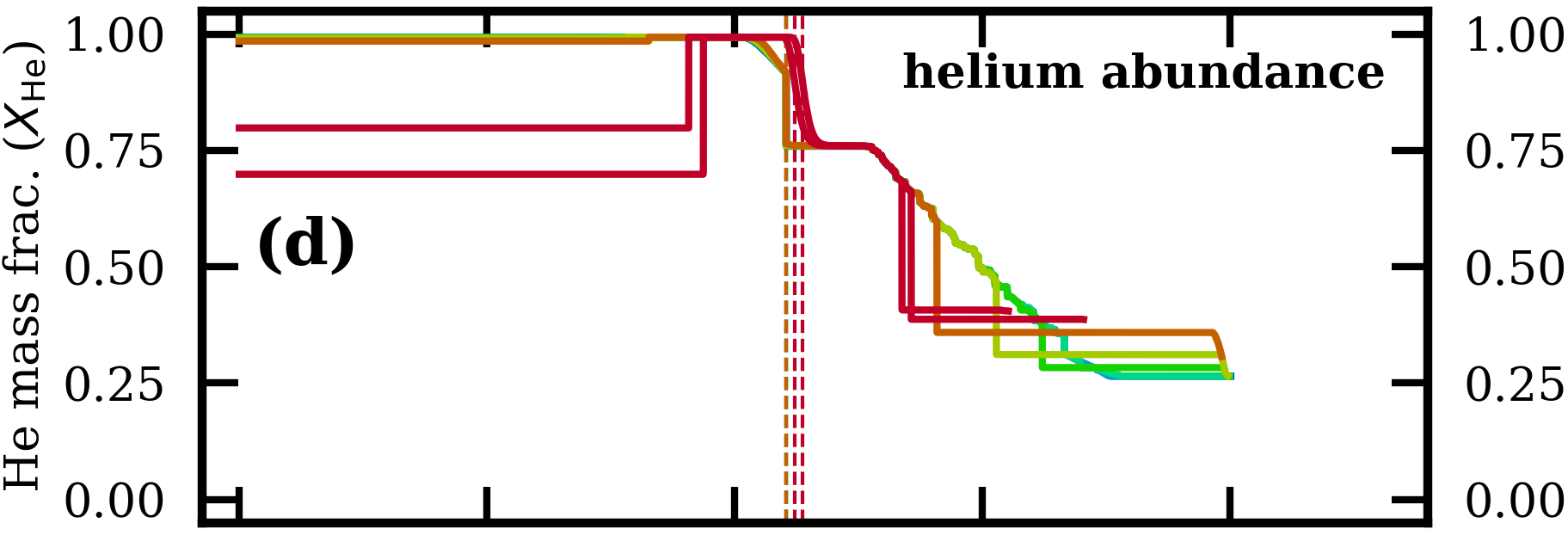}
    \phantomsubcaption
    \vspace{-1.2\baselineskip}
    \label{fig.Ebind_detail_16_04_d}
    \end{subfigure}
    \begin{subfigure}{1.0\columnwidth}
    \includegraphics[width=1.0\columnwidth]{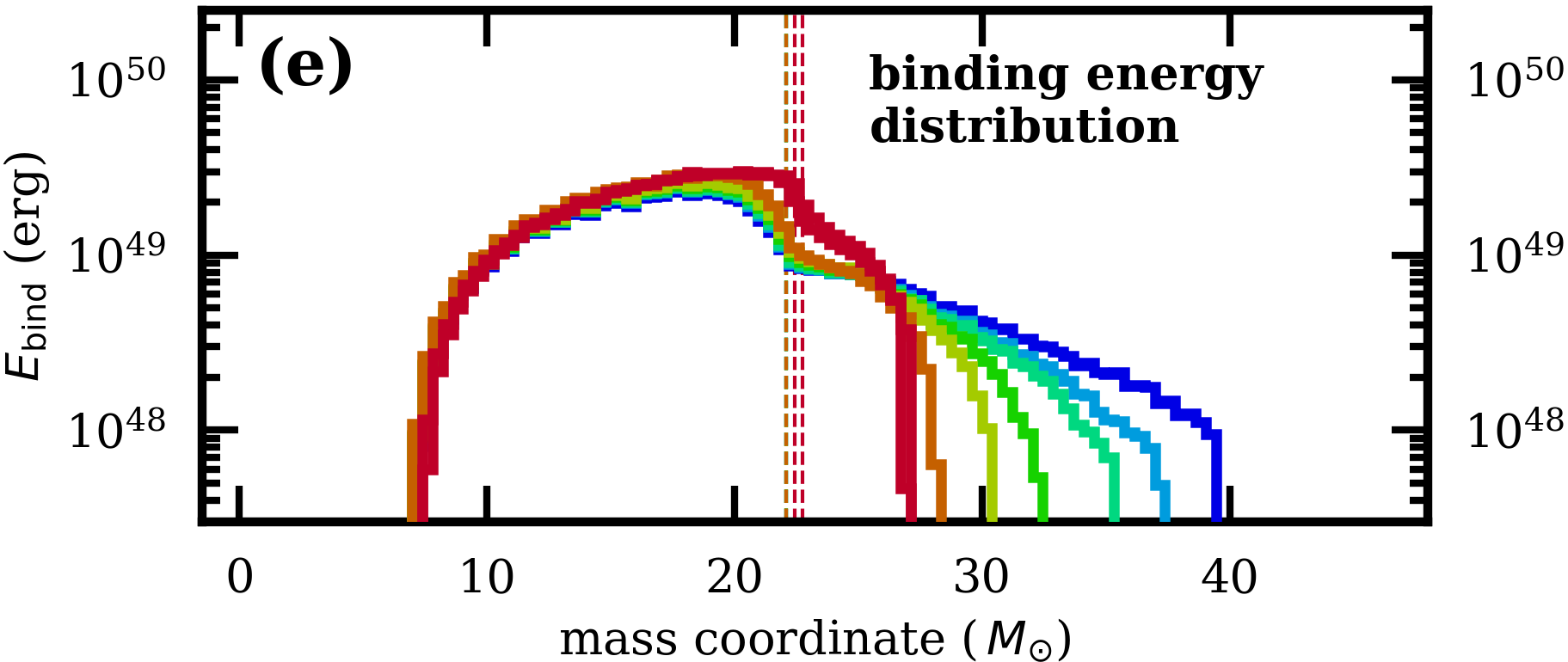}
    \phantomsubcaption
    \label{fig.Ebind_detail_16_04_e}
    \end{subfigure}
    \caption{Similar to Fig.~\ref{fig.Ebind_detail_16_01} but for a model
    with $M_{\rm ZAMS} = 47.5 \msun$ at $Z = 0.4\zsun$ metallicity.
    The two upper panels \ref{fig.Ebind_detail_16_04_a} and \ref{fig.Ebind_detail_16_04_b} are centered
    on the evolutionary phase during which the radius substantially increases, which is mainly the phase of rapid HG expansion.
    The different structural response of the envelope to the outer convective zone (panel~\ref{fig.Ebind_detail_16_04_c}), 
    associated with differences in the abundance profile (panel~\ref{fig.Ebind_detail_16_04_d}), lead to a much smaller impact 
    of the increasing $f_{\rm conv}$ on the envelope binding energy $E_{\rm bind}$ (see text for details).
    }
     \label{fig.Ebind_detail_16_04}
\end{figure}

In the next three panels, we take a closer look 
at the internal structure of the star at several selected moments during its evolution.
Panel \ref{fig.Ebind_detail_16_01_c} shows the internal mass-radius profiles, with 
the outer convective zone being marked in thicker lines.
The bifurcation point of each profile, i.e. the point that divides between the ejected envelope and the remaining 
remnant of the donor in a CE event, 
is marked with a solid vertical line located at $X_{\rm H} = 0.1$.
With time (i.e. profile colors changing from blue to red) 
the inner part of the star contracts as a result of continuous burning in the core.
On the other hand, as the envelope becomes more and more convective, the outer 
layers in the star move to larger radial coordinates. The point in between
the contracting and the expanding layers of the star, 
located at the mass coordinate of about $\sim 23 \msun$ in panel~\ref{fig.Ebind_detail_16_01_c},
is an important divergence point.
The helium abundance profiles in panel~\ref{fig.Ebind_detail_16_01_d}
reveals that the divergence point coincides with a steep composition jump at the boundary 
between the helium core (where $X_{\rm He} \approx 1$ in the helium shell) and the H-rich envelope. 
Importantly, it also closely coincides with the assumed location of the bifurcation point, 
marked with dashed vertical lines. In other words, the part of the giant above the bifurcation point, 
which would need to be ejected in a successful CE phase, is also the part that expands to larger radii coordinates as the outer envelope becomes 
more and more convective. As a result, the entire envelope becomes less gravitationally bound 
(see panel~\ref{fig.Ebind_detail_16_01_e}) and $E_{\rm bind}$ steeply decreases when $f_{\rm conv}$ increases.

\setlength{\footskip}{20pt}

As signaled before, giants which expand to develop a deep outer convective envelope already during the HG phase do not see
their envelope binding energies decrease in a significant way as a result of that. In Fig.~\ref{fig.Ebind_detail_16_04} 
we illustrate an example of such a case, a model with $M_{\rm ZAMS} = 47.5 \msun$ and $Z = 0.4\zsun$. In panels a-e we 
plot the same quantities as in the corresponding panels in Fig.~\ref{fig.Ebind_detail_16_01}. Note that the two upper panels 
\ref{fig.Ebind_detail_16_04_a} and \ref{fig.Ebind_detail_16_04_b} have a different scaling on the x-axis (linear) than panels 
\ref{fig.Ebind_detail_16_01_a} and \ref{fig.Ebind_detail_16_01_b} (logarithmic) as they only focus on the part of the evolution during
which the radius substantially increases (the part relevant for mass transfer interactions). 
The decrease of $E_{\rm bind}$ around the time $4.54$ Myr in panel~\ref{fig.Ebind_detail_16_04_b} is initiated 
by a contraction of the inner parts of the star after the end of MS and the associated expansion of the outer layers 
(before the outer envelope becomes convective), as also seen at the very beginning of $E_{\rm bind}$ evolution in panel~\ref{fig.Ebind_detail_16_01_b} for the $Z = 0.1\zsun$ giant. 
During the development of an outer convective envelope ($f_{\rm conv} \approx 0.6$) the binding energy decreases by less than a factor of 2.

The bottom three panels in Fig.~\ref{fig.Ebind_detail_16_04} illustrate the essential difference in the structure of a convective HG star 
compared to a more evolved convective giant from Fig.~\ref{fig.Ebind_detail_16_01}. As the outer convective zone extends deeper and deeper inside the envelope 
(panel~\ref{fig.Ebind_detail_16_04_c}), the outer layers of the star expand to larger radial coordinates, while the inner part of the star slightly contracts. 
This is a similar behavior to the one in panel~\ref{fig.Ebind_detail_16_01_c}. However, in the case of the HG giant in panel~\ref{fig.Ebind_detail_16_04_c},
the divergence point between the contracting and the expanding part of the star is no longer located very close to the helium-core boundary 
at a mass coordinate $\approx 23 \msun$ (where $X_{\rm He} \approx 1$) but it appears higher up inside the star at $\approx 26 \msun$. 
We find that such a different location of the divergence point in HG giants is associated with a distinctively different internal abundance profile, 
see panel~\ref{fig.Ebind_detail_16_04_d}.
The composition jump near the helium core where $X_{\rm He} \approx 1$ is smaller than the one in panel~\ref{fig.Ebind_detail_16_01_d}:
the helium abundance drops only down to $X_{\rm He} \approx 0.75$, forms a plateau, and then starts to gradually decrease again at a mass coordinate of 
about $26 \msun$. One can notice that the divergence point is roughly associated with the secondary drop when $X_{\rm He}$ decreases below $\sim 0.75$.
Consequences of the structural change on the binding energy distribution within the envelope 
can be seen in panel~\ref{fig.Ebind_detail_16_04_e}. The $\sim 2-3 \msun$ of helium-dominated material ($X_{\rm He} \approx 0.75$) at 
the bottom of the envelope move slightly inward inside the star, their binding energy increases, which counteracts the decreasing binding energy of the outer 
convective layers. As a result, the overall binding energy of the envelope stays roughly unaffected.

We find that the example in Fig.~\ref{fig.Ebind_detail_16_04} is representative for models of giants
that reach the convective-envelope stage during HG expansion. 
The reason for the crucial difference in helium abundance profiles between panels \ref{fig.Ebind_detail_16_01_d} and \ref{fig.Ebind_detail_16_04_d} 
has to do with the size of an intermediate convective zone (ICZ)\footnote{Note that the term intermediate convective zone sometimes also refers 
to a region of semiconvective and convective mixing that can take place above the H-burning core already during the MS, see for instance \citet{Farmer2016}.
Here, we follow a slightly different nomenclature \citep[e.g.][]{Langer1985,Kaiser2020}.}
that develops in a star on top of the hydrogen-burning shell 
right after the end of MS, i.e. during the phase of rapid HG expansion when the star is out of thermal equilibrium \citep[e.g.][]{Langer1985,Schootemeijer2019,Kaiser2020}. 

In models that already at this stage expand all the way to the red-giant branch and develop an outer convective envelopes the ICZ tends to quickly 
disappear, as in the model in Fig.~\ref{fig.Ebind_detail_16_04} and as also found by previous authors \citep[e.g.][]{Langer1985,Schootemeijer2019,Davies2019,Kaiser2020}.
In contrast, in models that expand less during the HG stage and stably burn helium as blue or yellow supergiants the ICZ does not disappear quite as quickly and becomes
more extended (in mass coordinate), as in the model in Fig.~\ref{fig.Ebind_detail_16_01}.
A more extended ICZ means that more hydrogen is mixed from the H-rich envelope into the He-rich layers above the 
helium core and, as a result, the composition jump at the core-envelope boundary is larger and the helium abundance at the bottom of the envelope is smaller,
as can be seen by comparing Fig.~\ref{fig.Ebind_detail_16_01_d} with Fig.~\ref{fig.Ebind_detail_16_04_d}. Another difference between the models in Fig.~\ref{fig.Ebind_detail_16_01} 
and Fig.~\ref{fig.Ebind_detail_16_04} is that in the $0.1 \zsun$ case the H-burning shell remains convective throughout all of core-helium burning
while in the $0.4 \zsun$ case the ICZ disappears and the H-burning shell becomes radiative. It turns out that, compared to the role played by the ICZ 
which is present in both cases, the nature of the H-burning shell is not 
an important factor for the abundance profiles shown in Fig.~\ref{fig.Ebind_detail_16_01_d} with Fig.~\ref{fig.Ebind_detail_16_04_d}: 
these abundance patters are already determined by the ICZ.

The reasoning presented above explains the relation between the degree of expansion during the HG phase, the size of the ICZ, and the abundance profile of deep envelope layers.
However, the exact extent of the ICZ in a given model has to be considered highly uncertain. It is sensitive to the assumed efficiency of semiconvective mixing \citep{Schootemeijer2019},
convective overshooting \citep{Kaiser2020}, 
numerical accuracy of stellar models \citep[see App.~\ref{sec.App_CE_acc} and][]{Farmer2016}, or treatment of convective boundaries in regions with steep composition changes
(see for example the \texttt{convective premixing}
scheme introduced in MESA \citet{Paxton2019}). But perhaps most importantly, the extend of the ICZ likely depends on
the abundance profile above the H-burning shell left behind by the previous evolution. It is therefore sensitive to rotational mixing \citep{Georgy2013} and 
to (semi)convective mixing above the H-burning core already during the MS \citep{Farmer2016}. Additionally, in the case of a past accretor from previous mass transfer phases
(which is most definitely relevant for the case of companions in BH binaries studied here), the abundance profile above the H-burning shell can be significantly different compared to 
an unperturbed single stellar model due to rejuvenation and growth of the H-burning core of the mass gainer \citep{Hellings1983,Hellings1984,Braun1995,Wellstein2001,Dray2007}.

It is also worth mentioning that the relation between the degree of HG expansion and the degree of mixing in the ICZ operates both ways: if the mixing is not 
very efficient then a certain model may expand to the RSG stage whereas the same model but with efficient mixing (for instance by semiconvection 
or rotationally induced mixing) will in some cases expand much less and burn helium as a blue supergiant \citep[see App.~B of][]{Klencki2020}. 
This allows for a calibration of models by the observed ratio of blue to red supergiants. For example, the post-MS expansion of the models used in this 
study was calibrated to match the ratio observed in the SMC.

In summary, the internal structure and chemical profiles presented here suffer from 
uncertainties in various internal mixing processes, numerical challenges associated with the ICZ, and are likely affected by a previous accretion phase. 
In the same time, however, it seems reasonable to expect some systematic differences in the envelope structure (and the binding energy) between stars that become convective 
during the HG phase and those that only do so much later during there evolution. The examples shown in Fig.~\ref{fig.Ebind_detail_16_01} and Fig.~\ref{fig.Ebind_detail_16_04}
offer some guidance on how these differences could look like.


It is interesting to speculate about the consequences of different envelope structures of giants seen in
Fig.~\ref{fig.Ebind_detail_16_01} and Fig.~\ref{fig.Ebind_detail_16_04} for the CE evolution itself, 
in particular given that the location of the divergence point in the $Z = 0.4 \zsun$ model is substantially
different than that of the bifurcation point (under usual assumptions). We discuss this 
further in Sec.~\ref{sec.disc_composite_scenario}. The examples given in Fig.~\ref{fig.Ebind_detail_16_01} 
and Fig.~\ref{fig.Ebind_detail_16_04} show that the internal envelope structure of a convective-envelope supergiant 
can be significantly depend on the evolutionary stage of the star.

\section{Discussion}
\label{sec:discussion}

\subsection{How robust is the prediction that CE evolution in BH binaries requires RSG donors?}

\label{sec.disc_uncert}

In Sec.~\ref{sec.res_CEeject} we found that the parameter space for CE ejectability in BH/NS binaries is practically 
limited to donors with outer convective envelopes, i.e. cool supergiants (RSGs, see Fig.~\ref{fig.HRD}).
This conclusion was reached by pursuing a very optimistic case of the energy budget, with multiple assumptions in favor
of an easier CE ejection (Sec.~\ref{sec.model_summary}). In particular, we assumed that the entire energy input from 
orbital shrinkage, internal energy of the envelope, and recombination energy, as well as an energy input from accretion 
can be transferred into kinetic energy and used to help unbind the envelope. We also assumed no energy loss in any form
(i.e. $\alpha_{\rm CE} = 1.0$), neglecting energy sinks such as radiation from the surface or excess kinetic energy in outflows.
These assumptions are clearly extreme. For instance, in view of hydrodynamic simulations, a more realistic assumption
would be $\alpha_{\rm CE} \lesssim 0.7$ or even $\alpha_{\rm CE} \sim 0.1$ during the possibly reached self-regulated phase of
the CE evolution (see Sec.~\ref{sec.method_energy_budget} for details).
These results suggest that the parameter space for a successful CE ejection in Figs.~\ref{fig.sep_6} 
and \ref{fig.HRD} is most likely an upper limit on the true (realistic) CE ejectability, see for e.g. Fig.~\ref{fig.sep_6_real}. 

From the observational side, various attempts have been made in the past to constrain the CE ejection efficiency $\alpha_{\rm CE}$
by linking post-CE systems to their reconstructed pre-CE progenitors \citep[e.g.][]{Nelemans2000,Zorotovic2010,Zorotovic2011,deMarco2011,Davis2012,PortegiesZwart2013}.
While most of these results suffer from uncertainties in the reconstruction technique of pre-CE parameters and the $\alpha_{\rm CE}$ values they yield
are degenerated with the binding energy parameter $\lambda_{\rm CE}$, for the majority of the post-CE systems the inferred 
values of $\alpha_{\rm CE}$ are clearly below $1.0$, although in some cases only when the internal envelope energy is included
\citep[see for instance ][for a discussion of the impact of thermal energy on the inferred $\alpha_{\rm CE}$ values]{deMarco2011}.
It is worth pointing out, however, that the formation of three double helium white dwarfs analyzed by \citet{Nelemans2000} as well as 
one of the post-CE systems analyzed by \citet{deMarco2011} seemingly cannot be reconstructed without $\alpha_{\rm CE} > 1$. 
This might be a signal of the limitations of the energy budget formalism, some of which we will discuss below. This might also 
be an indication that CE evolution does not always take place when we expect it to. Another type of systems for which 
$\alpha_{\rm CE} > 1.0$ values were advocated, perhaps more relevant in the context of massive stars and BBH merger progenitors, 
are BH low-mass X-ray binaries \citep[BH-LMXBs, e.g.][]{Kalogera1999,Podsiadlowski2003}.
In Sec.~\ref{sec.disc_BHLMXB} we analyze their case in more detail and argue that $\alpha_{\rm CE} \leq 1.0$ could potentially be sufficient.

There are two caveats of our method: the possible contribution of other energy sources in CE ejection (unaccounted for 
in our energy budget) and lower envelope binding energies.
Apart from the energy terms considered in this work, other energy sources have been discussed in the literature. 
One possible addition is the energy from nuclear burning \citep[e.g.][]{Ivanova2003}. However, as pointed out by 
\citet{Ivanova2013}, when the donor's envelope is lifted from the core, the nuclear energy input is most likely going
to decrease with respect to radiative losses from an expanding emitting area.\footnote{An exception could be non-degenerate
accretors, e.g. low-mass MS stars \citep{Podsiadlowski2010}.}

Another addition to the energy budget that has been proposed is an enthalpy term $P/\rho$ \citep{Ivanova2011_enthalpy}. 
While enthalpy is primarily responsible for energy redistribution \citep[rather than being a new energy source][]{Ivanova2013},
it leads to solutions with quasi-steady outflows from envelopes even before their total energy becomes positive. However,
this requires the CE ejection to happen on a long thermal timescale (e.g. after a self-regulated spiral-in phase), at which
point radiative losses increase significantly and our assumptions with $\alpha_{\rm CE} = 1.0$ are extremely optimistic 
(e.g. \citealt{Clayton2017} find $\alpha_{\rm CE}$ between 0.046 and 0.25).

Here, we considered energy input from accretion $\Delta E_{\rm acc}$ estimated as Eddington luminosity of the compact 
accretor times 1000 yr, similarly to \citet{Voss2003} and \citet{Kruckow2016}. It should 
be noted that the accretion rate does not need to be Eddington limited due to neutrino cooling and photon trapping 
inside the accretion flow inside the CE \citep[e.g.][]{Houck1991,Edgar2004,MacLeod2015a}. In fact, 
in rare instances super-Eddington accretion during a CE phase could produce BHs with masses in the PISN mass gap \citep{vanSon2020}.
While the luminosity that is radiated away and absorbed by the surrounding envelope is still likely of the order of the Eddington luminosity, 
the total accretion luminosity estimated as $\eta \dot{M} c^2$ (with $\eta \sim 0.1$) could be in some cases much 
higher \citep{MacLeod2015b,MacLeod2017,De2020}. If the accretion is taking place through accretion disk then polar 
outflows or jets could serve as a way of transferring this super-Eddington power to the envelope, thus helping in 
its ejection \citep{Armitage2000,Soker2015}. However, among other uncertainties of this process, it is unclear whether 
a persistent accretion disk forms around an embedded compact object. Hydrodynamic simulations by \citet{Murguia-Berthier2017} 
suggest that accretion through a disk is a rare and transitory phase, which may occur when a BH or a NS passes through zones 
of partial ionization in the outer envelope layers. 

Apart from the uncertainties of the CE energy budget discussed above, there are caveats associated with calculations 
of envelope binding energies $E_{\rm bind}$ of massive giant donors. 
First, the value of $E_{\rm bind}$ can be very sensitive to the location of the bifurcation point, i.e. the lower 
limit in the integral in Eqn.~\ref{eq.Ebind}. We discuss this further as a partial envelope ejection scenario in Sec.~\ref{sec.disc_composite_scenario}.
Second, we used single stellar models to infer binding energies of stars which have most likely been mass gainers in 
a mass transfer phase in the past. While this is a common practice, the envelope structure of a rejuvenated star could be quite different from that of a 
single unperturbed star of the same mass \citep{Braun1995,Wellstein2001,Dray2007}.
Third, $E_{\rm bind}$ values computed from single stellar models represent the binding energy at the point of RLOF 
rather then at the onset of the CE phase. In reality, before the mass transfer rate increases sufficiently and the system
becomes dynamically unstable, a giant donor is likely going to lose a sizable part of its envelope as a result of 
pre-CE mass transfer \citep{MacLeod2020}. This could mean that $E_{\rm bind}$ values relevant for the CE evolution 
are smaller than calculated in Sec.~\ref{sec.res_Ebind}. On the other hand, because most of the binding energy is
located in deep envelope layers (those close to the helium core), a loss of even $30\%$ of mass from the outermost
layers of a giant star does not necessarily have any significant impact on the total binding energy 
(Klencki et al., in prep.). Additionally, a pre-CE mass transfer from a donor that is more massive then the accretor is 
going to shrink the orbital separation even before the onset of the CE phase.

In summary, unless jets play a crucial role in driving CE ejection \citep[e.g.][]{Soker2015,Shiber2019}, our model for the CE outcome and binary survival is
optimistic with respect to more realistic assumptions. As such, the expectation that CE ejection in BH/NS binaries 
requires RSG donors at the moment appears quite robust. A possible alternative scenario of a partial envelope ejection 
followed by an immediate phase of stable mass transfer is suggested in Sec.~\ref{sec.disc_composite_scenario}.

\subsection{Upper luminosity limit of RSGs: indication of a maximum mass beyond which the CE channel does not operate?}

\label{sec.disc_Lrsgmax}

We have shown that according to the current understanding of the CE evolution 
and based on unperturbed single star models of the donor star, a CE ejection in a BH binary is only 
possible if the CE phase was initiated by a convective-envelope donor. Observationally, such donors are cool
supergiants (RSGs), see Fig.~\ref{fig.HRD}. Interestingly, there appears to be an upper RSG luminosity limit of 
$\logL_{\rm RSG;max} \approx 5.8$, which corresponds to $M_{\rm ZAMS} \approx 40 \msun$, above which no RSGs are
observed in the Milky Way or several other local galaxies of different types and compositions (see Sec.~\ref{sec.HRD}
and references therein). This leads to the question whether stars with $M_{\rm ZAMS} \gtrsim 40 \msun$ ever expand 
to become RSGs with significant outer convective envelopes, or whether the apparent limit $\logL_{\rm RSG;max}$ indicates 
a maximum donor mass of about $\sim 40\msun$ beyond which the CE channel does not operate. 
Such a limit would correspond to an upper limit on the secondary BH mass at about $\sim 20 \msun$ (mass of the
helium core of a $40 \msun$ giant) as well as imply an upper limit on the primary BH mass at 
$M_{\rm BH;max} = M_{\rm RSG;max} / q_{\rm crit;conv} \approx 25$-$30\msun$ for $M_{\rm RSG;max} = 40 \msun$ and $q_{\rm crit;conv} = 1.5$. 
This is in tension with observations as about half of the BBH merger detections reported to date are consistent
with total binary masses beyond $50 \msun$ \citep{LIGO_o1o2sum}.
We not that this concern was first raised by \citet{Mennekens2014} who argued that because stars with masses above $40 \msun$
do not become RSGs, the formation of the most massive BBH mergers from isolated binary evolution is rather unlikely.

On one hand, it is possible that radial expansion of the most massive stars is quenched due to extensive mass loss
before they reach the RSG stage \citep[e.g.][]{Vanbeveren1991,Vanbeveren1998,Smith2014}. This is in line with the fact that the empirical HD limit in the
upper right corner of the HR diagram \citep{Humphreys1979,Ulmer1998}, beyond which almost no stars are observed in
the Milky Way and in the LMC, coincides with a significant increase in stellar winds and the occurrence of the luminous
blue variable phenomenon \citep{Humphreys1994,Smith2011,Grafener2012,Jiang2015,Jiang2018}. Alternatively, the upper right corner of the 
HR diagram might be unpopulated because of a currently unknown process that would prevent the formation of extended
superadiabatic layers and density inversions in radiation-dominated envelopes of massive giants that evolve close to 
the Eddington limit, see App.~A of \citet{Klencki2020}. Stellar engineering solutions which effectively mimic such a
process in 1-D stellar codes (motivated to some extend by numerical difficulties that occur otherwise) produce evolutionary
tracks with no RSGs above a certain luminosity limit. In the case of stars that were mass transfer accretors in the 
past (which is likely the case for stellar companions in BH binaries), they were shown to avoid the RSG stage in the 
non-rejuvenated cases, which could be relevant for stars that accreted during late MS stages or during their post-MS evolution \citep[e.g.][]{Braun1995},
similarly to products of case B stellar mergers \citep{Vanbeveren2013,Justham2014}.

On the other hand, it is also possible that stars with $M_{\rm ZAMS} \gtrsim 40 \msun$ do expand to the RSG stage but
their RSG lifetime is very short (e.g. several hundred years), which is why they are not observed. This could be due 
to high mass loss rates from loosely bound and turbulent convective envelopes of RSGs and a blueward evolution once 
most of the envelope is lost, as could be obtained in stellar tracks \citep[e.g.][]{Chen2015}. It is also possible, 
especially at low metallicity \citep[e.g.][]{Klencki2020}, that the RSG stage is reached very late into the evolution 
of a massive star, once most or all of the helium has been burned in the core, resulting in a short RSG lifetime 
\citep{Higgins2020}.

At the moment, it remains unclear whether or not stars above $M_{\rm ZAMS} \approx 40 \msun$ expand to the RSG stage
during their evolution, especially at very low metallicity environments for which the observations are sparse. Therefore,
the CE evolution channel could in principle produce BBH mergers with both component masses all the way up to the 
PISN mass gap. However, the observational lack of RSGs above a certain luminosity limit
is a warning sign which ultimately needs to be addressed in detail. Alternatively, if one becomes convinced 
that most of the BBH merger population originates from the CE evolution channel, then their rate might be an indication 
that at least some of the stars more massive than $40 \msun$ do reach the RSG stage but have so far eluded observation.

\subsection{Implications for population synthesis and merger rates of compact binaries}

\label{sec.disc_popsynth}

Here we discuss the implications of our findings for the formation rate of compact binary mergers through the CE evolution 
channel. Given that the parameter space for mass transfer from convective-envelope donors is very small and limited to 
a narrow range in orbital periods (see Fig.~3 in \citealt{Klencki2020}), one may wonder whether it is possible for the 
CE evolution channel alone to explain the BBH merger rate inferred from observations, as reported by many authors in
population synthesis calculations. At face value, this is not necessarily a problem because only a small fraction of
all massive binaries need to finish their evolution as merging BBH systems in order to explain the local merger rate 
\citep[roughly 1 every 100 systems, e.g.][]{Chruslinska2019,Neijssel2019,Giacobbo2020}. In practice, however, compact
binary mergers in population synthesis models originate from a much larger binary parameter space, with many systems 
evolving through a CE phase initiated by a non-convective donor \citep{deMink2015,Klencki2018}. For instance, in a
recent state-of-the-art population synthesis model by \citet{VignaGomez2020} only $\sim 20\%$ of CE episodes leading
to the formation of a BNS merger were initiated by a cool supergiant with an outer convective envelope (in their 
dominant channel for the BNS formation).

There are two main reasons for why, according to our findings, population synthesis models tend to overpredict the number of systems that evolve 
through and survive a CE phase. First, following the \textsc{BSE} code \citep{Hurley2002}, it is customary for its 
successors to rely on evolutionary type of the donor \citep[as defined in][]{Hurley2000}, rather then the type of
its envelope, in order to choose the appropriate critical mass ratio or the value of $\zeta_{\rm ad}$ to determine 
mass transfer stability. In some cases this has led to a mistake of applying stability criteria for 
convective-envelope donors (i.e. more prone to evolve through a CE phase) to all stars of the core-helium burning CHeB type:
notably in the \text{StarTrack} population synthesis code \citep{Belczynski2008,Dominik2012,Belczynski2016} as well as 
in the \textsc{COMPAS} code \citep{Stevenson2017,VignaGomez2018}. Donors of the CHeB type are especially common at 
subsolar metallicity environments, however, most of them have outer radiative envelopes and only those that have expanded
to the red-giant branch can be treated as convective-envelope donors \citep[see Fig.~3 and Sec.~3.4 of][]{Klencki2020}.
This mistake has led to an artificially increased number of CE evolution cases from CHeB donors in population synthesis,
especially at low metallicity and in the mass regime of BH progenitors. Besides having an impact on the merger rates, it 
could be partly responsible for the significant preference for low metallicity in the formation of BBH mergers predicted by 
\citet{Belczynski2010} as well as more recent population synthesis models.\footnote{The low metallicity is likely favored 
regardless, due to weaker stellar winds, but possibly to a lesser extent than in most population synthesis models.} 

Secondly, the envelope binding energies of massive radiative-envelope giants could be underestimated in some population
synthesis calculations, which tend to assume $\lambda_{\rm CE} = 0.05$
\citep[e.g.][]{Dominik2012,Belczynski2016,Stevenson2017,Neijssel2019} or $0.1$ \citep{Mapelli2017,Giacobbo2018},
while lower values of the order of $\sim 0.01$ would be more appropriate for stars with $M_{\rm ZAMS} \gtrsim 30 \msun$ 
at the giant stage before an outer convective-envelope is 
formed\citep[see Fig.~\ref{fig.App_lambda_6} as well as][]{Podsiadlowski2003,Loveridge2011,Wang2016,Kruckow2016}.
\footnote{Whether $\lambda_{\rm CE}$ increases again once the outer convective envelope develops may depend 
on the evolutionary stage (see Sec.~\ref{sec:res_details}), with convective giants above a certain metallicity-dependent
mass possibly maintaining $\lambda_{\rm CE} \approx 0.01$.}
The assumption of $\lambda_{\rm CE} = 0.05-0.1$ is usually justified by the fits obtained by \citet{Xu2010}. However,
these authors only computed models with masses up to $M_{\rm ZAMS} = 20 \msun$. In a later paper from the same group, 
\citet{Wang2016} extended the fits to higher masses and showed a further decrease of $\lambda_{\rm CE}$ down to $\sim 0.01$ 
for more massive and larger radiative-envelope giants, similar to other studies as well as in line with the results presented here in
Sec.~\ref{sec.res_Ebind}. In the case of convective-envelope giants, also those with $M > 20 \msun$, the binding energies can be much smaller
(reaching $\lambda_{\rm CE}$ values as large as $\sim 0.2-0.3$), although that could strongly depend on the evolutionary 
stage of the star (see Sec.~\ref{sec:res_details}). 

It should be noted that for the lowest metallicity studied here ($Z = 0.01\zsun$), we find a significant parameter space 
for CE ejectability also in the case of non-convective donors, although only those with masses $M_{\rm ZAMS} < 50 \msun$
(see Fig.~\ref{fig.sep_6} and Fig.~\ref{fig.sep_6_real}). One may wonder whether that could be enough to reproduce the BBH merger rate.
While in the current population synthesis models the formation of BBH mergers is indeed the most efficient at very low metallicity $Z \approx 0.01 \zsun$
\citep[e.g.][]{Klencki2018,Neijssel2019,Graziani2020}, its contribution has to be weighted by the corresponding fraction of the star 
formation rate (SFR) at $Z \approx 0.01 \zsun$. \citet{Chruslinska2019b} shows that this fraction is negligible in the local Universe and that it is still
small within redshifts $z < 3$ (see their Fig. 8 and C.1). 
As a result, metallicity of the BBH merger population may actually peak at $Z > 0.01 \zsun$ (Broekgaarden et al. in prep).

In summary, if our approach accurately predicts the limits on CE ejectability in BH/NS binaries and the CE survival 
is indeed only possible in the case of convective-envelope donors (see Sec.~\ref{sec.disc_uncert} for caveats), then the formation of compact binary mergers could be significantly
overestimated in some population synthesis models (i.e. by a factor of a few or more), especially at low metallicity.
Additionally, it remains an open question whether or not massive stars ($\gtrsim 40 \msun$) ever expand to become
RSGs with outer convective envelopes during their evolution (sec.~\ref{sec.disc_Lrsgmax}). If the answer is negative
then that would further reduce the BBH merger population that could originate from the CE evolution channel
as well as indicate an upper limit on the primary BH mass at $\sim 25$-$30\msun$ and the total BBH merger 
mass at about $\sim 50 \msun$, in tension with the most massive BBH mergers detected to date  \citep[see Sec.~\ref{sec.disc_Lrsgmax} for details and also][]{Mennekens2014}.
It is worth noting that in most population synthesis models stellar tracks for $M_{\rm ZAMS} > 40 \msun$ do expand to reach the RSG stage,
with RSG luminosities above the observational upper limit at $\logL_{\rm RSG;max} \approx 5.8$.

\subsection{Tidal spin-up in close BH-WR binaries}

\label{sec.BH_WR_spinup}

In the previous sections we discussed how the formation of BBH mergers through the CE evolution channel is most likely 
only possible if the CE phase is initiated by a massive convective-envelope donor, i.e. a cool supergiant. 
Interestingly, low metallicity models of massive stars ($Z \lesssim 0.2 \zsun$) often expand to the convective-envelope 
stage very late in their evolution, during advanced burning stages after core-helium depletion, just several thousand years
away from the core-collapse \citep[see Fig.~\ref{fig.sep_6} as well as Fig.~3 in][]{Klencki2020}. 

\begin{figure}
      \includegraphics[width=\columnwidth]{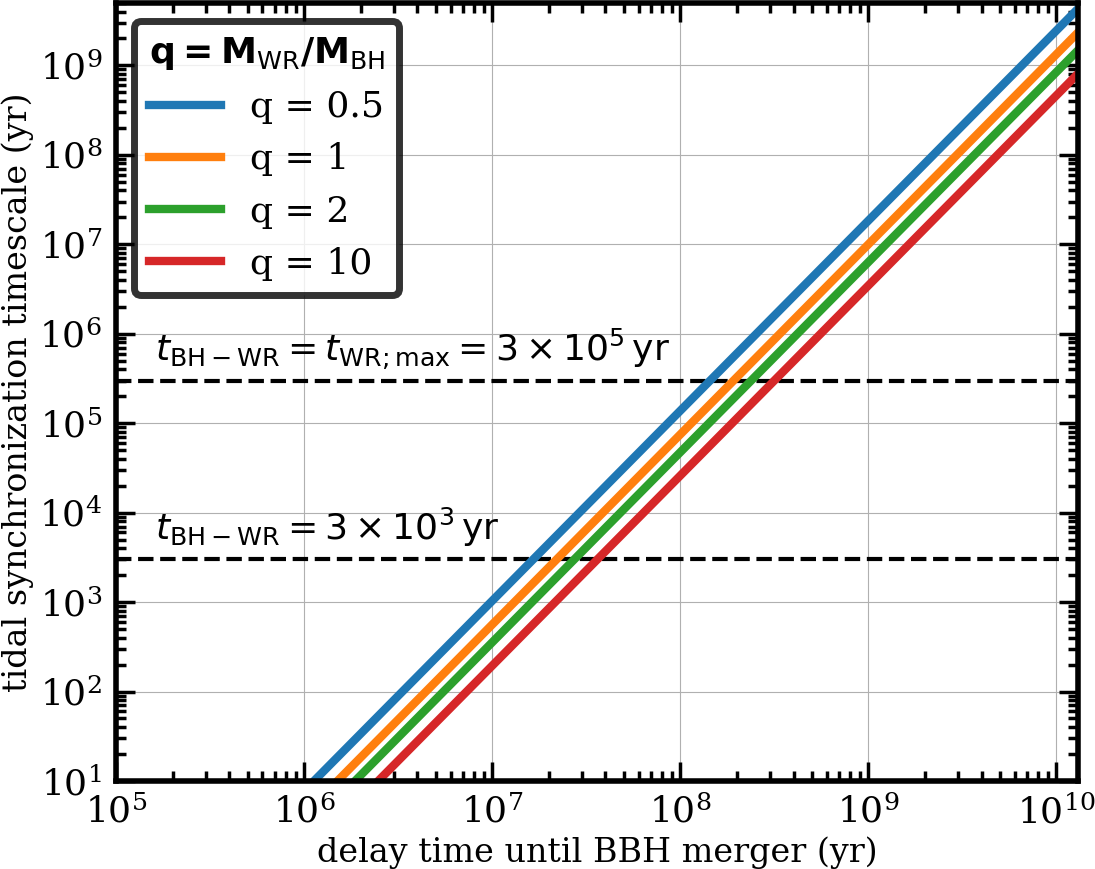}
    \caption{Relation between the tidal synchronization timescale in a BH-WR binary and the delay time (between the 
    formation and the merger) of the corresponding BBH system, after \citet[][see their Eqn.~14.]{Kushnir2016}. 
Plotted for several different mass ratios $q = M_{\rm WR} / M_{\rm BH}$. Tidal interactions can efficiently spin-up
the WR star only if the synchronization timescale is shorter than the BH-WR binary lifetime $t_{\rm BH-WR}$. The two 
dashed horizontal lines mark two different values of $t_{\rm BH-WR}$: roughly the maximum WR lifetime $\sim 3 \times 10^5 \rm \, yr$,
assumed by several authors \citep[eg.][]{Kushnir2016,Hotokezaka2017a,Piran2018}, and a significant smaller value
$t_{\rm BH-WR} = 3 \times 10^3 \rm \, yr$ which is more realistic in the case of the CE channel at low metallicity.}
    \label{fig.disc_kushnir}
\end{figure}

The remaining lifetime of the donor star is a strict upper limit on the duration of the subsequent BH-WR stage, 
i.e. a stage which likely follows right after the CE is ejected. Close BH-WR binaries are speculated to be the most 
immediate progenitors of BBH systems that merge within the Hubble time. In fact, several examples of such systems are known and observable as bright X-ray 
sources \citep[eg.][]{Carpano2007,Bulik2011,Belczynski2013,Liu2013}. It was proposed that the interplay between tidal 
interactions (spin-up) and the wind mass-loss (spin-down of the WR star) during the BH-WR stage is crucial for 
the final WR spin value and that in systems with a sufficiently small separation the WR star should 
always be critically spinning at core-collapse \citep{Kushnir2016}. Several authors predict that if  some of the BBH
LIGO/Virgo sources were formed by binary evolution then tidal spin-up during the BH-WR stage has to be responsible 
for a sub-population of BBH mergers in which the secondary BH is formed with the maximum spin ($a=1$) and the effective 
spin $\chi_{\rm eff}$ is positive \citep[][]{Hotokezaka2017a,Hotokezaka2017b,Zaldarriaga2018,Piran2018}. 
The recently announced discovery of GW190412 \citep{LVK_GW190412}, a BBH merger with $\chi_{\rm eff} = 0.17-0.59$
or $\chi_{\rm eff} = 0.64-0.99$, depending on the choice of the priors \citep{Mandel2020}, might be the first representative of this sub-population.

However, if the duration of the BH-WR stage $t_{\rm BH-WR}$ at low metallicity is typically very short (several thousand 
years), as suggested by the considerations above, then the BH-WR binary needs to be extremely compact for any significant
spin-up of the WR star. Binary separation at the BH-WR stage can be related to the separation of the descendant BBH system,
which in turn is the main factor determining the delay time until the BBH merges due to GW emission. 
\citet{Kushnir2016} derived an approximate relation between the tidal synchronization timescale of a BH-WR binary and
the delay time of a corresponding BBH system $t_{\rm delay}$ (see their Eqn.~14). In Fig.~\ref{fig.disc_kushnir} we plot
this relation for several different mass ratios. We also mark two values of $t_{\rm BH-WR}$: the approximate CHeB duration
of massive stars, i.e. the maximum lifetime of a WR star, 
$\sim 3 \times 10^5 \rm \, yr$, as assumed by several authors \citep[eg.][]{Kushnir2016,Hotokezaka2017a,Piran2018}, and 
a significant smaller value $t_{\rm BH-WR} = 3 \times 10^3 \rm \, yr$ which is more realistic in the case of CE evolution at low metallicity.
Fig.~\ref{fig.disc_kushnir} shows that for short BH-WR lifetimes of several thousand years only BBH mergers with very short
delay times ($\lesssim 30 \rm \, Myr$) can have the spin of the secondary BH affected by tidal spin-up. 
It is currently unknown whether BBH mergers with delay times this short can form in the CE evolution channel.
If such systems are non-existent or very rare then our results indicate that tidal spin-up during the BH-WR stage is unlikely
to have any significant effect on the $\chi_{\rm eff}$ distribution of local BBH mergers, especially these formed at low metallicity.

\subsection{CE ejection in BH-LMXB progenitors}

\label{sec.disc_BHLMXB}

Most of the known Galactic BHs reside in BH-LMXBs in which the companion is typically 
a MS star with $M \lesssim 1 \msun$ and the orbital period is very short (usually $< 1$ day) \citep{Casares2014,Tetarenko2016}. 
The formation of such systems has been a long-standing puzzle. Given the extreme mass ratio and small separation of BH-LMXBs,
their progenitors have most likely evolved through a CE phase. However, it was argued by several authors on the basis of energy 
budget considerations that a secondary of only $\lesssim 1.5 \msun$ would not be able to provide enough energy to unbind the 
envelope of a massive BH progenitor  (\citealt{PortegiesZwart1997,Kalogera1999,Podsiadlowski2003,Justham2006,Yungelson2008}, 
although see \citealt{Yungelson2006}). This led to a conjecture of an extremely high CE efficiency, i.e. a possibility of
$\alpha_{\rm CE}$ larger than 1 (even as high as a few tens in the case of BH-LMXBs). This apparent violation of energy conservation
could be interpreted as some unknown mechanism, unaccounted for in the standard energy budget \citep[e.g.][]{Podsiadlowski2010}. 
The assumption of $\alpha_{\rm CE}$ larger than 1 has made its way also to population synthesis models of compact binary mergers \citep[e.g.][]{Giacobbo2018b,Evans2020}.

\begin{figure}
      \includegraphics[width=\columnwidth]{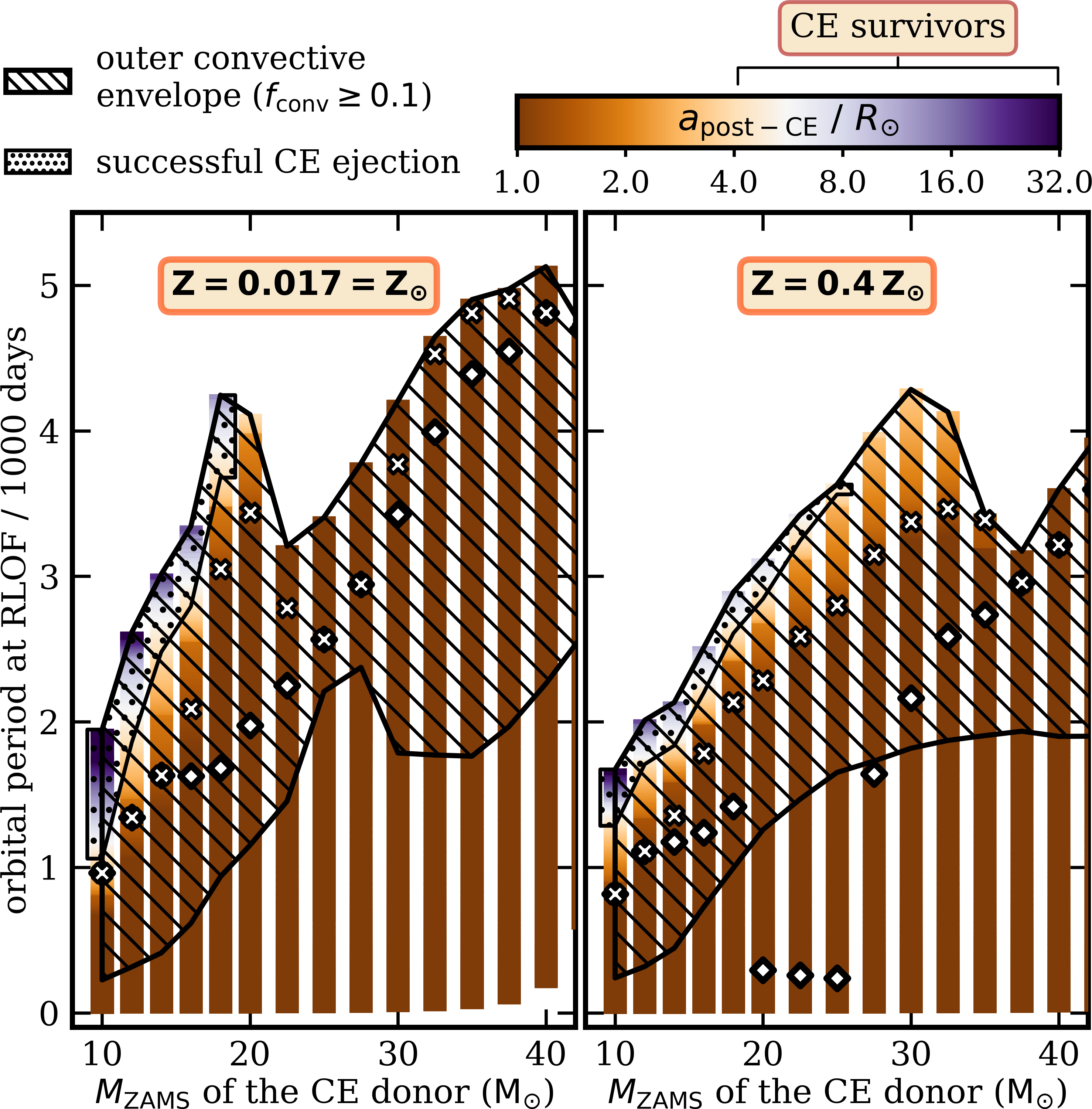}
    \caption{CE ejectability in binaries with a massive donor ($M_{\rm ZAMS} = 10-40 \msun$) and a low-mass secondary of $1.0 \msun$,
    shown for two metallicites that are the most relevant for the Milky Way. Notation same as in Fig.~\ref{fig.sep_6}, except that
    the Y-axis shows orbital period instead of the donor radius and the color indicates the post-CE orbital separation $a_{\rm post-CE}$ 
    estimated from the energy budget (Sec.~\ref{sec.method_energy_budget}) without the accretion term $\Delta E_{\rm acc}$. 
    Separations $a_{\rm post-CE} \approx 4-10 \rsun$ are in line with the formation of the observed population of short-period
    BH-LMXBs \citep{Kalogera1999,Repetto2015}.}
    \label{fig.BHLMXB}
\end{figure}

Interestingly, the binding energies of convective-envelope supergiants at advanced evolutionary stages (Fig.~\ref{fig.Ebind_6}) 
described in detail in Sec.~\ref{sec:res_details} are low enough to allow for a CE ejection even in the case of low-mass secondaries,
without the need for $\alpha_{\rm CE} > 1$. Fig.~\ref{fig.BHLMXB}, similar to Fig.~\ref{fig.sep_6}, shows CE ejectability in 
binaries with massive CE donors ($M_{\rm ZAMS} = 10-40 \msun$) in which the companion is a $1.0 \msun$ star, i.e. potential 
BH-LMXB progenitors (depending on the minimum mass required for the formation of a BH). Post-CE separations $a_{\rm post-CE}$,
color-coded in the figure, follow from the energy budget described in Sec.~\ref{sec.method_energy_budget} without the energy 
term from accretion (due to stellar nature of the secondary). Separations within $a_{\rm post-CE} \approx 4-10 \rsun$ agree well 
with the estimated separations of short-period BH-LMXB progenitors at the moment of the BH formation \citep{Kalogera1999,Repetto2015}.
Note that at solar metallicity (the left panel) the decrease in envelope binding energy of convective-envelope giants is quenched for
masses $M_{\rm ZAMS} \gtrsim 20 \msun$ due to mass loss in our models \citep[see also ][and Fig.~\ref{fig.Ebind_6}]{Podsiadlowski2003}.
If $M_{\rm ZAMS} \approx  20\msun$ is the minimum mass required for the formation of a BH then none of the survivors in the left
panel of Fig.~\ref{fig.BHLMXB} could be progenitors of BH-LMXBs. However, at the lower metallicity of $Z = 0.4 \zsun = 0.0068$ the
range of survivors extends to $M_{\rm ZAMS} = 25\msun$ (the right panel). Given that the average metallicity of the Milky Way stars 
is subsolar \citep[e.g. $Z = 0.0088$ with the assumptions taken by][]{Brott2011}, Fig.~\ref{fig.BHLMXB} suggests that at least some
of the Galactic BH-LMXBs could be formed through CE evolution without $\alpha_{\rm CE}$ exceeding $1$. Even more so, as pointed out 
by \citet{Yungelson2006}, because mass-loss rates from massive stars are uncertain, it is possible that for a different set of 
mass-loss assumptions the range of CE survivors at the Solar metallicity would extend to $M_{\rm ZAMS} > 20 \msun$ as well \citep{Smith2014,Renzo2017}. 
Finally, there might be islands of non-explodability in the lower ZAMS mass range where some progenitors with $M_{\rm ZAMS} < 20 \msun$ 
end their lives in failed explosions and produce BHs as well \citep{OConnor2011,Ugliano2012,Ertl2016,Ertl2020,Couch2020,Patton2020}.

A separate question is whether the parameter space for CE ejection in Fig.~\ref{fig.BHLMXB} is large enough to produce the entire
Galactic population of BH-LMXBs, something that population synthesis models have been struggling with \citep[e.g.][]{Wiktorowicz2014}.
Notably other formation channels for BH-LMXBs have also been suggested, for example intermediate-mass X-ray binaries with magnetic donors \citep{Justham2006}, 
formation in hierarchical triples \citep{Naoz2016}, or dynamical formation through tidal capture in the field \citep{Michaely2016,Klencki2017}.

\subsection{Alternative scenario: partial envelope ejection followed by stable mass transfer}

\label{sec.disc_composite_scenario}

It is well-known that binding energy of a giant star is very sensitive to the exact location 
of core-envelope boundary for CE ejection, i.e. the bifurcation point 
\citep[][]{Dewi2000,Tauris2001,Ivanova2011b,deMarco2011,Kruckow2016}.
This is because most of the binding energy is stored is deep envelope layers, close to the helium core 
(see Fig.~\ref{fig.Ebind_detail_16_01_e} and Fig.~\ref{fig.Ebind_detail_16_04_e} as well as Fig.~3 in \citealt{Kruckow2016}).
As a result, moving the bifurcation point upwards by even a few percent of the total mass of a giant may reduce the envelope 
binding energy by an order of magnitude or even more. Here, following \citet{Ivanova2011b}, the bifurcation point is assumed
to be the maximum compression point $M_{\rm cp}$ within the H-burning shell (i.e. a local maximum of $P/\rho$), which in massive
evolved stars corresponds very well to the point of steep composition gradient at the hydrogen abundance $X_{\rm H} \sim 0.05-0.15$.
This is a standard assumption made across the recent literature \citep{Xu2010,Loveridge2011,Wang2016,Kruckow2016}.

Interestingly, in the envelope structure of a $\sim 35\msun$ convective-giant at $Z = 0.4\zsun$ in Fig.~\ref{fig.Ebind_detail_16_04}
(originally $M_{\rm ZAMS} = 47.5 \msun$) layers above the bifurcation point, located at $M_{\rm cp} \approx 23 \msun$, are still 
compressed at very compact radial coordinates $R < 5 \rsun$, see Fig.~\ref{fig.Ebind_detail_16_04_c}. These layers are
helium-rich ($X_{\rm He} \approx 0.75$) and extend out to mass coordinate $M \approx 26 \msun$ at which point a steep density
and composition gradient separates them from the loosely bound outer convective zone. 
As the model evolves from yellow to red profiles in Fig.~\ref{fig.Ebind_detail_16_04_c}, when a significant amount of mass
($\sim 10\msun$) is being lost in winds and the envelope becomes increasingly convective, the entire inner part of the star
out to $M \approx 26 \msun$ contracts slightly while the layers above are expanding.
We find that such an envelope structure is characteristic of massive giants that become convective as a result of a rapid 
expansion during the HG phase. It is distinctively different from the structure of low-mass red giants or massive giants 
that develop outer convective zones at a more advanced evolutionary stage (e.g. Fig.~\ref{fig.Ebind_detail_16_01}), in which
the convective zone can extend deep down close to the helium core and the maximum compression point is also a clear divergence
point between the expanding and contracting parts of the star.

Given the compact size of layers above the helium core in the model from  Fig.~\ref{fig.Ebind_detail_16_04}, it is possible 
to imagine a partial envelope ejection in which only the extended and loosely bound outer part of the envelop is ejected during
the CE phase (at mass coordinates $> 26\msun$). This would imply a bifurcation point located $\sim 3 \msun$ above the standard
core-envelope boundary and a much lower envelope binding energy. 
For such a partial envelope ejection during a CE inspiral to be possible one likely requires the helium-rich layers at the 
bottom of the envelope to remain compact during mass loss from the outer envelope layers taking place on the short CE timescale.
Whether or not that it is likely to happen needs to be confirmed with detailed stellar models. The first step has recently 
been carried out by \citet{Fragos2019}, who in their hydrodynamic 1-D model of a NS inspiral inside the envelope of a $12 \msun$ giant found evidence for a CE ejection 
at a point when some of the envelope layers ($\sim 0.3-0.5 \msun$ with $X_{\rm He} \approx 0.7$ ) still remain on top of the helium core.

A separate but important question is what would happen with the system right after such a CE phase. One possibility is that the 
partially-stripped donor star would re-expand again on the thermal timescale ($\sim 100-1000$ yr) leading to RLOF and another phase 
of mass transfer, as also discussed by \citet{Kruckow2016}. Such a behaviour was found already by \citet{Ivanova2011b} in models
of stripped giants with remaining masses above the maximum compression point.
Interestingly, \citet{Quast2019} have recently shown that mass transfer from a massive giant that has been stripped down to its 
deep helium-enriched layers could be stable for significantly unequal mass ratios and proceed on a long nuclear timescale of 
core-helium burning ($\sim 10^5$ yr) with a super-Eddington mass transfer rate.
Taken together, these results support the idea of an alternative scenario to the classical evolution through a CE phase:
a scenario in which a partial envelope ejection during the CE phase is followed by a long-lasting phase of stable mass transfer 
from a helium-enriched star, likely a blue supergiant, during which the system appears as a ULX.

It should be noted that the high mass transfer stability in models by \citet{Quast2019} was obtained only if the giant was to 
lose most but not exactly all of its envelope during the CE phase ($\gtrsim 90\%$ in mass), at which point the helium-rich layers 
close to the helium core would be brought to the surface. At a first glance, this suggests that a certain level of fine-tuning is
required. On the other hand, \citet{Quast2019} only explored synthetic models with simple internal abundance profiles. In real 
supergiants, depending on the size of the ICZ above the H-burning shell during the previous evolution, 
helium-rich layers ($X_{\rm He} \sim 0.5-0.7$) might extend to layers located significantly higher up inside the envelope 
(e.g. Figs~\ref{fig.Ebind_detail_16_01_d} and \ref{fig.Ebind_detail_16_04_d}). This could also be the case for accretors 
from a past mass transfer phase, especially if a significant amount of He-rich material was accreted. Such a structure could possibly support stable
mass transfer evolution also when a different, smaller part of the H-rich envelope is ejected in the CE phase, especially at 
lower metallicity where stripped stars are relatively more compact \citep{Gotberg2017,Gotberg2018,Klencki2019,Laplace2020}.

The final separation of a binary system once both a partial envelope ejection and a further mass transfer phase are concluded 
depends on the mass of the compact accretor, how conservative the second stable mass transfer is, and how much angular momentum 
is carried away in the non-accreted matter. For instance, \citet{Fragos2019} estimated that a potential post-CE 
mass transfer phase would further decrease the separation in their binary of a $1.4 \msun$ NS and a $3.2 \msun$ stripped donor
by a factor of $\sim 1.5$. 
This, combined with the orbital shrinkage during the CE phase, can be translated into a single effective $\alpha_{\rm CE}$ value. 
In the case of the system computed by \citet{Fragos2019}, they estimated an effective ratio
$\Delta E_{\rm grav} / \Delta E_{\rm orb} = \alpha_{\rm grav} \approx 5$, which roughly translates to $\alpha_{\rm CE} \approx 2$, 
essentially indicating an easier CE ejection and the binary survival compared to the classical picture of CE evolution in which the entire envelope is ejected straight away.
\footnote{Note that $\alpha_{\rm CE}$ from the notation of \citet{Fragos2019} is the same as $\alpha_{\rm grav}$ in ours.}
In the case of more massive accretors, e.g. typical stellar BHs, the orbital separation would likely decrease less or even 
increase as a result of a more equal mass ratio during the stable mass transfer phase, thus producing intermediate rather than 
short period systems. The fact that some of the recently ejected matter from the CE phase would likely still remain in the proximity
of the system (for example as a circumbinary disk) makes predictions for the evolution of orbital parameters rather uncertain.
The alleged non-interacting BH binary system with orbital period of $\sim 83$ days and a companion $\sim 3 \msun$ star discovered 
by \citet[][although see \citealt{vdHeuvel2020,Thompson2020}]{Thompson2019} might be a product of such an evolution if a CE phase was initiated by the BH massive progenitor.

In summary, a different location of the bifurcation point in supergiant donors would result in lower envelope binding energies 
and could in principle allow for CE ejections even in the case of non-convective donors. It would also lead to another phase of 
mass transfer after the end of the CE phase. It seems unlikely that close (merging) BBH systems could be the final product of
such an evolution but if they were then the corresponding BBH merger rate could possibly be linked to the number density of
observed ULX sources \citep{Inoue2016,Finke2017,Klencki2019,Mondal2020}.

\section{Conclusions}
\label{sec:conclusions}

We studied what kind of binaries with BH or NS accretors and supergiant donors could realistically be expected to
evolve through and survive a CE phase, thus making them potential progenitors of compact binary mergers. In pursue 
of the most optimistic case, we assumed an extreme version of the CE energy budget, in which all the energy sources
are fully efficient (orbital shrinkage, internal energy, recombination energy, energy from accretion), energy sinks
are neglected (e.g. no energy loss in radiation), and the envelope acceleration is perfectly fine-tuned to the local 
escape velocity. We computed envelope binding energies from detailed stellar models of massive stars ($M_{\rm ZAMS}$ 
between $10$ and $80 \msun$) at 6 different metallicities from $Z = 0.017$ to $0.00017$ and taking 
$X_{\rm H} = 0.1$ criteria for the bifurcation point.
We assumed that the CE evolution occurs in BH/NS binaries with mass ratios $q = M_{\rm donor}/M_{\rm accretor}$ larger
than $q_{\rm crit;rad} = 3.5$ for radiative-envelope donors and $q_{\rm crit;conv} = 1.5$ for convective-envelope donors. 
This choice might be optimistic in view of the recently found increased stability of mass transfer from massive giant 
donors \citep{Ge2015,Pavlovskii2017}. Our findings are summarized below.

 $-$ Envelope binding energy of a supergiant depends very strongly not only on the envelope type (convective vs radiative)
 but also on the evolutionary stage, which in turn is related to metallicity. The binding energy can decrease very 
 significantly (by more then an order of magnitude) when the envelope becomes convective during core-helium burning or 
 at a later evolutionary stage. However, a transition from a radiative to a convective state in an envelope of a rapidly
 expanding HG giant does not affect the binding energy nearly as significantly. As a result, through its impact on the 
 radial expansion of massive giants, metallicity could have an indirect but very strong effect on the binding energies of 
 convective-envelope giants and may be a crucial factor determining the fraction of systems surviving the CE phase.
 This conclusion depends on presence of the ICZ during the HG phase.
 In App.~\ref{app:fits} we provide fits to $\lambda_{\rm CE}$ values which can be used in population synthesis calculations. 
 
$-$ Survival of the CE phase in BH/NS binaries is strongly limited to cases in which the donor star is a
convective-envelope red supergiant. This is the case even under several optimistic assumptions working in favor 
of an easier CE ejection. ULXs with RSG donors might be the immediate progenitors of such CE events and, later on, compact binary mergers.

$-$ If stars above $\sim40 \msun$ never expand to the RSG stage, as suggested by the empirical upper luminosity
limit of RSGs at $\logL_{\rm RSG;max} \approx 5.7$, then our result indicates an upper limit on the 
total BBH merger mass at about $\sim 50 \msun$ and the primary BH mass at $\sim 25$-$30 \msun$. 
This would be in tension with the CE channel being the origin of about half of the BBH mergers detected to date.

$-$ Merger rates of compact binaries formed through the CE evolution might be severely overestimated in some 
of the recent population synthesis models due to (a) extrapolation of $\lambda_{\rm CE}$ values fitted to
$M_{\rm ZAMS} \leq 20\msun$ models to obtain binding energies of much more massive stars (with $M_{\rm ZAMS}$ 
as high as $100-150\msun$) and (b) a practice of treating all core-helium burning stars as convective-envelope 
giants. The latter issue is likely partially responsible for the significantly higher formation rate of BBH mergers
at low metallicity in said models. 

$-$ Lifetimes of close BH-WR systems (i.e. CE survivors) formed at low metallicity are typically very short
(several thousand years), especially at $Z \leq 0.1 \zsun$. Only those with smallest orbital separations, 
producing BBH mergers with short delay times ($\lesssim 30$ Myr), are likely to experience any significant tidal spin-up of the WR star.

$-$ Models at $M_{\rm ZAMS} \lesssim $ 20$-$40$\msun$ (depending on $Z$) remain RSGs at advanced evolutionary 
stages after the end of core-helium burning. Their envelopes can become deeply convective, leading to exceptionally
small binding energies ($\lambda_{\rm CE} \sim 1.0$). For such donors, CE ejections appears energetically possible even
for low-mass ($1 \msun$) MS accretors, making them potential progenitors of BH-LMXBs without the need for $\alpha_{\rm CE} > 1.0$.

$-$ Envelopes of more massive stars ($M_{\rm ZAMS} \gtrsim 40\msun$), even if they reach the RSG stage, 
never become fully convective as their $M_{\rm conv}/M_{\rm env}$ ratio does not exceed $\sim 0.7$. In some models
layers above the helium core remain very compact ($< 5 \rsun$) all the way up to a steep density gradient near the bottom 
of the convective envelope, which revives the question of core-envelope boundary in CE evolution. Eventual partial envelope 
ejection during the CE phase would likely lead to another, possibly stable and long-lasting phase of mass transfer.

$-$ Because of its influence on the radial expansion of massive giants, metallicity affects the degree of internal 
mixing in deep envelope layers of supergiants. As a result, two RSGs similar in terms of mass, radius, and luminosity 
but of different metallicities can have very different chemical profiles, density structures, and envelope binding energies.
This might need to be taken into account when constructing RSG models for an input to hydrodynamic simulations of the CE phase. 

\begin{acknowledgements}
We would like to thank the anonymous referee for a very thorough review and multiple useful comments and suggestions 
that helped to improve this work. 
The authors acknowledge support from the Netherlands Organisation for Scientific Research (NWO). It is a pleasure to acknowledge 
valuable discussions and suggestions from Alejandro Vigna-G{\'o}omez, Natasha Ivanova, Nadia Blagorodnova, Morgan MacLeod, 
Onno Pols, Kristhell L{\'o}pez, Marianne Heida, Karel Temmink, Glenn-Michael Oomen, Raphael Hirschi, Stephen Justham, Selma de Mink, 
and Ilya Mandel.
\end{acknowledgements}

\bibliographystyle{aa}
\bibliography{ULX_bib.bib}

 \begin{appendix}
 
 \section{Common envelope binding energy parameter fits}
\label{app:fits} 

\begin{figure}
      \includegraphics[width=\columnwidth]{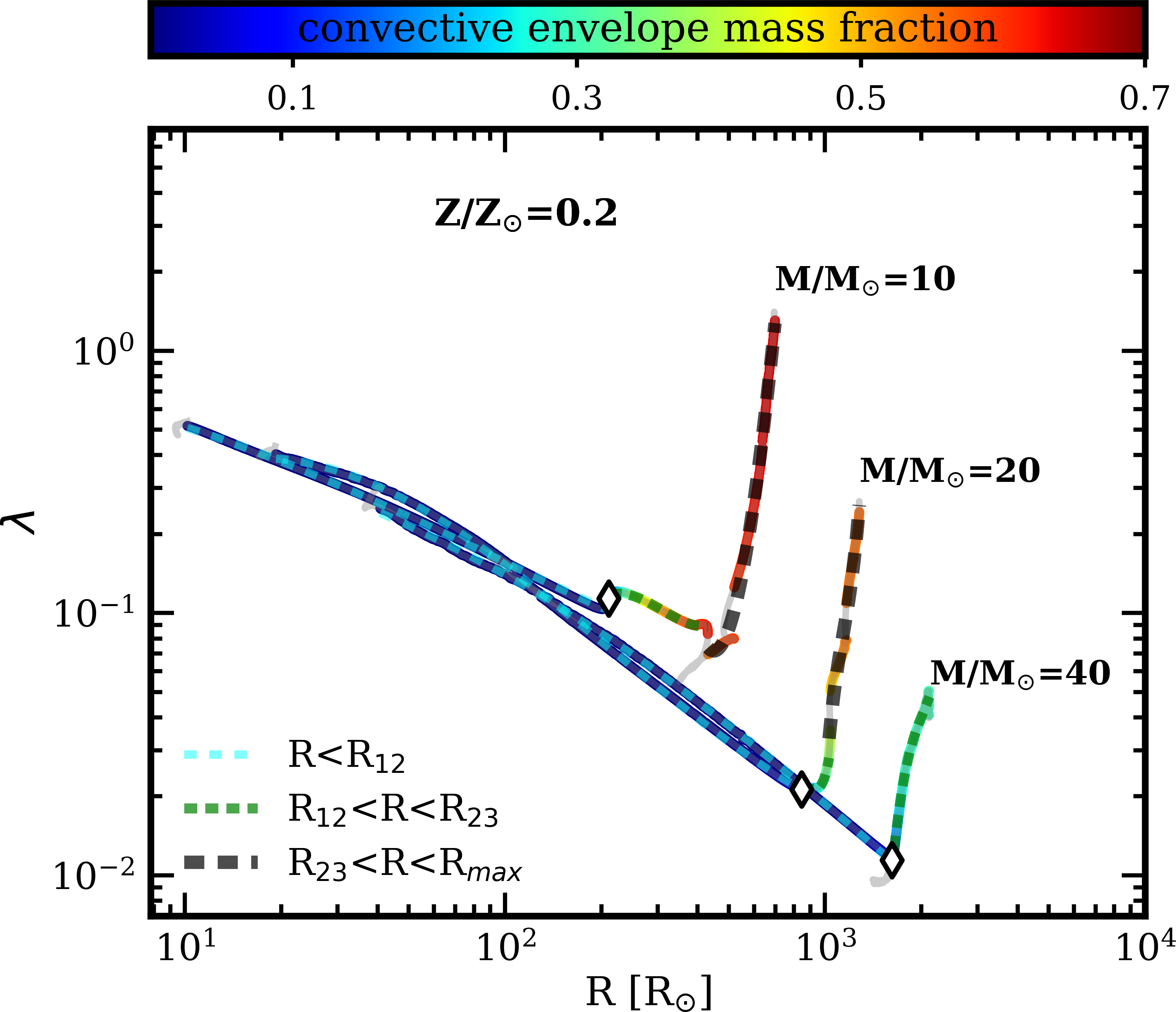}
    \caption{Polynomial fits to CE binding energy parameter ($\lambda$)
    as a function of radius shown for three example stellar models.
    The dashed lines in different colors indicate different ranges
    in which the fits have been made. The cyan diamond indicates
    the radius at which at least 10\% of the mass of the envelope
    becomes convective, separating the first and the second
    fitted ranges. The background lines are colored according to the convective envelope mass fraction.
    The gray parts, where the radius
    was smaller than the maximum 
    radius reached during the previous evolution, were
    not included in the fit. }
    \label{fig.fit_example}
\end{figure}

To facilitate the use of our results for instance 
in population synthesis studies,
we provide fits to $\lambda$ parameter
that is commonly used to describe the binding energy of the envelope 
and relate it to the properties of the donor star in CE phase
(see Eq. \ref{eq: binding energy}).
We fit $\rm log_{10}\left(\lambda \right)$ as a function of the radius of the giant donor star $\rm log_{10}\left(R\right)$ with a third order polynomial
\begin{equation}
\rm log_{10}\left(\lambda \right) = a_{i} log_{10}\left(R/R_{\odot} \right)^{3} + b_{i} log_{10}\left(R/R_{\odot}  \right)^{2} + c_{i} log_{10}\left(R/R_{\odot}  \right) + d_{i}
\end{equation}
The fit is divided in three ranges in R: coefficients with i=1 are applicable for R$<\rm R_{12}$, those with i=2 describe
$\lambda$ between $\rm R_{12}<$R$<\rm R_{23}$ and those with i=3 describe $\lambda$ between $\rm R_{23}<$R$<\rm R_{max}$.
The fitted coefficients and ranges depend on the stellar mass and metallicity and are given for each of the stellar models 
considered in this study under this url: \url{https://ftp.science.ru.nl/astro/jklencki/}.
The example fit for 10, 20 and 40 $M_{\odot}$ models at Z=0.2 $Z_{\odot}$
is shown in Fig. \ref{fig.fit_example}.
Below $R_{12}$ (cyan diamond in Fig. \ref{fig.fit_example}), $\lambda$ decreases monotonically as a function of radius. 
The relation starts to bend around $R_{12}$, when the mass fraction in the convective envelope increases
(except for the highest stellar masses $\gtrsim 50 \ M_{\odot}$ at the highest metallicities considered in this study, 
see Sec. \ref{sec.res_Ebind}).
In our fits $R_{12}$ corresponds to the radius at which the effective temperature reaches the value below which at 
least 10\% of the outer mass in a star becomes convective (see Fig. 6 in \citealt{Klencki2020}).
The third range $\rm R_{23}<$R$<\rm R_{max}$ was introduced to improve the quality of the fit at the highest radii, 
where $\lambda$ rapidly increases. For the most massive donors the third range is not necessary (in those cases $\rm R_{23}=R_{max}$).
Only the parts where the radius of the star increases beyond the largest radius reached during its previous evolution
(relevant for the onset of the mass transfer and common envelope) are included in the fit.

\section{Numerical robustness of models}

\label{sec.App_CE_acc}

\subsection{The setup}

 \begin{figure*}[!h]
\centering
    \includegraphics[width=480px]{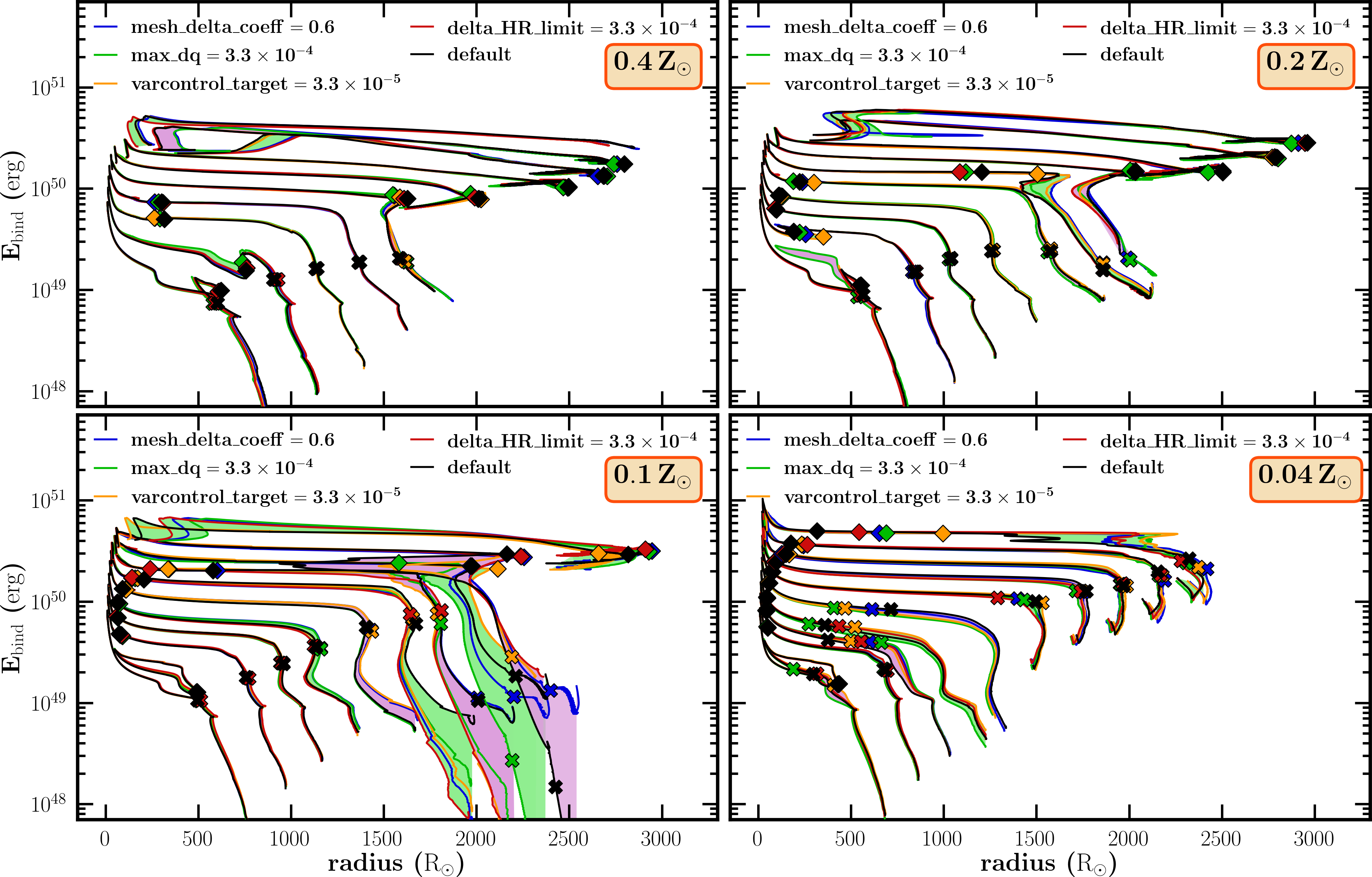}
    \caption{Comparison between the envelope binding energies $E_{\rm bind}$ of selected stellar models computed with the default assumptions (black line) and 
    in four model variations with increased spacial or temporal resolution (see text for details), plotted as a function of the stellar radius. Selected stellar masses 
    are $12$, $16$, $20$, $25$, $32.5$, $40$, $47.5$, $55$, $65$, and $80 \msun$. Each panels shows different metallicity. Diamonds mark the onset of core-helium burning, 
    crosses indicate central helium depletion. Shaded regions mark the area of uncertainty span by model variations,
    with interchanging colors to increase readability. Note that only the post-MS evolution is included in the figure. Differences between the models are likely mainly 
    due to envelope inflation and density inversions in the outer envelopes, see text for details. None of the numerical difficulties is affecting the conclusions of this study.}
    \label{fig.App_CEacc_Ebind}
\end{figure*}

In this section, we explore the numerical robustness of models from \citet{Klencki2020} (analyzed in this work) 
with respect to variations in spacial and temporal resolution. 
As an example of how an extended analysis of this kind could look like, we refer the reader to \citet{Farmer2016}.
Here, we mainly put emphasis on these model properties that are the most relevant 
in the context of CE evolution: the internal envelope structure and the envelope binding energy. 
MESA offers various controls to specify the mass (spacial) resolution of a model. One such parameter is \texttt{mesh\_delta\_coeff} which 
controls the global factor limiting the allowed change in stellar structure quantities between two adjacent cells. Lowering the value of 
\texttt{mesh\_delta\_coeff} leads to a higher number of cells (roughly by the same factor). The default value in MESA is $\texttt{mesh\_delta\_coeff} = 1.0$, 
whereas the default value for the models used in this work is $\texttt{mesh\_delta\_coeff} = 0.8$. 
Here, we also test $\texttt{mesh\_delta\_coeff} = 0.6$ (variation a).
Another parameter controlling mass resolution is \texttt{max\_dq}, which sets the maximum cell mass in units of the star's mass. 
The default MESA value is $\texttt{max\_dq} = 10^{-2}$, whereas the value used by \citet{Klencki2020} is $\texttt{max\_dq} = 10^{-3}$. 
Here, we also test $\texttt{max\_dq} = 3.3 \times 10^{-4}$ (variation b). 
Apart from the two mesh controls mentioned above,
models from \citet{Klencki2020} were computed with increased spacial resolution in regions with high H/He abundance gradients, 
see \url{https://zenodo.org/record/3740578#.X1s21nVfguw} for all the input MESA files.

As in the case of spacial resolution, MESA offers a large variety of parameters to control the timestep. One such parameter is 
\texttt{varcontrol\_target}, which controls the maximum allowed relative change in stellar structure quantities between two subsequent timesteps 
(if the change is greater then the timesteps needs to be reduced). The default value in MESA as well as in the models used in this work is 
$\texttt{varcontrol\_target} = 10^{-4}$. Here, we also explore a variation with $\texttt{varcontrol\_target} = 3.3 \times 10^{-5}$ (variation c). 
Another relevant parameter is \texttt{delta\_HR\_limit}, which limits the change in position in the HR diagram between two subsequent timesteps. 
This is particularly important during the phase of evolution right after the end of MS when a star is rapidly expanding on a short thermal timescale, 
the core is contracting, and a convective zone is typically forming above the hydrogen burning shell \citep[e.g.][]{Langer1985,Schootemeijer2019,Klencki2020}.
In default MESA this control is disabled, whereas the models used in this work take $\texttt{delta\_HR\_limit} = 10^{-3}$. Here, we also test the value 
$\texttt{delta\_HR\_limit} = 3.3 \times 10^{-4}$ (variation d).

To get a better idea of the numerical robustness of our models, we computed four variations a-d described above, two with increased spacial and two with increased temporal resolution. 
For each variation we computed stellar models at 10 different masses: $12$, $16$, $20$, $25$, $32.5$, $40$, $47.5$, $55$, $65$, and $80 \msun$. We explored 
four different metallicities: $0.4$, $0.2$, $0.1$, and $0.04 \zsun$ (with $\zsun = 0.017$). 

\subsection{Numerical robustness: envelope binding energies} 

\label{sec.App_CE_acc_Ebind}

In Fig.~\ref{fig.App_CEacc_Ebind}, similarly to Fig.~\ref{fig.Ebind_6}, we compare the envelope binding energies (plotted as a function of the stellar radius) 
in four numerical variations with the default models used throughout this study. We plot selected models at ten different masses and four different metallicities as listed above. 
Diamonds mark the onset of core-helium burning ($Y_{\rm C}$ drops below $0.975$), 
crosses indicate central helium depletion ($Y_{\rm C} < 0.001$). Shaded regions mark the area of uncertainty span by model variations,
with interchanging colors to increase readability. Note that only the post-MS evolution is included in the figure. For simplicity, the end of MS 
is determined based on the helium core mass becoming non-zero in MESA but it coincides quite well with a condition $X_{\rm C} < 10^{-6}$ 
(i.e. core-hydrogen depletion).

There are three ways in which the numerically-varied models are different from the default ones. First, there are clear discrepancies 
in the evolution of the most massive models right after the end of MS (except for the lowest metallicity $0.04 \zsun$). We find that these models 
are numerically not robust (not converged) in terms of their stellar radii at terminal-age MS (TAMS) and also during the final MS stages.
For instance, numerical variations in the three most massive models at $0.4 \zsun$ ($55 \msun$, $65 \msun$, and $80 \msun$) show changes in the TAMS radius 
as large as $100-200 \rsun$. Even larger is the numerical noice in the TAMS radius of the $80 \msun$ models at $0.1 \zsun$ (spread from $\sim 130$ to $550 \rsun$) 
and, especially, the $80 \msun$ models at $0.2 \zsun$ metallicity ($\sim 1000-1200 \rsun$ for models with increased spacial resolution, $\sim 550 \rsun$ otherwise).
The problems arise due to 
envelope inflation which occurs in these models during late stages of MS \citep[e.g.][]{Grafener2012,Sanyal2015,Sanyal2017}
and poses a well-known numerical challenge for 1D stellar codes of computing extremely diluted radiation-dominated envelopes with a density inversion 
(see Sec.~7 in \citealt{Paxton2013} and App.~A in \citealt{Klencki2020}). Notably, it was shown that in 3D simulations 
the envelopes do not become as inflated as in 1D stellar models \citep[i.e. the radius is smaller][]{Jiang2018}, which shows the limitations of 1D codes in 
computing the radii of very massive MS stars. In the context of the present study and CE ejectability, uncertainty in the final MS radii leads to uncertainty in the exact size and shape 
of the excluded MS-area in Fig.~\ref{fig.sep_6}. 

Second, Fig.~\ref{fig.App_CEacc_Ebind} shows significant differences between the stellar radii of some of the models at the RSG stage (typically the more massive ones), 
when the star is already expanded to its almost largest radius and the outer convective zone is growing. Once again, during this part of the evolution 
a density inversion is particularly large, which leads to numerical difficulties. This is also when the superadiabatic layer, where the convection becomes supersonic and 
the mixing-length theory is outside its region of applicability, is the most extended. As such, paraphrasing \citet{Paxton2013}, 
stellar radii of massive RSGs from any 1D stellar evolution calculations should be considered highly uncertain. 
Notably, much of the radius uncertainty is associated with the period of temporary contraction in models that reach the RSG stage before the onset of core-helium burning. In 
the context of binary evolution and a possibility of RLOF, this phase is not very relevant and does not affect out results. 
The more important is the uncertainty in the maximum radius reached during the phase of rapid HG expansion, i.e. just before
the above-mentioned occasional contraction. In this case the numerical noise is less consequential: the typical radius variations
are smaller than $5\%$ with an exception of the $55 \msun$ models at $0.1 \zsun$ ($\sim 10\%$ uncertainty).

Apart from discrepancies in the stellar radius, some of the models at $Z = 0.1 \zsun$ show large numerical noise 
in the envelope binding energy $E_{\rm bind}$ during late evolutionary stages, when the 
envelope becomes deeply convective and the value of $E_{\rm bind}$ is decreasing significantly. This is due to variations in the
exact location of the bifurcation point between these models 
(i.e. the boundary between the ejected envelope and the remaining remnant of the donor star during a CE phase), as we will demonstrate later in this Appendix. 
This means that while the prediction of very lower binding energies of deeply convective RSGs at advanced stages of core-helium burning appears robust (and so does
the prediction that CE ejection is possible for these donors), the exact values of $E_{\rm bind}$ of these stars loosely 
bound envelopes should be considered highly uncertain in our models. 

Third, as indicated by different locations of diamonds and crosses between some of the model variations in Fig.~\ref{fig.App_CEacc_Ebind}, stellar radius of a model during
a transitory phase (onset or end of core-helium burning) is in some cases not numerically robust in our models. This likely has to do with the fact that stellar radii 
of blue supergiants are very sensitive to the exact abundance profile in the bottom envelope layers surrounding the hydrogen burning shell
\citep[e.g.][]{Georgy2013,Klencki2020,Kaiser2020}. The abundance pattern in this part of the star is a result of internal mixing which takes place shortly after the 
end of MS and is notoriously difficult to compute in stellar models due to a abundance discontinuities (a step-like profile) left by semiconvection \citep{Farmer2016}.

It is important to stress that while the models used in this work are in some aspects not numerically robust, as revealed by Fig.~\ref{fig.App_CEacc_Ebind}, the overall 
behavior of the binding energy as a function of radius (and therefore the convective-envelope mass fraction $f_{\rm conv}$) is robust. 
In particular, numerical variations in the value of $E_{\rm bind}$ for radiative-envelope donors are never larger than $15 \%$ and typically below $10 \%$, 
which is rather insignificant in view of various physical uncertainties in the energy-budget formalism for CE evolution (see Sec.~\ref{sec.disc_uncert}).
The numerical uncertainty of 
our models does not affect any of the results presented in Sec.~\ref{sec.res_Ebind} and 
Sec.~\ref{sec.res_CEeject} (at least within the scope explored above).

\subsection{Numerical robustness: detailed internal structure}

\label{sec.App_CE_acc_prof}

 \begin{figure}
\centering
    \includegraphics[width=1.0\columnwidth]{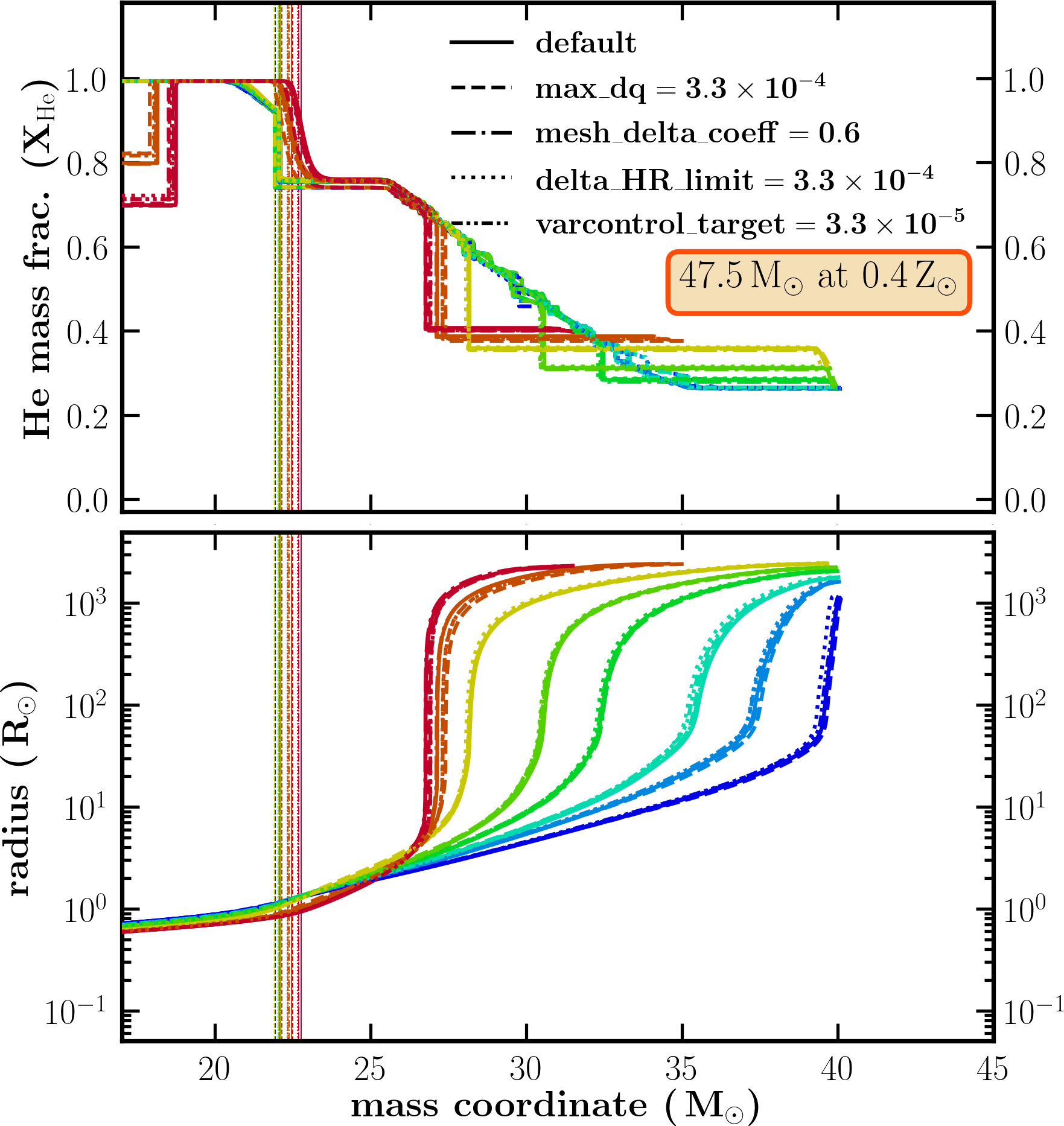}
    \caption{Comparison of the internal helium abundance and mass-radius profiles between the standard model (solid lines) and numerical variations with increased 
    resolution (other lines), plotted for the case with $M_{\rm ZAMS} = 47.5 \msun$ and $0.4 \zsun$ analyzed in Sec.~\ref{sec:res_details}. As in Fig.~\ref{fig.Ebind_detail_16_04}, 
    different colors correspond to different internal profiles of a model selected based on the depth of the outer convective zone (increasing as the color changes from blue to red).
    Vertical lines show the location of the bifurcation point for CE evolution. }
    \label{fig.App_CEacc_prof_04}
\end{figure}

 \begin{figure}
\centering
    \includegraphics[width=1.0\columnwidth]{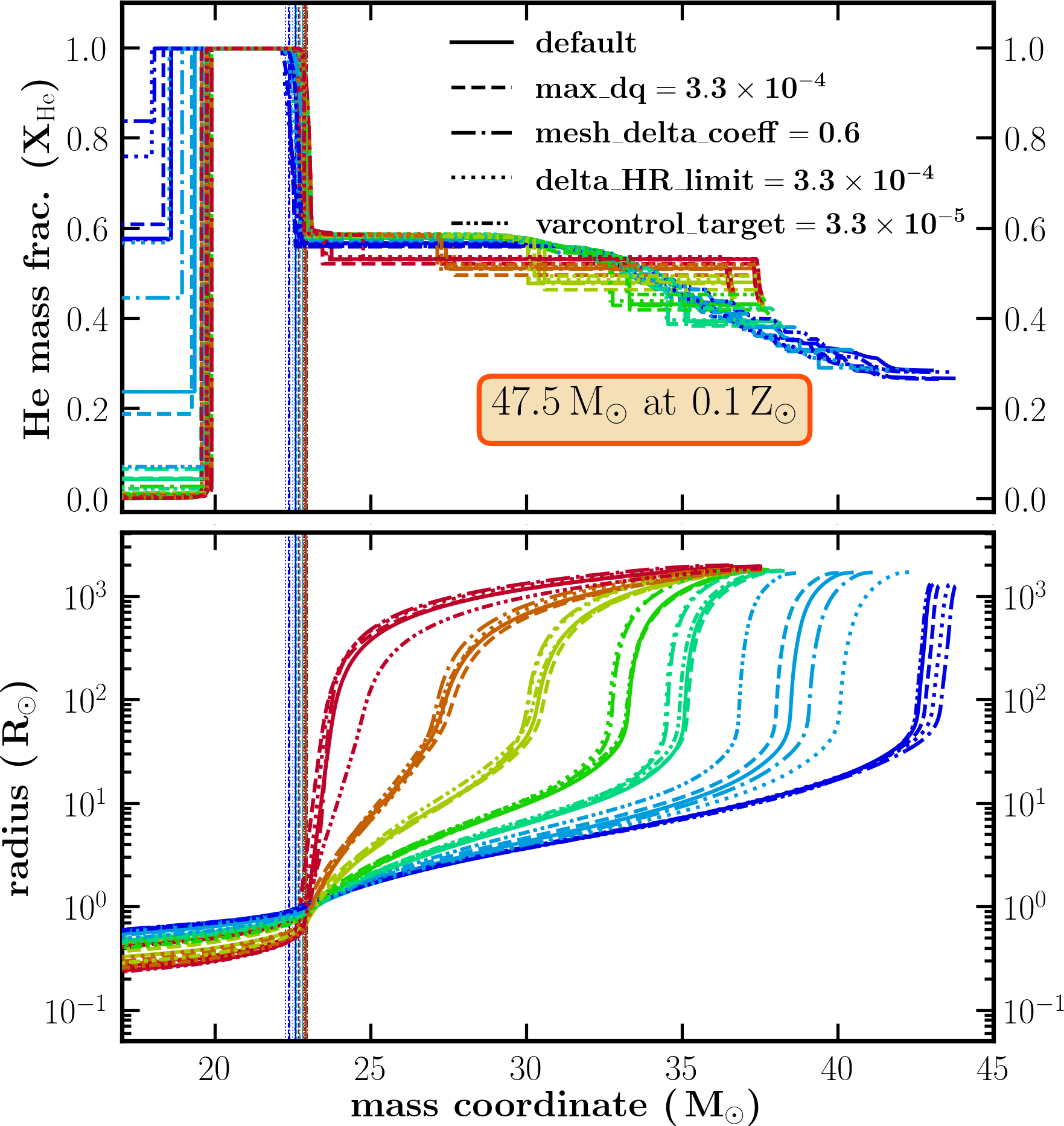}
    \caption{Same as Fig.~\ref{fig.App_CEacc_prof_04} but for a model at $Z = 0.1 \zsun$ metallicity (see also Fig.~\ref{fig.Ebind_detail_16_01} in the main text).}
    \label{fig.App_CEacc_prof_01}
\end{figure}

In Fig.~\ref{fig.App_CEacc_prof_01} and Fig.~\ref{fig.App_CEacc_prof_04} we take a look at numerical robustness of the internal structure 
of the two $M_{\rm ZAMS} = 47.5 \msun$ models studied in detail in Sec.~\ref{sec:res_details}. Similarly to Fig.~\ref{fig.Ebind_detail_16_01} and Fig.~\ref{fig.Ebind_detail_16_04}, with 
different colors we plot internal profiles selected based on their increasing depth of the outer convective zone $f_{\rm conv}$ (going from blue to red).
We compare the default model (solid line) with numerical variations, investigating the helium abundance and the mass-radius profile. 
Vertical lines mark the bifurcation point for CE evolution. 

None of the numerical variations in Fig.~\ref{fig.App_CEacc_prof_04} show any significant change with respect to our standard model. 
The largest variation in the bifurcation point location between models with different resolution is about $1.4 \%$ in mass and about $2 \%$ in radius.

The lower metallicity model $Z = 0.1 \zsun$ in Fig.~\ref{fig.App_CEacc_prof_01}, one the other hand, reveals two uncertainties. First, because of the uncertain 
radius of a core-helium burning blue/yellow supergiant for a given central helium abundance as discussed at the end of Sec.~\ref{sec.App_CE_acc_Ebind}, different models 
in Fig.~\ref{fig.App_CEacc_prof_01} reach the same outer convective zone depth at slightly different times. This is the most evident with the light-blue profiles, 
for which the total stellar mass is clearly different between model variations. The overall shape of each profile is unaffected. 

Second, in the case of internal profiles of very deeply convective envelopes (the orange and red lines), there are some differences in the slope of the mass-radius profile near 
the bifurcation point (especially the model with $\texttt{varcontrol\_target} = 3.3 \times 10^{-5}$ compared to all the others).
While these variations do not lead to any significant uncertainty in the mass coordinate of the bifurcation point (which does not vary by more 
than $1.5 \%$) they do result in the radial coordinate being up to $\sim 30\%$ larger in the variation with $\texttt{max\_dq} = 3.3 \times 10^{-4}$ ($0.9 \rsun$ 
instead of $0.7 \rsun$, roughly). This propagates to very noticeable differences in the 
late envelope binding energy evolution of some of the $Z = 0.1 \zsun$ models (see Fig.~\ref{fig.App_CEacc_Ebind}) when 
their envelopes become the most expanded, the most deeply convective, and the least gravitationally bound.

Most importantly, the overall shape of internal profiles in Fig.~\ref{fig.App_CEacc_prof_01} and Fig.~\ref{fig.App_CEacc_prof_04} is unaffected by an increase in resolution, 
and the results presented in Sec.~\ref{sec:res_details} are numerically robust.

\subsection{Numerical robustness: HR diagram at $Z = 0.2 \zsun$}

\begin{figure}
\centering
    \includegraphics[width=1.0\columnwidth]{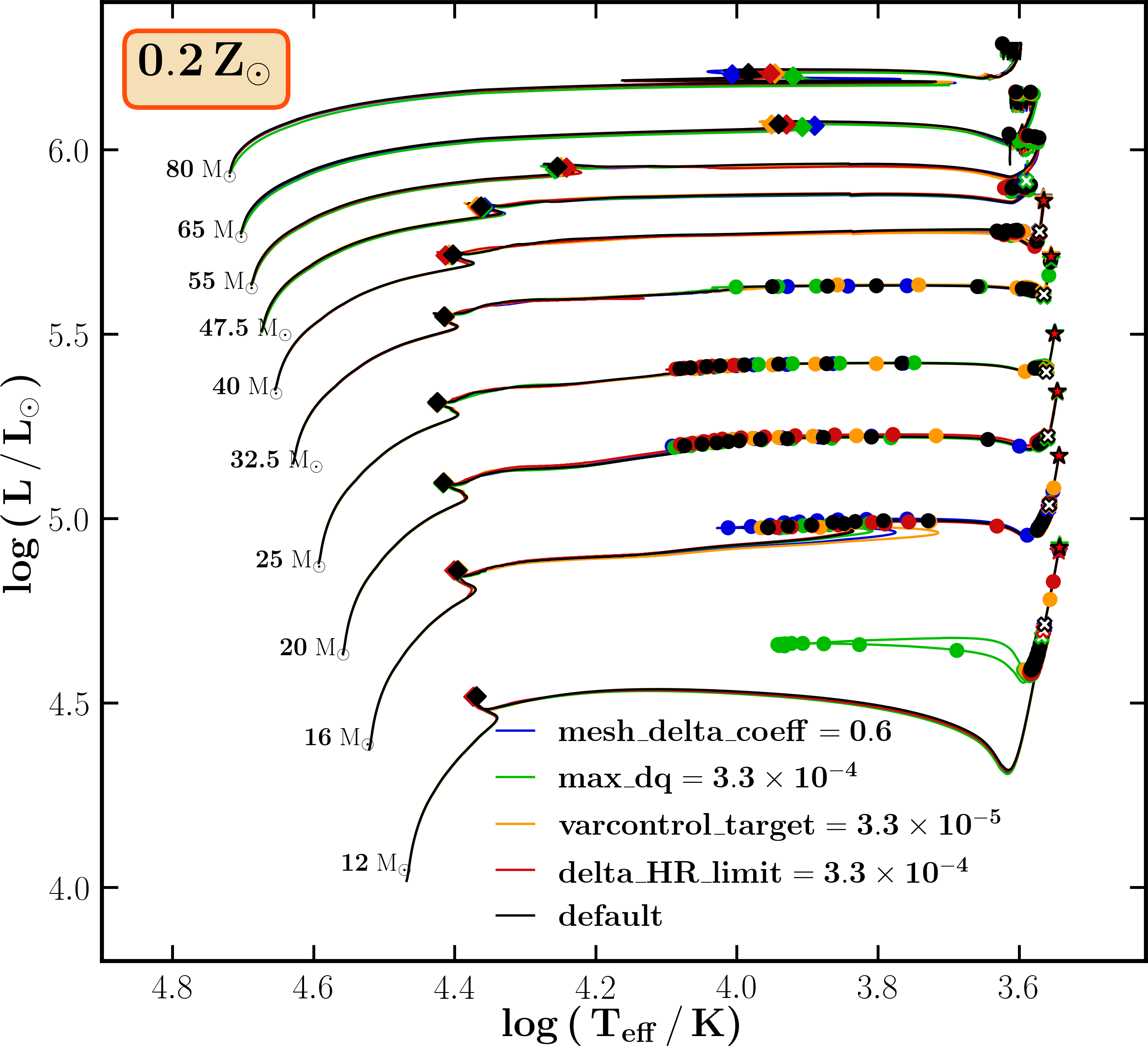}
    \caption{HR diagram with selected models at $0.2 \zsun$ metallicity (SMC-like). We compare
    models computed with increased resolution (various colors) to the standard models used in this work (in black). Diamonds mark the terminal-age MS, 
    dots mark the position of the star taken every 50,000 years during its post-MS evolution, white crossed mark the end of core-helium burning, 
    whereas red stars the end of central carbon burning. }
    \label{fig.App_CEacc_HRD}
\end{figure}

In Fig.~\ref{fig.App_CEacc_HRD}, similarly to Fig.~\ref{fig.HRD}, we plot an HR diagram with selected models at $0.2 \zsun$ metallicity (SMC-like). We compare
models computed with increased resolution (various colors) to the standard models used in this work (in black). See the figure caption for a description of the symbols. 
None of the numerical variations show very significant discrepancies with respect to the default model. One noticeable difference is in the location of the core-helium burning phase 
in the HR diagram (marked with circular dots). In particular, the $12 \msun$ model experiences a blue loop in a variation with $\texttt{max\_dq} = 3.3 \times 10^{-4}$ 
(increased minimal cell size). These differences do not affect the envelope binding energies and the results on CE ejectability. As discussed in Sec.~\ref{sec.App_CE_acc_Ebind}, 
numerical tests reveal that the sizes of MS stars with inflated envelopes are not a robust prediction of our models, showing variations for the most massive models in metallicities $Z \geq 0.1 \zsun$.
This can also be seen in Fig.~\ref{fig.App_CEacc_HRD} in the form of a chaotic behavior of the $80 \msun$ model in the final stages of the MS. 

\label{sec.App_CE_acc_HRD}

 \section{Additional figures}
 
 \label{sec:App_additional_figures}
 
 \begin{figure*}
\centering
    \includegraphics[width=0.8\textwidth]{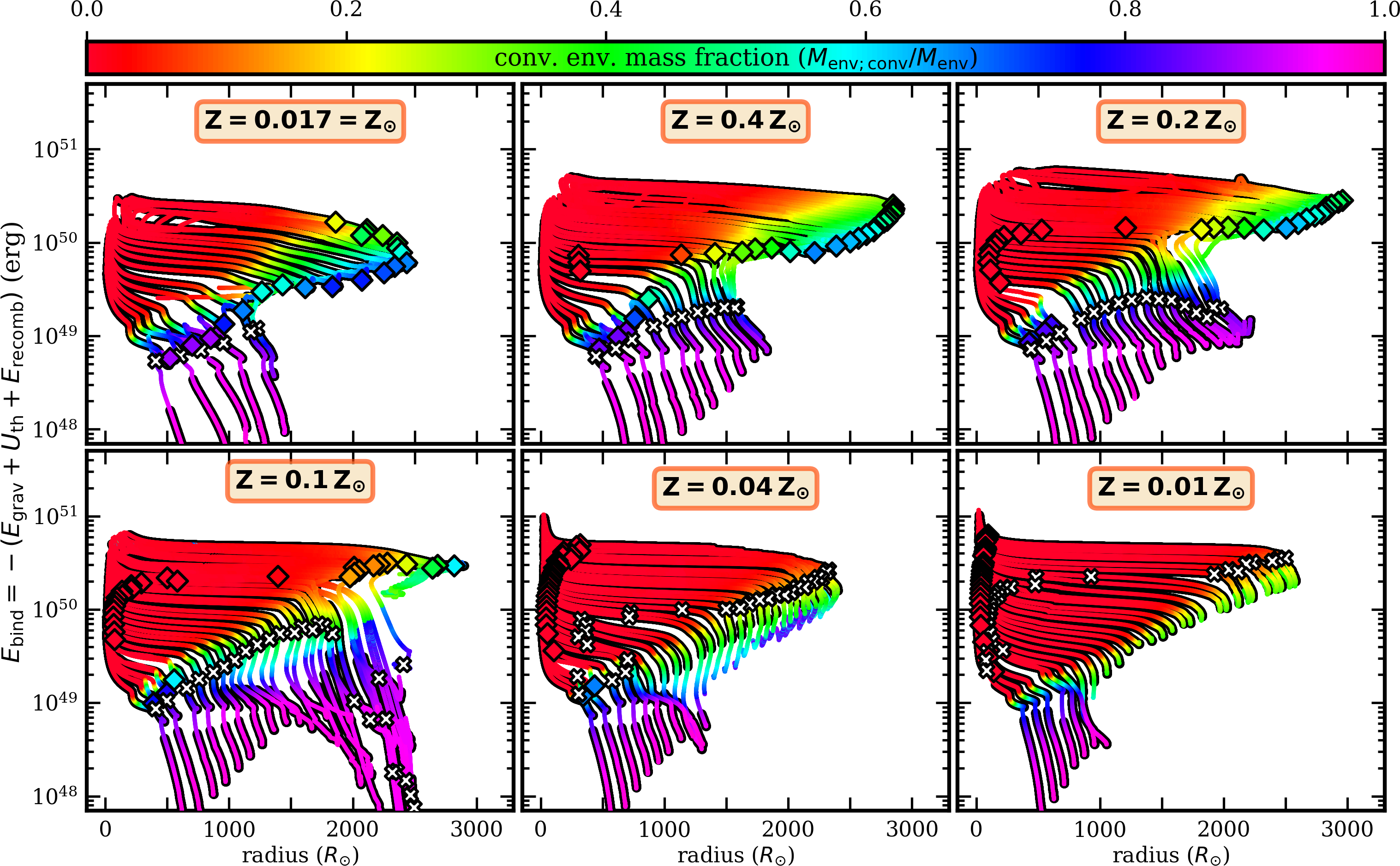}
    \caption{Same as Fig.~\ref{fig.Ebind_6} but with all the masses in our grid (30 models between $10$ and $80 \msun$ for each metallicity).}
    \label{fig.Ebind_6_all}
\end{figure*}

\begin{figure*}
\centering
    \includegraphics[width=0.8\textwidth]{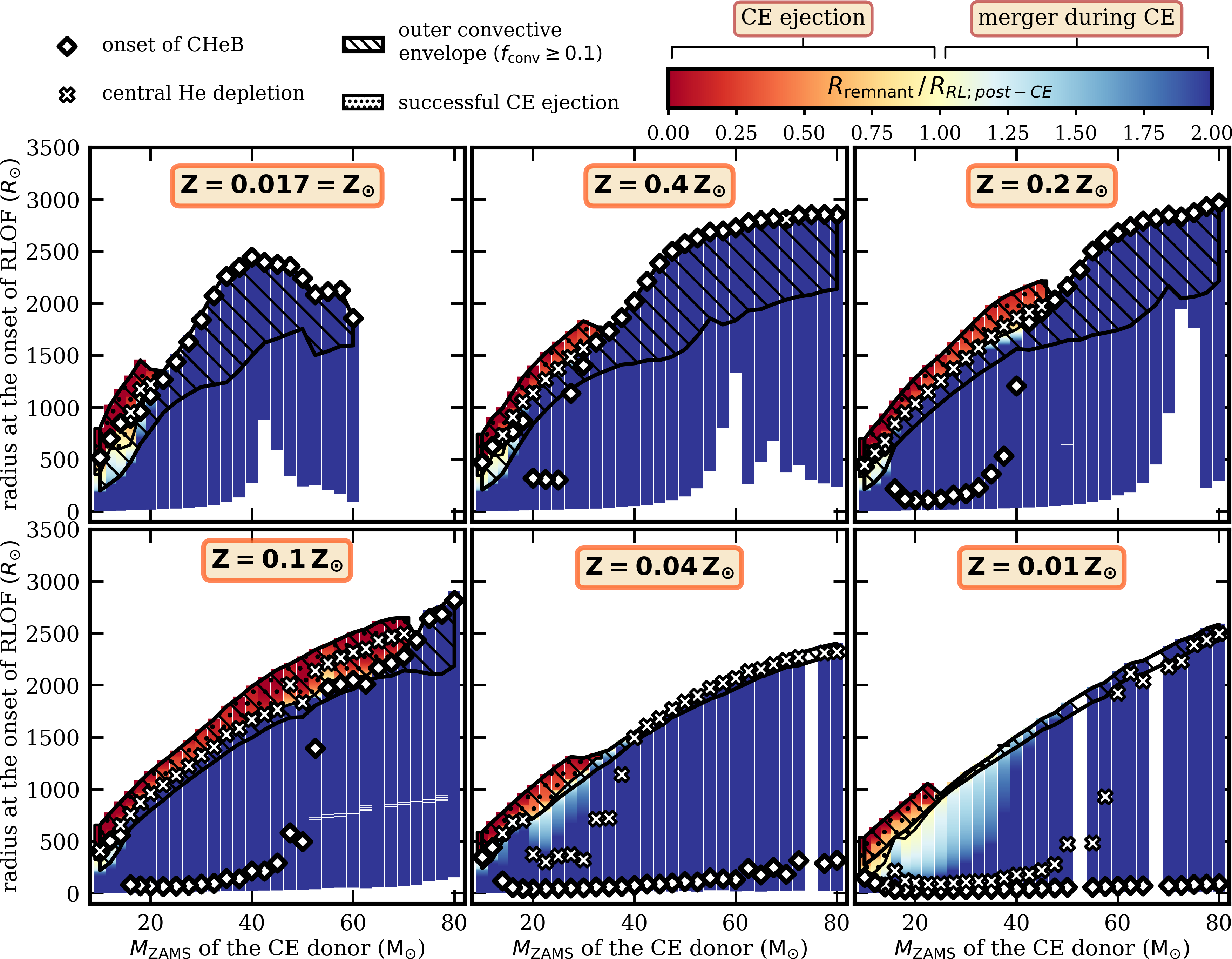}
    \caption{Same as Fig.~\ref{fig.sep_6} with the companion being a $2 \msun$ NS. The parameter space for CE ejections is 
    limited to cases in which the envelope binding energies have decreased significantly due to the donor becoming a convective-envelope giant at an advanced evolutionary stage.}
    \label{fig.sep_6_NS}
\end{figure*}

 \begin{figure*}
\centering
    \includegraphics[width=0.8\textwidth]{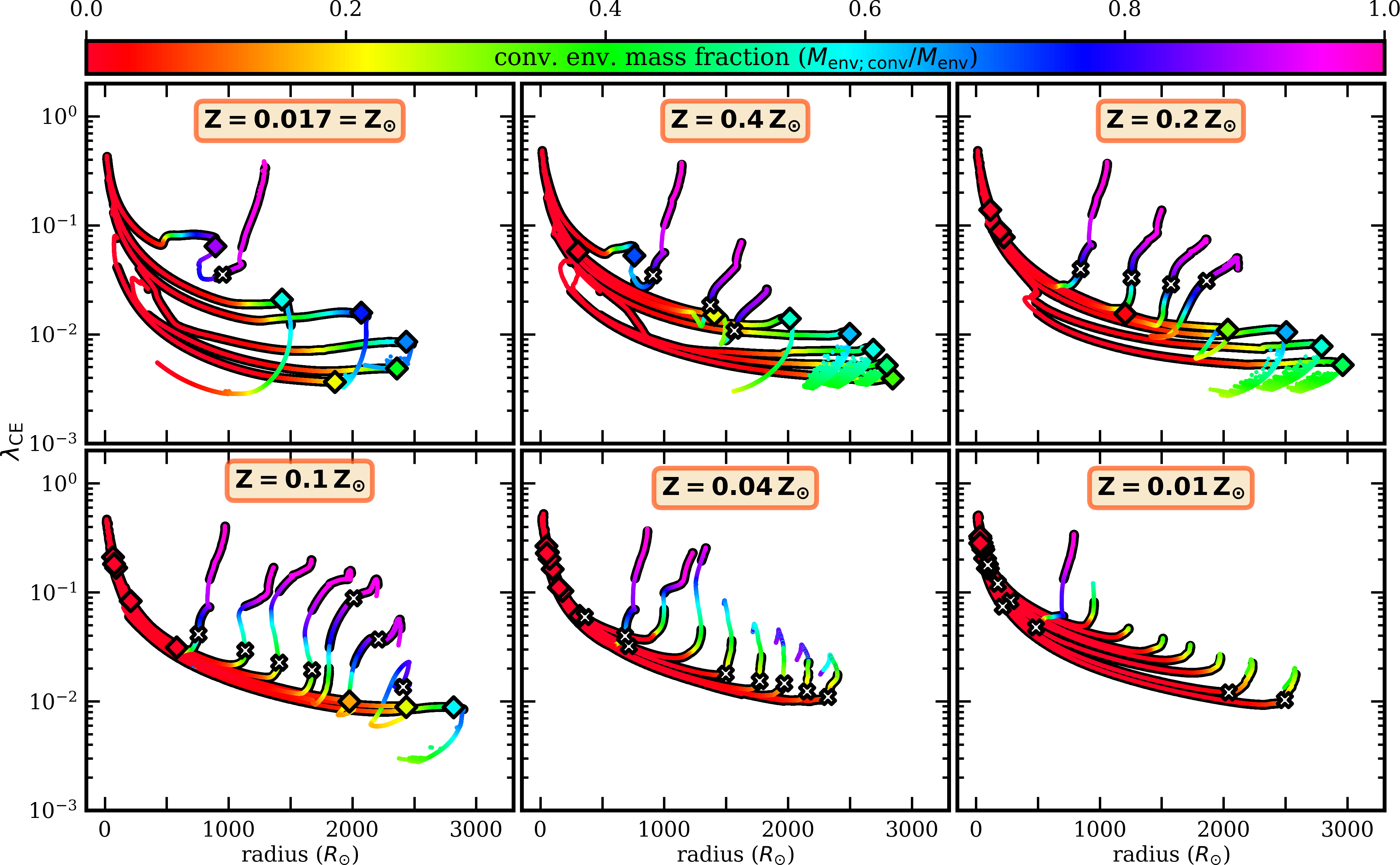}
    \caption{Same as Fig.~\ref{fig.Ebind_6}, but showing the evolution of  $\lambda_{\rm CE}$ parameter instead of the envelope binding energy. Diamonds indicate 
    the onset of core-helium burning, crosses indicate the end of core-helium burning.}
    \label{fig.App_lambda_6}
\end{figure*}

 \end{appendix}

\end{document}